\documentclass[10pt,twoside]{amundi_article}

\usepackage[table]{xcolor}
\usepackage{lmodern}
\usepackage{amssymb}
\usepackage{amsfonts}
\usepackage{amsmath}
\usepackage{amsbsy}
\usepackage{fancybox}
\usepackage{fancyhdr}
\usepackage{graphicx}
\usepackage[hyphens]{url}
\usepackage{arydshln}
\usepackage{multirow}
\usepackage{pdflscape}
\usepackage{tikz}
\usepackage{comment}
\usepackage{nicefrac}
\usepackage{dsfont}
\usepackage{algorithmic}
\usepackage{algorithm}
\usepackage{lipsum}

\newlength{\figurewidth}
\newlength{\figureheight}
\DeclareMathAlphabet\mathbfcal{OMS}{cmsy}{b}{n}
\DeclareFontFamily{OT1}{pzc}{}
\DeclareFontShape{OT1}{pzc}{m}{it}{<-> s * [1.3] pzcmi7t}{}
\DeclareMathAlphabet{\mathpzc}{OT1}{pzc}{m}{it}

\usetikzlibrary{positioning}
\usetikzlibrary{patterns,arrows,decorations.pathreplacing}
\usetikzlibrary{backgrounds}

\usepackage[
        left=3.0cm,
        right=3.0cm,
]{geometry}

\newlength\FHoffset
\fancyheadoffset{\FHoffset}

\usepackage[authoryear]{natbib}

\usepackage{ragged2e}

\usepackage{subfigure}
\usepackage{booktabs}

\definecolor{amundi_blue}{RGB}{0,176,240}
\definecolor{amundi_dark_blue}{RGB}{0,28,75}
\definecolor{darkblue}{rgb}{0.0, 0.0, 0.55}
\usepackage{hyperref}
\hypersetup{
    colorlinks=true,
    citecolor=darkblue,
    linkcolor=darkblue,
    filecolor=darkblue,      
    urlcolor=darkblue
}

\usepackage{pdfpages}
\usepackage{bookmark}

\usepackage{enumitem}
\usepackage{balance}

\usepackage{tabularx}

\begin{document}

\setcounter{page}{1}

\title{\textbf{\color{amundi_blue}Portfolio Risk Measurement Using a Mixture Simulation Approach}}

\author{
\hspace{-2.5cm}
{\color{amundi_dark_blue} Seyed Mohammad Sina Seyfi} \\
\hspace{-2.5cm} Khatam University 
% Visual Computing Master \\
% Ecole Polytechnique \\
% \texttt{sarah.perrin@polytechnique.edu}
% Amundi Asset Management \\
% Paris \\
% \texttt{thierry.roncalli@amundi.com}
\and
{\color{amundi_dark_blue} Azin Sharifi} \\
Sharif University \\
of Technology
\and
{\color{amundi_dark_blue} Hamidreza Arian} \\
Sharif University \\
of Technology
}

\date{\color{amundi_dark_blue} September 2020}

\maketitle

\begin{abstract}
\noindent
Monte Carlo Approaches for calculating Value-at-Risk (VaR) are powerful tools widely used by financial risk managers across the globe. However, they are time consuming and sometimes inaccurate. In this paper, a fast and accurate Monte Carlo algorithm for calculating VaR and ES based on Gaussian Mixture Models is introduced. Gaussian Mixture Models are able to cluster input data with respect to market's conditions and therefore no correlation matrices are needed for risk computation. Sampling from each cluster with respect to their weights and then calculating the volatility-adjusted stock returns leads to possible scenarios for prices of assets. Our results on a sample of US stocks show that the Gmm-based VaR model is computationally efficient and accurate. From a managerial perspective, our model can efficiently mimic the turbulent behavior of the market. As a result, our VaR measures before, during and after crisis periods realistically reflect the highly non-normal behavior and non-linear correlation structure of the market. 
\end{abstract}



\section{Introduction}

The stock market behaves like a black box, so that the exact causal relationship between causes and values is ambiguous. The only accurate insights we can make on the black box are market data such as stock price and trading volume \citep{sui2003value}. We can consider two main types of risk in portfolio management: risk that can be eliminated by investment diversification, and risk that cannot be avoided by diversification, or often referred to market risk. Risk management provides mathematical machinery to measure the latter type of risk \citep{klaassen2009economic}. 

JP Morgan Chase Bank, first proposed the VaR approach and is widely used for risk estimation. As the VaR approach is becoming increasingly popular in financial risk management, numerous methods for estimating the VaR have been introduced \citep{peng2019modeling}. VaR is a quantile estimation technique for measuring portfolio downside risk over a defined horizon. For this calculation, the expected loss and the expected excess loss beyond some threshold are integral. The main challenge in the estimation of these types of risk measures is the measurement of profit and loss distribution itself, particularly the left tail associated with large losses of this distribution \citep{glasserman2002portfolio}. Quantitative approaches for estimating the distribution of changes in portfolio value depend on two forms of modeling considerations: estimates of changes in the underlying risk factors to which a portfolio is exposed, and a method for interpreting these changes in risk factors into changes in portfolio value \citep{glasserman2002portfolio}. 

The variance-covariance method made popular by \cite{morgan1996riskmetrics} is the simplest and perhaps the most commonly used approach to modeling changes in portfolio value. This approach is based on two main assumptions: the distribution of portfolio profit and loss is conditionally normal, and the standard deviation can be measured from the covariance matrix of the underlying risk factors. Another important line of research is focused on relaxing the assumption that portfolio value changes linearly with changes in risk factors, which results in methods commonly called delta-gamma developed in \citep{britten1999non, duffie2001analytical,wilson1999value}. {Offering a similar computational cost as the delta-gamma approach, \cite{date2016measuring} propose a novel heuristic methods by mapping non-normal marginal distributions to normal distributions using a probability conserving transformation. Their approach is specially designed for portfolios exposed to non-linear functions of non-normal risk factors. Another approach for measuring VaR is the Extreme Value Theory (EVT) approach which models the extreme portion of a general distribution. After the financial subprime crisis of 2007-08, EVT has gained considerable attention for modelling fat tail properties of return distributions \citep{stoyanov2011fat, furio2013extreme, Rossignolo2012, Danielsson2016}. Other than the above developments on extreme value modeling, quantile regression frameworks such as conditional auto-regressive Value-at-Risk (CAViAR) model support both return asymmetry and time-varying uncertainty \citep{engle2004caviar}. The main idea behind CAViaR is that the VaR, like stock market volatility, can be predicted by auto-regressive models which are able to capture the volatility clustering phenomenon. Several authors have proposed extensions and modifications for CAViaR to adapt it to real market conditions such as intraday price ranges, trading-loss limits, overnight returns and close-to-close ranges \citep{Chen2012, Fuertes2013, Meng2018}.
} 

Risk factors are believed to be normally distributed in the traditional models of risk. Because of this assumption of normality, the factor dependency structure is fully explained by the correlation between factors using the statistics of second order \citep{haugen2001modern}. Some researchers propose to construct more realistic models of portfolio factor to ease the normality assumption. The stock returns have been shown to be non-normal with high peak values and heavy tails \citep{fama1965behavior}. \cite{chin1999computing}, for example, suggests an efficient solution to the VaR portfolio problem calculation. Their approach imposes independence between factors by using the independent component analysis (ICA) and captures the non-normality of the return distribution by Gaussian Mixture Model (Gmm). 

In the risk management community, VaR calculation methods are also divided into full and local valuation categories. Local valuation methods produce one-time portfolio value, and then calculate VaR based on the underlying risk factors model. The delta-normal model, based on a linear approximation of changes in asset values due to the underlying factors, is an example of the local valuation method. On the other hand, by re-profiling the portfolio over a range of scenarios, the full evaluation methods measure risk. Monte Carlo simulation methods are among full valuation techniques \citep{jorion2000value}. 

Monte Carlo simulation (MCS) methods are statistical techniques to solve complex mathematical and statistical problems for capturing the uncertain conditions by considering many different states that each of which is probable to take place. As \cite{reddy2016simulating} mentions, simulating an assets price is to generate some paths which the asset can follow in the future. As for \cite{sengupta2004financial}, if the price of an asset was predictable, simulation was unnecessary, for there was only one possible future price. Needless to say that simulating would become worthy only when the generated paths are produced based on some rules about historical data and other dynamic information about the prices and market factors. Furthermore, Monte Carlo simulation generates sequences of risk factors which are then used to estimate future behavior of portfolios value \citep{rollett2004monte}. There are three steps in implementation of any Monte Carlo simulation engine: first it randomly generates sample paths, second evaluates the payoff along each path, and finally calculates an average along all possible trajectories to attain an estimation \citep{chen2007monte}. 

One of the traditional models for simulating stock price, Geometric Brownian Motion (GBM), models stochastic price movements of financial assets by log-normal dynamics \citep{abidin2014forecasting}. GBM makes two strong assumptions about the price process. First, the price returns, follow a normal distribution, and second, the past information about financial instruments fully represents the future behavior of the portfolio. Despite popularity of the GBM process, it has many weaknesses, including that it dose not capture market extreme behaviors such as non-normality. 

The emergence of data-driven analyses have led to the increased use of generative modelling to learn the probabilistic dependencies between random variables. Although their apparent use has largely been limited to image recognition and classification in recent years, generative machine learning algorithms can be a powerful tool for predicting behavior of financial markets \citep{wong2019combined}.

A generative algorithm used for predicting stock market and generating new data is Gaussian Mixture Model (Gmm). Gmms have several distinct advantages that make them suitable for modeling stock returns. With enough number of components, they can model future outcomes to any probable scenario when they are carefully fit to train data using the Expectation-Maximization algorithm. Furthermore, they by no means make any normality assumptions about return distribution when the number of components is equal or greater than two \citep{bishop2006pattern}. \cite{kon1984models, tan2007approximation} studies statistical properties of the Gaussian mixture distribution and found that it could provide a reliable approximation of the probability distribution function for data, capturing heavy tails and excess kurtosis specially in multi-modal situations.

As stated in \cite{tan2005modeling}, statistical characteristics, such as skewed distributional shape or heavy-tailed behavior, are easily modeled using a class of Gaussian mixture distributions. A mixture distribution has the ability to approximate different distribution shapes by changing the component weights and other distribution parameters such as mean and variance. \cite{glasserman2013monte} proposes an effective method to estimate the tail distribution related to the rare events through a simulation approach by using Importance Sampling (IS). Their application of importance sampling not only reduces the size of simulation samples, but also improves the accuracy of the estimated probabilities. However, in their framework, it is difficult to identify the optimal parameter used for importance sampling. Nonetheless, under the Gmm framework, importance sampling is implicitly used in generation of new samples as the model assumes a number of components weighted to produce extreme data. Using importance sampling by Gaussian mixture models can substantially improve the effectiveness of simulation approaches compared to the standard Monte Carlo simulation.

{In this paper, a new price simulation model which maintains the second assumption of GBM (future is made up by the past) and relaxes the first one (Normality assumption) is proposed. Even though we apply our model to stock market, it can easily be used for simulating other asset classes which exhibit fat tail and non-normal properties. Our model provides the advantage that it does not assume the stock market returns conform to a Normal distribution.} We then calculate an efficient and fast method for estimating VaR and ES that is brought from the generated scenarios from a number of Gaussian interdependent joint distributions. Our model proposes an efficient Monte Carlo approach for simulating stock prices in a non-normal framework where return distributions follow mixture of Gaussian. 

In order to improve robustness of our VaR engine, we have re-scaled simulated returns based on the ratio of short term versus longer term volatility levels. We calculate VaR of an investment portfolio, compared the performance of the Gmm-based approach with and without re-scaling, and observed that re-scaling of the generated returns yields much better results. Our method has a number of advantages. First, it is highly accurate and superior to some other well-known VaR methods based on back-testing routines. Second, compared to other Monte Carlo applications for measuring risk, our model is quite time efficient. 
% Third, our approach passes the Monte Carlo robustness tests {\color{red} this is not done yet, but considering the superior performance of our model compared to others, should be a straightforward implementation}. 
Last but not least, our model is capable of capturing conditional correlation phenomenon and therefore, its performance during financial crisis periods is quite impressive.

% The rest of this paper is organized as follows.  

% {\color{red}
% \cite{kim2003financial} applies Support Vector Machines methods, to predict the daily Korean composite stock price index. Their method uses a risk function consisting of the empirical error and a regularized term which is derived from the structural risk minimization principle. They compare it with back-propagation neural networks and case-based reasoning. Their empirical results show that SVM provides a promising alternative for stock market prediction. \cite{pan2010performing} proposes a hybrid model to enhance prediction ability. Their approach merges the principal component regression (PCR) model and the general regression neural network (GRNN) to resolve both multi-collinearity problems and non-linearity problems at the same time. The empirical results show that the prediction power of the hybrid model is more powerful than each method separately. 
% }
\section{Gmm as a Conditional Generative Model}

A generative model, as its name suggests, has this capacity to learn the structure of the data, say, their distribution or statistical properties and thus create new unseen samples. The new samples are generated in a way that they have the same features as those by which the model is trained. Generative models also are able to deal with intractable distributions. No matter how much the data is complicated, the algorithm will learn to mimic those structures. One of the most used methods for generating new data is Gaussian Mixture Models (Gmm) which divides the data into segments and simulates members of each segment independently from others. By the way Gmm segments the train data, it can capture the non-normality of stock market, and hence it outperforms models with normality assumptions like GBM during crisis periods.

\subsection{Model Setup}

Gaussian Mixture Models (Gmms) are probabilistic mixture models with density functions combined of a number of Gaussian densities. More precisely, Gmm's density function is the weighted sum of a finite number of Normal probability density functions:
\begin{equation}
    P\left( x \right)=~\underset{i=1}{\overset{{{N}_{c}}}{\mathop \sum }}\,{{\omega }_{i}}\text{ }\!\!\Phi\!\!\text{ }(x|{\boldsymbol{{\mu }_{i}}},~{{\text{ }\!\!\Sigma\!\!\text{ }}_{i}}),
\end{equation}
where $N_c$ is the number of components, $\omega_i$s are the weights adding up to one, $\underset{i=1}{\overset{{{N}_{c}}}{\mathop \sum }}\,{{\omega}_{i}}=1$, $\boldsymbol{\mu_i}$s are mean vectors and $\boldsymbol{\Sigma_i}$s are the covariance matrices for $i = 1, 2, ..., N_c$. Moreover, the Normal density function $\Phi\!\!\text{ }(x|{\boldsymbol{{\mu }_{i}}},~{{\text{ }\!\!\Sigma\!\!\text{ }}_{i}})$ is given by
\begin{equation}
\text{ }\!\!\Phi\!\!\text{ }\left( x\text{ }\!\!|\!\!\text{ }{\boldsymbol{{\mu }_{i}}},~{{\boldsymbol{\Sigma }}_{i}} \right)={{\left( 2\pi  \right)}^{-\frac{k}{2}}}\det {{\left( {{\boldsymbol{\Sigma }}_{i}} \right)}^{-\frac{1}{2}}}\text{ }\!\!~\!\!\text{ exp}\left\{ -\frac{1}{2}{{\left( x-~{\boldsymbol{{\mu }_{i}}} \right)}^{'}}\text{ }\!\!\boldsymbol{\Sigma_{i}^{-1}}\left( x-{\boldsymbol{{\mu }_{i}}} \right) \right\} ,
\end{equation}
with $\boldsymbol{x} \in \mathbb{R}^k$ being the input data.  

\subsection{Parameter Estimation}

For estimating the parameters of the Gaussian Mixture Model we need to find the parameters $\boldsymbol{\theta} =\left\{ \boldsymbol{\omega },~\boldsymbol{\mu},~\boldsymbol{\Sigma} \right\}$. Assume $p(z_{j}=1|x_i)$ denotes the probability that $x_i$ belongs to the $j$th Gaussian distribution, and the sequence of $z=\{z_1\, \dots, z_{N_c}\}$ are set of latent variables following a binary distribution, $z_i\in \{0,1\}$, with ${\omega}_{j}=p(z_j = 1)$. Gmm in addition assumes each data sample can belong to only one cluster, so only one $z_i$ can take the value one, at a time. We also have the probability measure on the vector of latent variables
\begin{equation}
p(\boldsymbol{z})=\prod_{j=1}^{N_{c}} p\left(z_{j}=1\right)^{z_{j}}=\prod_{j=1}^{N_{c}} \omega_{j}^{z_{j}} ,
\end{equation}
and the conditional probabilities of samples conditioned on latent variables
\begin{equation}
p\left(x_{i} | \boldsymbol{z}\right)=\prod_{j=1}^{N_{c}} p\left(x_{i} | z_{j}=1\right)^{z_{j}}=\prod_{j=1}^{N_{c}} \Phi\left(x | \mu_{j}, \Sigma_{j}\right)^{z_{j}} .
\end{equation}
From $p(x_i , \mathbf{z})= p(x_i|\mathbf{z})p(\mathbf{z})$, the above yields
\begin{equation}
p(x_i)=\sum_{j=1}^{N_c}p(x_i|\mathbf{z})p(\mathbf{z})=\sum_{j=1}^{N_c}\omega_i \Phi(x_i |\mu_{i},\Sigma_{i}) .
\end{equation}
According to the Bayes rule, we obtain the responsibility function as
\begin{equation} \label{hhh}
p\left(z_{j}=1 | x_{i}\right)=\frac{\omega_{j} \Phi\left(x_{i} | \mu_{j}, \Sigma_{j}\right)}{\sum_{i=1}^{N_{c}} \omega_{i} \Phi\left(x_{i} | \mu_{i}, \Sigma_{i}\right)}=r_{{ij}} .
\end{equation}
In order to calibrate the parameters of Gmm, the Expectation-Maximization algorithm is used as follows

\begin{enumerate}[leftmargin=3\parindent]
	\item \textbf{Initializing step:} parameters $\boldsymbol{\theta} =\left\{ \boldsymbol{\omega },~\boldsymbol{\mu},~\boldsymbol{\Sigma} \right\}$ are randomly initialized.
	\item \textbf{Expectation step:} by taking an expectation from $z_{j}$ we calculate the responsibility functions $$p(z_{j}|x_i, \boldsymbol{\theta}) = r_{ij}.$$
	\item \textbf{Maximization step:} update parameters as by
	$$\begin{array}{c}{\omega_{j}=\frac{\sum_{i = 1}^{N} r_{i j}}{N}}, \\ {\mu_{j}^{\text {new}}=\frac{\sum_{i = 1}^{N} r_{i j} x_{i}}{\sum_{i = 1}^{N} r_{i j}}}, \\ {\Sigma_{j}^{\text {new}}=\frac{\sum_{i = 1}^{N} r_{ij}\left(x_{i}-\mu_{j}\right)\left(x_{i}-\mu_{j}\right)^{\prime}}{\sum_{i = 1}^{N} r_{ij}}}\end{array},$$
	where $N$ is the number of all data used as input of the model. 
    \item Repeat step 2 and 3 until convergence. 
\end{enumerate}

\subsection{Generating new samples}

A generative model is linked to a conditional probability function $p(X|Y)$, where $X$ is sample and $Y$ is the target. These models are able to produce new instances with respect to $Y$. Gmm is used for fitting a PDF over the data, and using that PDF will create new samples. A generative model should at first learn the exact structure of data in order to get to generate new ones. As sampling from a joint distribution is a challenging, the Gmm can be used for scenario generation (see \cite{wang2018conditional}). For generating new data, Gmm simply samples separately from every component by $\Phi(x|\mu_i,\Sigma_i)$. The proportion of the numbers of samples from each distribution $\Phi(x|\mu_i,\Sigma_i)$ to all samples is equal to $\omega_i$. In other words, Gmm samples from each class with respect to its weight and finally integrates all samples. In this way, the generated data has the joint distribution of $P(x)= \sum_{i=1}^{N_c}\omega_i\Phi(x|\mu_i,\Sigma_i)$  which guarantees that the new data resembles the historical one. 

For any given probability density function $p(x)$, assume $N_T$ represents the total number of generated samples, and $N_0 \leq N_T$ indicates the number of samples less than an arbitrary value $x_0$, by infinitely increasing the number $N_T$
\begin{equation}
   \lim _{N_{T} \rightarrow \infty} \frac{N_{0}}{N_{T}}=\int_{-\infty}^{x_{0}} p(x) dx .
\end{equation}    
In the case of a Gmm, $N_0^*$ the number of samples from Gmm that are less than $x_0$, will be equal to the summation of all $N_0^i$, $i=1,2, ... ,N_c$, number of samples less than $x_0$ in each of normal distribution. Then we have
\begin{equation}
\begin{aligned} 
\lim _{N_{T} \rightarrow \infty} \frac{N_{0}^{*}}{N_{T}}&=\lim_{N_{T} \rightarrow \infty} \frac{\sum_{i=1}^{N_{c}} N_{0}^{i}}{N_{T}}=\sum_{i=1}^{N_{c}} \lim_{N_{T} \rightarrow \infty} \frac{N_{0}^{i}}{N_{T}}=\sum_{i=1}^{N_{c}} \omega_{i} \int_{-\infty}^{x_{0}} \Phi\left(x | \mu_{i}, \sigma_{i}\right) d x\\
&=\sum_{i=1}^{N_{c}} \int_{-\infty}^{x_{0}} \omega_{i} \Phi\left(x | \mu_{i}, \sigma_{i}\right) d x=\int_{-\infty}^{x_{0}} \sum_{i=1}^{N_{c}} \omega_{i} \Phi\left(x | \mu_{i}, \sigma_{i}\right) d x .
\end{aligned}
\end{equation}
The equation above mathematically shows that new samples conform to a Gaussian Mixture distribution. Therefore, the simulation algorithm for a mixture of Gaussians can be planned in three steps. First for $i=1,..., N_c$, the number of samples for each component $\Phi(x|\mu_i,\sigma_i )$, is chosen as $\omega_i N_T$. Second, we generate $\omega_i N_T$ samples for $i$th component, and finally we integrate all the generated samples from all the components \citep{wang2018conditional}.
\section{Stock Price Simulation}

\subsection{Limitations of Geometric Brownian Motion}

One of the most commonly used processes in finance is Geometric Brownian motion \citep{black1973pricing}. However, GBM suffers from the assumption that stock prices are log-normally distributed, while it has been observed as an stylized fact that in many cases, stock returns have fat tail and are skewed. Symbolically, GBM offers the following model for stock prices: $r_t=  \ln (  \frac{S_{t}}{S_{t-1}}) \sim \Phi(\mu,\sigma)$, where $r_t$ and $S_t$ refer to return and price processes, respectively. $\Phi$ is also a Normal distribution with parameters $\mu$ and $\sigma$ calibrated from historical data. This assumption leads to 
\begin{equation}
    \ln \left(\frac{S_{t}}{S_{t-1}}\right)=\mu \delta t+\sigma \epsilon \sqrt{\delta t},
\end{equation}
where $\delta t$ is the period of time and $\epsilon \sim \Phi(0,1)$. The first term of the price dynamics introduces the drift, and the second term presents the diffusion effect. Rearranging the equation above gives
\begin{equation}\label{eq:gbmsim}
\begin{aligned}
\frac{S_{t}}{S_{t-1}}=\exp (\mu \delta t+\sigma \epsilon \sqrt{\delta t}), \\
S_{t}=S_{t-1} \exp (\mu \delta t+\sigma \epsilon \sqrt{\delta t}).
\end{aligned}
\end{equation}
Therefore, the equation \eqref{eq:gbmsim} enables us to simulate stock paths as many times as needed, by sampling $\epsilon$ from a standard normal distribution. 

However, as \cite{morone2008financial} mentions, there are some stylized facts in the financial markets, which GBM is not able to produce them, such as fat-tail distributions in the stocks' returns. Volatility clustering, also, is another phenomenon that GBM cannot provide any explanation for that. \cite{takahashi2019modeling} mentions other stylized facts, and gain and loss asymmetric for financial time series, which GBM is unable to explain, but some of the generative models can capture the financial market's pattern. Considering the stylized facts directly affects the performance of simulated data, and accordingly, the VaR.

\subsection{Gmm as an Alternative to GBM} \label{sec:4.2}

In this section, we propose a simulation approach for modeling the non-normality features of stock return processes. Gaussian Mixture models have the advantage of capturing fat tail effects in returns of stocks and investment portfolios. Gmm's probability density function (PDF) follows 
\begin{equation}
    p\left( x \right)=~\underset{i=1}{\overset{{{N}_{c}}}{\mathop \sum }}\,{{\omega }_{i}}\text{ }\!\!\phi\!\!\text{ }(x|{{\mu }_{i}},~{{\text{ }\!\!\Sigma\!\!\text{ }}_{i}}),
\end{equation}
where $\phi\!\!\text{ }(x|{{\mu }_{i}},~{{\text{ }\!\!\Sigma\!\!\text{ }}_{i}})$ denotes a Normal PDF with parameters $\mu_i$ and $\Sigma_i$ as mean and covariance matrix. Therefore it is evident that every single point belongs to one of the $\Phi\!\!\text{ }(x|{{\mu }_{i}},~{{\text{ }\!\!\Sigma\!\!\text{ }}_{i}})$ for $i=1,..., N_c$. These are the number of classes that stock market experiences during a long period. Every class is assumed to be Normal. 

Generating a new sample, with any arbitrary number of data, say T, is carried out by sampling from components $\Phi\!\!\text{ }(x|{{\mu }_{i}},~{{\text{ }\!\!\Sigma\!\!\text{ }}_{i}})$ with corresponding number of samples proportional to $\omega_i$. It assures that if in a long time window, there exists a particular number of returns in a component, the same portion from those classes will appear in $T$ days. By using this approach for sampling a new path, the simulated returns resemble historical data in such a way that unlikely events are accounted for with the exact same probability of occurrence. In the GBM, however, the rare events are represented by a single term $\sigma$, and therefore using GBM to simulate stock prices no longer generates the rare historical events. Note that there is also no need for calculating statistical features manually as it is done in the GBM framework. 

In mathematical terms, to generate one path with $T$ new samples, the Gmm allocates $\omega_i T$ of samples to the class $i$. The new samples, $\{r_1^*,r_2^*,...,r_T^*\}$ are considered as a set of possible returns for the stock. The stock prices $\{S_1^*,S_2^*,...,S_T^*\}$ are created using generated returns by $S_t^*=S_{t-1}^* \exp{r_t^*}$ for $t=1,2,...,T$. Repeating this algorithm $m$ times, the model produces $m$ paths for stock price based on the same feature of historical data structure. Calibrating the Gmm is the most computationally time consuming part of our algorithm moving from one day to the next. We run the Expectation-Maximization (EM) algorithm daily, however, in each calibration step the parameter set $\boldsymbol{\theta}$ from previous step is used to initialize the EM calibration routine. We will show that this technique reduces the computational cost of our Gmm model calibration significantly. We also mention that, in the very first day of executing the EM algorithm, we use k-means initialization to start the calibration routine.
\section{Calculating Value at Risk (VaR) and Expected Shortfall (ES)}

VaR is about measuring the downside risk during a specific time frame and with respect to a significance level. Symbolically, $Pr(S\geq VaR_{\alpha, t}) = 1 - \alpha$. There are many methods for calculating VaR among which, Historical Simulation (HS), Parametric Method, and Monte Carlo simulation (MCS) are three of the most popular approaches used by academics and practitioners \citep{jorion2000value}. 
Historical Simulation approximates risk by the $\alpha$-percentile of historical returns and uses this quantile as a proxy for $\text{VaR}_{(\alpha,t)}$. This method presumes the historical returns will re-occur in the future with the same pattern from past. MCS applies the same technique for calculating VaR except that instead of historical returns, it re-generates many returns according to a particular dynamics like GBM. After option pricing and estimating derivatives sensitivities to their underlyings, estimating market risk is the most common usage of MCS \citep{dowd2007measuring}. It is not necessary to assume a closed-form representation for stock prices when applying Monte Carlo risk assessment. Monte Carlo simulations in general is based on the Law of Large Number (LLN), which suggests that if $X_1,X_2,...,X_N$ are i.i.d. random variables with mean $\mu$ and variance $\sigma^2$ then 
\begin{equation}
\operatorname{Pr}\left(\left|\frac{\sum_{i=1}^{N} X_{i}}{N}-\mu\right| \geq \epsilon\right) \leq \frac{\sigma^{2}}{N \epsilon^{2}}.
\end{equation}
As an application of LLN, $\frac{\sigma^{2}}{N \epsilon^{2}} \rightarrow 0$ when $N \rightarrow \infty$. Accordingly, if $X_i$ is the last generated price for $i$th path, then it should have the same statistical properties as the original data. 

As VaR is to measure the probability of loss in a particular time frame, MCS focuses only on the last returns simulated. Considering all last returns as $P^{*i}_T$, $i=1,... ,m$, the VaR becomes $\text{VaR}_{\alpha,T}= \text{Percentile} (P_T^i,1-\alpha\%)$, or the $\alpha$-percentile of all generated returns $P^{*i}_T$.
Using Gmm, the matrix of returns for $m$ paths, and $T$ evaluation days returns is 
\begin{equation}
\left[\begin{array}{ccc}{r_{1}^{*1}} & {\cdots} & {r^{*1}_T} \\ {\vdots} & {\ddots} & {\vdots} \\ {r_{1}^{* m}} & {\dots} & {r^{*m}_T}\end{array}\right] .
\end{equation}
From this matrix, stock price paths are generated according to $(S_t^{*j}=S_{t-1}^{*j} \exp{r_t^*})$
\begin{equation}
\left[\begin{array}{ccc}{S_{1}^{*1}} & {\cdots} & {S^{*1}_T} \\ {\vdots} & {\ddots} & {\vdots} \\ {S_{1}^{* m}} & {\dots} & {S^{*m}_T}\end{array}\right] .
\end{equation}
We then calculate the overall return for each path by $R_\text{Gmm}^i = \ln(\frac{S_T^*}{S_0})$. Finally, the $\alpha$-percentile of $R_\text{Gmm}^i$ is $VaR_{1-\alpha}$, and the corresponding expected shortfall is calculated according to
$$ES_{1-\alpha} = \frac{1}{n}\sum_{i=1} ^ m R^i_\text{Gmm} \times I_{\{R_\text{Gmm}^i \leq VaR_{1-\alpha}\}} . $$
where $I$ is an indicator function 
\begin{equation} \label{eq:indicator}
I_\text{A}=\left\{\begin{array}{l}{1, \quad \quad \quad \text{A,}} \\ {0, \quad \text{otherwise, }}\end{array}\right.
\end{equation}
and $n$ is the number of $1$s in $I_{\{R_\text{Gmm}^i \leq VaR_{1-\alpha}\}}$. Gmm guarantees that the new returns matrix has the same distribution as the historical returns up to the assumption of a mixture distribution. One of the main advantages of our approach in estimating Value-at-Risk using Gaussian Mixture distributions is that during crisis periods, our simulation algorithm learns quickly that clusters of correlated losses occur. Therefore, and by the fact that most models assume a linear constant correlation, our technique results in superior performance when compared to traditional VaR techniques. This is specially important since it is expected that a robust risk measurement tool is capable of being adapted to new market environments when going into and out of turbulent financial markets. It is also worthwhile to note that during crisis periods both correlation and volatilities increase significantly. In our algorithm, our model learns to allocate appropriate weights to various return clusters representing these periods. This way, when a crisis period starts higher weights are allocated to extreme market scenarios which results in a systematic increase in the risk measure result. As the market prepares to move out of the crisis period, the corresponding weight to such scenarios start to decline resulting in reducing the VaR result.  

\subsection{Adjusting the VaR and ES with Respect to Their Volatility and Ages} \label{adj_individual}

Gmms are powerful tools for capturing the entire distribution of returns, including their tails. But VaR depends not only on the returns levels, but also on the volatilities. The most well-known models for capturing volatility is GARCH \citep{engle2001theoretical}. An efficient model for estimating VaR and ES should be able to capture volatility clustering phenomenon. Traditional Monte Carlo risk measures assume input data is not ordered, while it is clear that the recent returns have more effects on the futures risk profile of investment portfolios, when compared to older returns. Here we propose a simple and fast method for adjusting the VAR and ES with respect to their fluctuations and their ages. 

Let $\sigma_{Long}$ be the standard deviation of returns series in a given period. For taking both age and volatility into account, we define $\sigma_{short}$ as the standard deviation calculated in a short and recent period, say two weeks to one month. Then, we update the calculated VaR and ES as follows
\begin{equation} \label{eq:adj}
\begin{split}
    \text{VaR}_{1-\alpha}^{adj} & = \text{VaR}_{1-\alpha} \times \frac{\sigma_{short}}{\sigma_{Long}}, \\
     \text{ES}_{1-\alpha}^{adj} & = \text{ES}_{1-\alpha} \times \frac{\sigma_{short}}{\sigma_{Long}}.
\end{split}
\end{equation}

By using the volatility term, $\sigma_{short}$ versus $\sigma_{long}$, older return patterns are updated to the current state of the market. The length of the short time period can be optimized by back-testing obtained results using different periods. Generally, based on our experience, the short period can be half or less than half of the long period. Equivalently, the above technique can be applied directly on the simulated matrix of returns
\begin{equation}
\left[\begin{array}{ccc}{r_{1}^{*1(adj)}} & {\cdots} & {r^{*1(adj)}_T} \\ {\vdots} & {\ddots} & {\vdots} \\ {r_{1}^{* m(adj)}} & {\dots} & {r^{*m(adj)}_T}\end{array}\right] =  \frac{\sigma_{short}}{\sigma_{Long}} \times \left[\begin{array}{ccc}{r_{1}^{*1}} & {\cdots} & {r^{*1}_T} \\ {\vdots} & {\ddots} & {\vdots} \\ {r_{1}^{* m}} & {\dots} & {r^{*m}_T}\end{array}\right] .
\end{equation}
and calculating VaR and ES according to adjusted matrix.
\section{Portfolio Value Simulation}

In this section, we explain our algorithm for using Gaussian Mixture distributions to estimate the risk level of an investment portfolio. To highlight the advantages Gmm makes in capturing the profile of the return distribution, we first provide the classical approach based on Geometric Brownian motion and then we discuss the improvements based on our Gmm-based approach.

\subsection{Simulating Returns based on Geometric Brownian Motion}

Monte Carlo simulation on the Geometric Brownian motion requires correlated random variables to generate future stock prices in a portfolio setting. Supposing that there are $N_s$ stocks in a portfolio. Then we can write the equation \eqref{eq:gbmsim} in the form
\begin{equation}
\left[\begin{array}{c}{S_{1_t}} \\ {\vdots} \\ {S_{N_{S_{t}}}}\end{array}\right]=\left[\begin{array}{c}{S_{1_{t-1}}\left(1+\mu_{1} \delta t\right)} \\ {\vdots} \\ {S_{N_{S_{t-1}}}\left(1+\mu_{N_{S}} \delta t\right)}\end{array}\right]+\left[\begin{array}{c}{S_{1_{t-1}} \sigma_{1} \epsilon_{1} \sqrt{\delta t}} \\ {\vdots} \\ {S_{N_{S t-1}} \sigma_{N_{S}} \epsilon_{N_{S}} \sqrt{\delta t}}\end{array}\right].
\end{equation}
Here $\xi=\{\epsilon_i\}_1^{N_s}$ is a matrix of random variables from a Normal standard distribution with $N_s$ columns and $T$ rows. As each row should be correlated based on the correlation structure of stocks returns, one may use Cholesky decomposition for creating $\xi$. This way, we can write $\rho$, correlation matrix as $\rho=AA'$, and put $\xi=A\epsilon$, with i.i.d. random variables $\epsilon$. However, as we are using a correlation matrix $\rho$ we are implicitly making an assumption that there exists a linear dependency structure between stocks. Repeating this process $m$ times generates $m$ different paths for each stock, and consequently $m$ different values for value of the portfolio. Portfolio's value with weights of $[w_1,... ,w_{N_s}]$ in the $i$th simulation path becomes

\begin{equation}
    S_{p_t} = [w_1,... ,w_{N_s}] \times \left[\begin{array}{c}{S_{1_t}} \\ {\vdots} \\ {S_{N_{S_{t}}}}\end{array}\right].
\end{equation}

Accordingly, the return from $i$th simulation is $r^{i*}_p= \ln( \frac{S^{i*}_{p,T}}{S_{P,0}})$. The $\alpha$-percentile of these generated returns is the risk measure $VaR_{T,1-\alpha}$ calculated based upon GBM assumptions. Considering a portfolio of varying assets, correlation between these assets is an important element from investment point of view. It is not rational to simulate all of the stocks independently due to the fact that returns of stocks are correlated with each other. The main point is that the correlations between stocks do not remain constant all along. Therefore, making assumptions on a single number as correlation between two stocks or equivalently a correlation/covariance matrix between $n$ stocks does not give realistic decision making results in most market environments. It has been observed that during crisis periods, stocks are highly correlated while in other conditions they are not \citep{ait2016increased, frank2009linkages}). This joint modelling of financial assets behavior during varying market environments has always posed great challenges to practitioners working in industry. A very well-known model for providing a structural framework for correlation between stocks and the market portfolio is the Capital Asset Pricing Model (CAPM) which considers one risk factor as the market risk and measures risk premium of all stocks according to a coefficient named $\beta$ \citep{merton1973intertemporal}. 

\subsection{Simulating returns based on Gaussian Mixture Model} \label{adj_portfolio}

One of the main advantages of using Gaussian Mixture models to simulate stock prices is that it does not require a correlation matrix driving the dynamics of stock returns. Instead, Gmm directly models joint distributions of stock returns by a finite number of Gaussian components. Allocating proper clusters to data with different parameters leads to different correlation structures according to the co-variation $\Sigma$ in each cluster. Therefore, there is no need for calculating correlation matrix since there are multiple covariance matrices with their associated weights in Gmm. By using a sampling method discussed in section \ref{sec:4.2}, every single sample $R^*$  will contain $N_s$ elements which denote each stock return. Sampling stock returns $m$ times will produce the $m \times N_s$ matrix
\begin{equation} \label{eq:simu_matrix}
    \left[\begin{array}{ccc}{r_{s_{1}}^{1*}} & {\cdots} & r_{s_{N_{S}}}^{1*} \\ {\vdots} & {\ddots} & {\vdots} \\ {r_{s_{1}}^{m*}} & {\cdots} & {r_{s_{N_{S}}}^{m*}}\end{array}\right] .
\end{equation}
As Gmm looks at all historical data through one lens, we suggest to adjust all simulated returns, with respect to their current volatility. We use the coefficient $\frac{\sigma_{short}}{\sigma_{long}}$ and adjust the returns by multiplying it with this coefficient. This results in a new matrix with elements
\begin{equation}
    r_{s_{j}}^{i*} :=  r_{s_{j}}^{i*} \times (\frac{\sigma_{short}}{\sigma_{long}})_{s_j} .
\end{equation}
Let $T$ be the time horizon at which we plan to simulate stock prices. We repeat our simulation algorithm for $T$ times, resulting in $T$ matrices as \ref{eq:simu_matrix}. The $i$th  simulated prices $(i = 1, \cdots, m)$ for stock $s_j$, $(j = 1, \cdots, N_s)$, at $T$ is
\begin{equation}
    S^{i*}_{s_j, T} = S_{s_j, 0} \times \exp{(\sum_{t=1}^T r^{i*}_{s_j,t})}.
    \end{equation}
The $T$-holding period return for the stock $s_j$ becomes $\sum_{t=1}^T r^{i*}_{s_j,t}$.  
The portfolio return considering the fixed weights $\left[w_{1}, \dots, w_{N_{s}}\right]^{\prime}$ is given by
\begin{equation}
\left[\begin{array}{c}{r^{1*}_p} \\ {\vdots} \\ {r_{p}^{m*}}\end{array}\right]=\left[\begin{array}{ccc}{\sum_{t=1}^T r^{1*}_{s_1,t}} & {\cdots} & {\sum_{t=1}^T r^{1*}_{s_{N_s},t}} \\ {\vdots} & {\ddots} & {\vdots} \\ \sum_{t=1}^T r^{m*}_{s_1,t} & {\cdots} & \sum_{t=1}^T r^{m*}_{s_{N_s},t}\end{array}\right] \times \left[w_{1}, \dots, w_{N_{s}}\right]^{\prime} ,
\end{equation}
and the $i$th simulated value for the portfolio in at $T$ is  
\begin{equation}
    S_{p, T}^{i^{*}}= S_{p, 0} \times \exp{{r_{p}^{i*}}}, \quad \text{for all} \quad i=1,2,\dots,m.
\end{equation}
Having the generated price series, the $\alpha$-percentile of portfolio returns is $\text{VaR}_{T,1-\alpha}$. Similarly, other risk factors can be calculated accordingly. For example, the expected shortfall, $ES_{T, 1-\alpha}$, can be calculated as $\frac{1}{n}\sum_{i=1} ^ m R^i_{Gmm} \times I_{\{R_{Gmm}^i \leq VaR_{1-\alpha}\}}$, where $I$ is the indicator function.
\section{Empirical Analysis}

In this section, we provide the numerical results based on our Python implementation for calculating the VaR for individual stocks and a portfolio including those stocks. We use Christoffersen back-testing for investigating validity of calculated VaR. We also use a quadratic loss function for making a comparison between our model and other models such as historical simulation, parametric (variance-covariance), Monte Carlo simulation and the GARCH methods. In addition, a Gaussian Copula Monte Carlo approach is employed for our selected portfolio and compared with our Gmm-based VaR model. We also compare our model with different types of VaR named CaViaR introduced by \cite{engle2004caviar}.

\subsection{Data}

The data for this study is from a number of S\&P500 stocks, namely 3M Company (MMM), American Express Co (AXP), Advanced Micro Devices Inc (AMD), American International Group (AIG), AFLAC Inc (AFL), Abbott Laboratories (ABT), Microsoft Corp (MSFT), Apple Inc. (AAPL), Amazon.com Inc. (AMZN), Bank of America Corp (BAC), JPMorgan Chase \& Co (JPM), Johnson \& Johnson (JNJ), Exxon Mobil Corp (XOM), Mastercard Inc (MA) and the S\&P500 index itself from Yahoo! Finance. The dates include the sub-prime crisis of 2007-2008. The start date for the stocks returns time series is $7/24/2006$. The time horizon for our VaR model, $T$, is 1, which means we have calculated daily VaR. We calculate the daily VaR for 1000 days. The logarithmic returns are used from $7/24/2006$ to $7/23/2007$. In the next step, the window is shifted one day till VaR is calculated up to $7/11/2011$ (For 1000 days). The descriptive statistic of returns series is illustrated in table \ref{tab_descriptive_statistic}. The jarque bera statistic has been rejected and shown with (R), when the associated p-value is under 0.01. As shown, no one of the stocks during a crisis follow a normal distribution. For calculating the VaR, Two confidence levels of $1\%$ and $5 \%$ are considered for daily VaRs. Furthermore, we calculate $1\%$ and $5 \%$ VaR of an equally weighted portfolio from these stocks. We used $N_c=3,4,5$, and $6$ for our Gmm's number of components. For the parameters $\sigma_{long}$ and $\sigma_{short}$, we choose 252 and 70 days, respectively. The results for different $\sigma_{short}$ is also shown in \ref{tab_sigma_s3}, \ref{tab_sigma_s4}, \ref{tab_sigma_s5} and \ref{tab_sigma_s6} in appendix. The number of simulations in our VaR model is 3000. However, considering 1000 number of simulations also gave us the same results. For calculating an accurate VaR result, the model should be able to capture the future distribution of the data. A comparison between Gmm and two other distributions, namely Normal and Student's t-distributions, on each stock is made. 
\begin{table}
  \centering
  \caption{Descriptive statistics of the return series}
\resizebox{\textwidth}{!}{
    \begin{tabular}{lcccccccc}
    % {\tblwidth}{@{}LCCCCCCCC@{}}
\toprule
      & \textbf{S\&P500} & \textbf{MMM} & \textbf{AXP} & \textbf{AMD} & \textbf{AIG} & \textbf{AFL} & \textbf{ABT} & \textbf{MSFTF} \\
\midrule
\textbf{skewness} & -0.247 & -0.115 & 0.093 & 0.024 & -2.150 & -1.507 & -0.174 & 0.350 \\
\textbf{kurtosis} & 8.417 & 5.193 & 6.605 & 3.702 & 41.435 & 28.815 & 6.095 & 9.001 \\
\multicolumn{1}{p{5.07em}}{\textbf{Jarque-Bera\newline{}statistic}} & 3671.8 (R) & 1394.3 (R)  & 2254.2 (R)  & 706.9 (R)  & 89721.2 (R)  & 43391.9 (R)  & 1924.3 (R)  & 4210.0 (R)  \\
\textbf{max} & 0.110 & 0.094 & 0.188 & 0.202 & 0.507 & 0.265 & 0.092 & 0.171 \\
\textbf{min} & -0.095 & -0.089 & -0.194 & -0.185 & -0.936 & -0.460 & -0.096 & -0.125 \\
\midrule
      & \textbf{AAPL} & \textbf{AMZN} & \textbf{BAC} & \textbf{JPM} & \textbf{JNJ} & \textbf{XOM} & \textbf{MA} & \multicolumn{1}{p{5.07em}}{\textbf{Portfolio\newline{}returns}} \\
\midrule
\textbf{skewness} & -0.452 & 0.694 & -0.179 & 0.350 & 0.660 & 0.105 & 0.189 & -0.404 \\
\textbf{kurtosis} & 5.793 & 12.413 & 13.613 & 9.838 & 15.655 & 13.216 & 13.529 & 8.474 \\
\multicolumn{1}{p{5.07em}}{\textbf{jarque bera\newline{}statistic}} & 1775.1 (R)  & 8061.8 (R)  & 9582.4 (R)  & 5025.4 (R)  & 12755.1 (R)  & 9026.7 (R)  & 9465.2 (R)  & 3743.2 (R)  \\
\textbf{max} & 0.130 & 0.239 & 0.302 & 0.224 & 0.115 & 0.159 & 0.189 & 0.115 \\
\textbf{min} & -0.197 & -0.246 & -0.342 & -0.232 & -0.080 & -0.150 & -0.133 & -0.129 \\
\bottomrule
    \end{tabular}
}
  \label{tab_descriptive_statistic}%
\end{table}%

\subsection{Investigating Goodness of Fit}

We use two goodness of fit approaches to verify how much the model is capable of being fitted on the data. Firstly we employ a root-mean-square error (RMSE) 
\begin{equation}
R M S E=\sqrt{\frac{1}{n} \sum_{i=1}^{n}\left(P D F_{i}^{\text {model }}-P D F_{i}^{\text {real}}\right)^2},
\end{equation}
between real PDF of data and various models, where $n$ is the number of points of PDF. The table \ref{tab_RMSE} shows the RMSE for different stocks during the period from $7/24/2006$ to $7/11/2011$. Since different number of components $N_c = 3, 4 , 5$ and $6$ have the same results we only show the Gmm for $N_c = 3$ components. As results show, it is clear that the RMSEs for Gmm is significantly lower than Normal and Student's t-distributions, which means that the Gmms with $N_c = 3,4, 5$ and $6$ are better models than traditional distributions.

Secondly, The log-likelihood function per sample is also shown in table \ref{tab_loglikelihood} which illustrates the greater amount for Gmms with different $N_c$ rather than Normal and Student's t-distribution. The results are showing the average of log-likelihoods for samples from $7/24/2006$ till $7/11/2011$. It can be seen that all Gmm log-likelihoods are higher than those for Normal and t student. On the other hand, table \ref{tab_loglikelihood} suggests that increasing the number of components from $3$ to higher numbers may lead to either a higher log-likelihood or to a same. 

Thirdly, we use a Kolmogorov-Smirnov (KS) test between real data and various models. Table \ref{tab_kstest} shows the p-values of testing the null hypothesis in the KS test, where the $H_0$ hypothesis tests whether the data conforms to the assumed distribution or not. The statistic for this test is $D_n = \sup |CDF_i^{model}-CDF_i^{real}|$ where CDF is the cumulative distribution function. In this test an instance randomly is sampled from the determined distribution and its CDF is presumed as $CDF_i^{model}$. If at the significance level of $1\%$, the p-value is higher than $1\%$, 
we fail to reject $H_0$, as shown in table \ref{tab_kstest}.

\begin{figure*}
\centering
\subfigure[]{{\includegraphics[width=5.0cm]{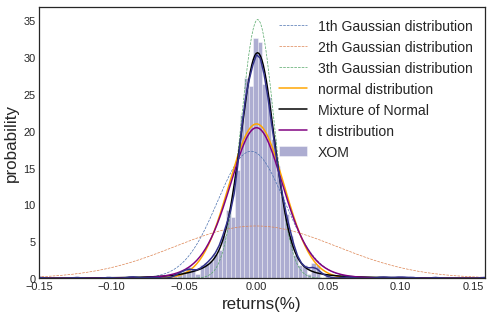}}}
\subfigure[]{{\includegraphics[width=5.0cm]{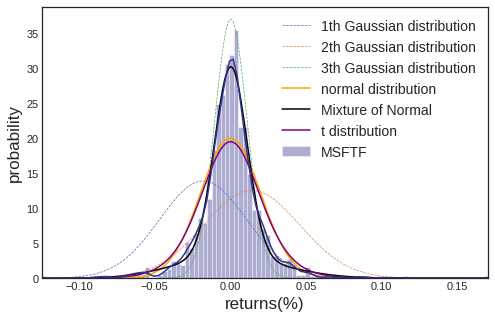}}}
\subfigure[]{{\includegraphics[width=5.0cm]{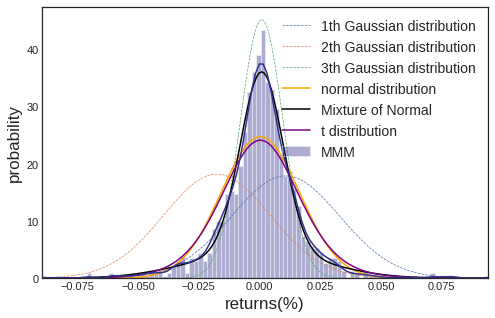}}}\\
\subfigure[]{{\includegraphics[width=5.0cm]{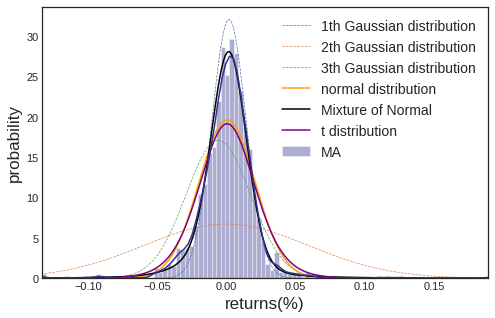}}}
\subfigure[]{{\includegraphics[width=5.0cm]{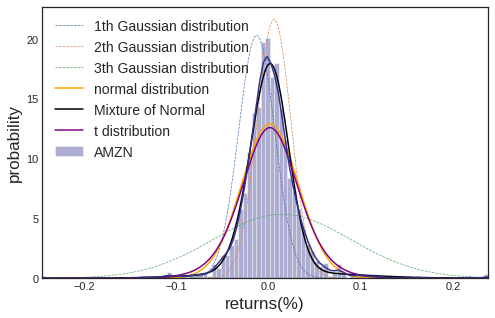}}}
\subfigure[]{{\includegraphics[width=5.0cm]{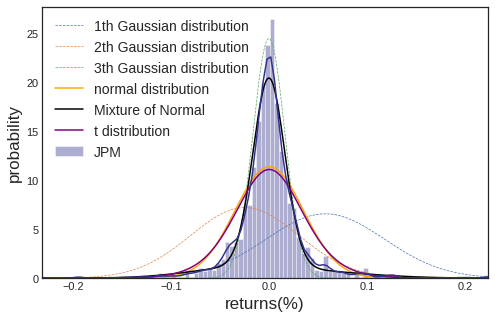}}}\\
\subfigure[]{{\includegraphics[width=5.0cm]{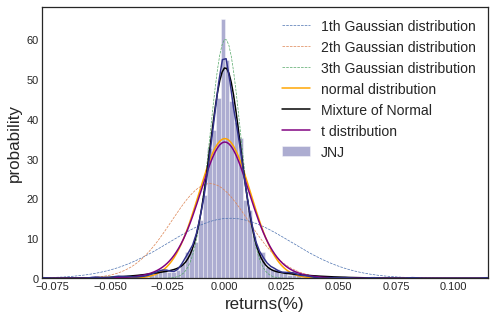}}}
\subfigure[]{{\includegraphics[width=5.0cm]{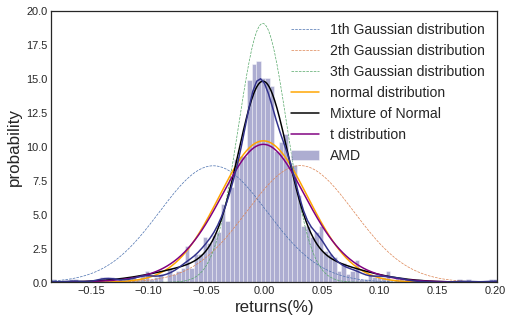}}}
\subfigure[]{{\includegraphics[width=5.0cm]{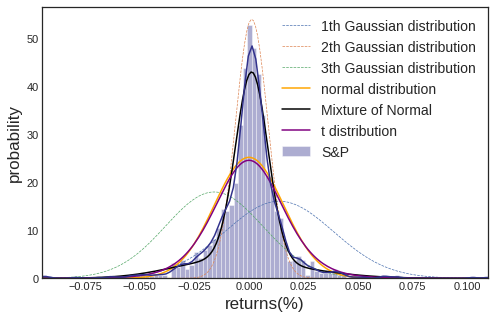}}}
\caption{Mixture of Normal distribution for individual stocks during a crisis period}
\label{fig:Nc3}
\end{figure*}

Figure \ref{fig:Nc3} including some of our stocks gives an illustration for a comparison between Normal and Student's t-distribution with Gmm's number of components $N_c = 3$. The Gmm fits a separate Normal distribution on the tails of the data in order to capture their fat-tail profile. The weights of these distributions are relatively low, due to the fact that they are less likely to occur and less data is available to fit their behavior. The graph for all stocks are shown in appendix in figures \ref{fig:Nc3_a} and \ref{fig:Nc3_b}. As table \ref{tab_kstest} shows there is no significant difference between efficiency of Gmm distribution with different number of components $N_c$, leading to the conclusion that the market can be explained well enough with $3$ different Normal distributions.
\begin{table}
  \centering
  \caption{RMSE comparison between different distributions}
\resizebox{\textwidth}{!}{
    \begin{tabular}{lccccccccccccccc}
    % {\tblwidth}{@{}LCCCCCCCCCCCCCCC@{}}
    \toprule
\multicolumn{1}{r}{} & \textbf{S\&P500} & \textbf{MMM} & \textbf{AXP} & \textbf{AMD} & \textbf{AIG} & \textbf{AFL} & \textbf{ABT} & \textbf{MSFTF} & \textbf{AAPL} & \textbf{AMZN} & \textbf{BAC} & \textbf{JPM} & \textbf{JNJ} & \textbf{XOM} & \textbf{MA} \\
\midrule
\boldmath{}\textbf{GMM ($N_C = 3$)}\unboldmath{} & 0.27  & 0.21  & 0.23  & 0.22  & 0.31  & 0.24  & 0.17  & 0.23  & 0.19  & 0.19  & 0.30  & 0.25  & 0.23  & 0.20  & 0.16 \\
\multicolumn{1}{l}{\textbf{Normal}} & 0.56  & 0.39  & 0.41  & 0.29  & 0.60  & 0.53  & 0.30  & 0.40  & 0.27  & 0.30  & 0.52  & 0.45  & 0.51  & 0.37  & 0.32 \\
t-student (d.o.f = 10) & 0.57  & 0.40  & 0.41  & 0.29  & 0.60  & 0.53  & 0.31  & 0.41  & 0.28  & 0.30  & 0.53  & 0.45  & 0.52  & 0.37  & 0.33 \\
\bottomrule

    % \toprule
    %       & \multicolumn{1}{c}{\textbf{GMM}} & \multicolumn{1}{c}{\textbf{GMM 4 Nc}} & \multicolumn{1}{c}{\textbf{GMM 5 Nc}} & \multicolumn{1}{c}{\textbf{GMM 6 Nc}} & \multicolumn{1}{c}{\textbf{normal}} & \multicolumn{1}{c}{\textbf{t}} \\
    %       & \multicolumn{1}{c}{\boldmath{}\textbf{$N_c = 3$}\unboldmath{}} & \multicolumn{1}{c}{\boldmath{}\textbf{$N_c = 4$}\unboldmath{}} & \multicolumn{1}{c}{\boldmath{}\textbf{$N_c = 5$}\unboldmath{}} & \multicolumn{1}{c}{\boldmath{}\textbf{$N_c = 6$}\unboldmath{}} & \multicolumn{1}{c}{\textbf{dist.}} & \multicolumn{1}{c}{\textbf{dist.}} \\
    % \midrule
    % \textbf{S\&P 500} & 0.27  & 0.29  & 0.22  & 0.21  & 0.56  & 0.57 \\
    % \textbf{MMM} & 0.22  & 0.24  & 0.20  & 0.20  & 0.39  & 0.40 \\
    % \textbf{AXP} & 0.23  & 0.23  & 0.20  & 0.20  & 0.41  & 0.41 \\
    % \textbf{AMD} & 0.22  & 0.22  & 0.21  & 0.21  & 0.29  & 0.29 \\
    % \textbf{AIG} & 0.31  & 0.30  & 0.30  & 0.32  & 0.60  & 0.60 \\
    % \textbf{AFL} & 0.25  & 0.23  & 0.22  & 0.22  & 0.53  & 0.54 \\
    % \textbf{ABT} & 0.17  & 0.18  & 0.18  & 0.18  & 0.30  & 0.31 \\
    % \textbf{MSFTF} & 0.23  & 0.24  & 0.21  & 0.22  & 0.40  & 0.41 \\
    % \textbf{AAPL} & 0.19  & 0.20  & 0.20  & 0.19  & 0.27  & 0.28 \\
    % \textbf{AMZN} & 0.19  & 0.19  & 0.18  & 0.18  & 0.29  & 0.30 \\
    % \textbf{BAC} & 0.30  & 0.28  & 0.26  & 0.24  & 0.52  & 0.53 \\
    % \textbf{JPM} & 0.25  & 0.25  & 0.22  & 0.22  & 0.44  & 0.45 \\
    % \textbf{JNJ} & 0.23  & 0.24  & 0.22  & 0.21  & 0.51  & 0.52 \\
    % \textbf{XOM} & 0.20  & 0.21  & 0.21  & 0.20  & 0.36  & 0.37 \\
    % \textbf{MA} & 0.16  & 0.16  & 0.16  & 0.17  & 0.32  & 0.33 \\
    
    % \bottomrule
    \end{tabular}%
}    
  \label{tab_RMSE}%
\end{table}%

\begin{table}
\tiny
% [width=1.0\linewidth,cols=7,pos=h!]
  \centering
  \caption{LogLikelihood per sample for different distributions}
% \resizebox{\textwidth}{!}{
\begin{tabular}{lcccccc}
    % \begin{tabularx}{\textwidth}{XXXXXXX}
    % {\tblwidth}{@{}LCCCCCC@{}}
    \toprule
       & \multicolumn{1}{p{4.215em}}{\boldmath{}\textbf{GMM\newline{}$N_C = 3$}\unboldmath{}} & \multicolumn{1}{p{4.215em}}{\boldmath{}\textbf{GMM\newline{}$N_C = 4$}\unboldmath{}} & \multicolumn{1}{p{4.215em}}{\boldmath{}\textbf{GMM\newline{}$N_C = 5$}\unboldmath{}} & \multicolumn{1}{p{4.215em}}{\boldmath{}\textbf{GMM\newline{}$N_C = 6$}\unboldmath{}} & \textbf{Normal\newline{}} &
       \multicolumn{1}{p{4.215em}}{\boldmath{}\textbf{t-student\newline{}d.o.f = 10}\unboldmath{}}
       \\
       \midrule
       \textbf{S\&P 500} & 2.91  & 2.90  & 2.93  & 2.93  & 2.73  & 2.79 \\
    \textbf{MMM} & 2.81  & 2.81  & 2.82  & 2.82  & 2.70  & 2.75 \\
    \textbf{AXP} & 2.18  & 2.19  & 2.20  & 2.20  & 2.01  & 2.07 \\
    \textbf{AMD} & 1.92  & 1.92  & 1.93  & 1.93  & 1.84  & 1.87 \\
    \textbf{AIG} & 1.82  & 1.83  & 1.83  & 1.82  & 1.26  & 1.41 \\
    \textbf{AFL} & 2.27  & 2.29  & 2.29  & 2.29  & 1.90  & 2.02 \\
    \textbf{ABT} & 2.95  & 2.95  & 2.95  & 2.96  & 2.87  & 2.91 \\
    \textbf{MSFTF} & 2.63  & 2.63  & 2.64  & 2.64  & 2.50  & 2.55 \\
    \textbf{AAPL} & 2.36  & 2.36  & 2.36  & 2.36  & 2.30  & 2.33 \\
    \textbf{AMZN} & 2.21  & 2.21  & 2.21  & 2.21  & 2.06  & 2.13 \\
    \textbf{BAC} & 2.02  & 2.03  & 2.04  & 2.05  & 1.65  & 1.75 \\
    \textbf{JPM} & 2.17  & 2.18  & 2.18  & 2.18  & 1.93  & 2.01 \\
    \textbf{JNJ} & 3.24  & 3.24  & 3.24  & 3.25  & 3.06  & 3.14 \\
    \textbf{XOM} & 2.71  & 2.71  & 2.71  & 2.71  & 2.54  & 2.62 \\
    \textbf{MA} & 2.64  & 2.64  & 2.64  & 2.65  & 2.48  & 2.55 \\
    \bottomrule
    % \end{tabularx}%
\end{tabular}
% }
  \label{tab_loglikelihood}%
\end{table}%

\begin{table}
% [width=1.0\linewidth,cols=13,pos=h!]
  \centering
  \caption{KS test's statistics and p-values}
\resizebox{\textwidth}{!}{
    \begin{tabular}{lcccccccccccc}
    % {@{}LCCCCCCCCCCCC@{}}
\toprule
      & \multicolumn{2}{p{7.07em}}{\boldmath{}\textbf{GMM\newline{}$N_C = 3$}\unboldmath{}} & \multicolumn{2}{p{7.07em}}{\boldmath{}\textbf{GMM\newline{}$N_C = 4$}\unboldmath{}} & \multicolumn{2}{p{7.07em}}{\boldmath{}\textbf{GMM\newline{}$N_C = 5$}\unboldmath{}} & \multicolumn{2}{p{7.07em}}{\boldmath{}\textbf{GMM\newline{}$N_C = 6$}\unboldmath{}} & \multicolumn{2}{c}{\textbf{Normal}} & \multicolumn{2}{p{7.07em}}{\textbf{t-student}\newline{}d.o.f = 10} \\
\midrule
      & p-value & statistic & p-value & statistic & p-value & statistic & p-value & statistic & p-value & statistic & p-value & statistic \\
\textbf{S\&P} & 0.149 & 0.046 & 0.513 & 0.033 & 0.679 & 0.029 & 0.865 & 0.024 & 0.00  & 0.119 & 0.00  & 0.122 \\
\textbf{MMM} & 0.341 & 0.038 & 0.149 & 0.046 & 0.865 & 0.024 & 0.451 & 0.034 & 0.00  & 0.086 & 0.00  & 0.090 \\
\textbf{AXP} & 0.394 & 0.036 & 0.513 & 0.033 & 0.838 & 0.025 & 0.451 & 0.034 & 0.00  & 0.111 & 0.00  & 0.115 \\
\textbf{AMD} & 0.136 & 0.046 & 0.293 & 0.039 & 0.890 & 0.023 & 0.679 & 0.029 & 0.00  & 0.074 & 0.00  & 0.079 \\
\textbf{AIG} & 0.084 & 0.050 & 0.341 & 0.038 & 0.149 & 0.046 & 0.102 & 0.049 & 0.00  & 0.186 & 0.00  & 0.191 \\
\textbf{AFL} & 0.316 & 0.038 & 0.513 & 0.033 & 0.271 & 0.040 & 0.913 & 0.022 & 0.00  & 0.160 & 0.00  & 0.164 \\
\textbf{ABT} & 0.865 & 0.024 & 0.367 & 0.037 & 0.612 & 0.030 & 0.777 & 0.026 & 0.00  & 0.062 & 0.00  & 0.069 \\
\textbf{MSFTF} & 0.316 & 0.038 & 0.230 & 0.042 & 0.976 & 0.019 & 0.712 & 0.028 & 0.00  & 0.090 & 0.00  & 0.097 \\
\textbf{AAPL} & 0.777 & 0.026 & 0.367 & 0.037 & 0.933 & 0.022 & 0.913 & 0.022 & 0.00  & 0.066 & 0.00  & 0.072 \\
\textbf{AMZN} & 0.394 & 0.036 & 0.865 & 0.024 & 0.482 & 0.034 & 0.950 & 0.021 & 0.00  & 0.085 & 0.00  & 0.091 \\
\textbf{BAC} & 0.032 & 0.058 & 0.093 & 0.050 & 0.149 & 0.046 & 0.394 & 0.036 & 0.00  & 0.158 & 0.00  & 0.163 \\
\textbf{JPM} & 0.124 & 0.047 & 0.341 & 0.038 & 0.422 & 0.035 & 0.578 & 0.031 & 0.00  & 0.124 & 0.00  & 0.128 \\
\textbf{JNJ} & 0.195 & 0.043 & 0.913 & 0.022 & 0.745 & 0.027 & 0.808 & 0.026 & 0.00  & 0.096 & 0.00  & 0.103 \\
\textbf{XOM} & 0.645 & 0.030 & 0.808 & 0.026 & 0.178 & 0.044 & 0.545 & 0.032 & 0.00  & 0.093 & 0.00  & 0.100 \\
\textbf{MA} & 0.712 & 0.028 & 0.513 & 0.033 & 0.712 & 0.028 & 0.777 & 0.026 & 0.00  & 0.082 & 0.00  & 0.092 \\
\midrule
results & \multicolumn{2}{c}{\textcolor[rgb]{ 0,  .635,  .314}{\checkmark}} & \multicolumn{2}{c}{\textcolor[rgb]{ 0,  .635,  .314}{\checkmark}} & \multicolumn{2}{c}{\textcolor[rgb]{ 0,  .635,  .314}{\checkmark}} & \multicolumn{2}{c}{\textcolor[rgb]{ 0,  .635,  .314}{\checkmark}} & \multicolumn{2}{c}{\textcolor[rgb]{ 1,  0,  0}{X}} & \multicolumn{2}{c}{\textcolor[rgb]{ 1,  0,  0}{X}} \\
\bottomrule
    \end{tabular}%
}
  \label{tab_kstest}%
\end{table}%

\begin{figure*}
\centering
\subfigure[]{{\includegraphics[width=5.0cm]{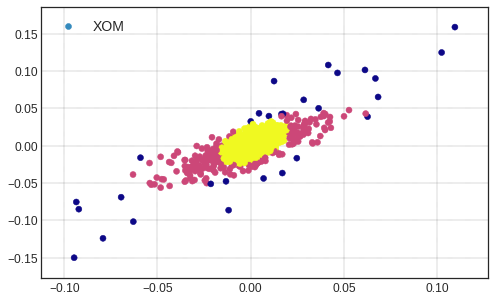}}}
\subfigure[]{{\includegraphics[width=5.0cm]{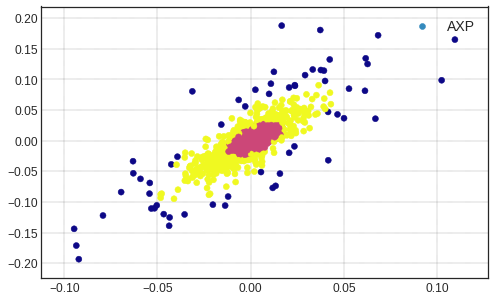}}}
\subfigure[]{{\includegraphics[width=5.0cm]{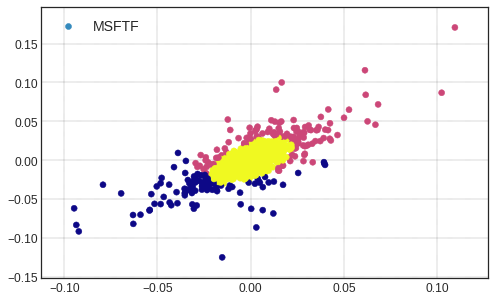}}}\\
\subfigure[]{{\includegraphics[width=5.0cm]{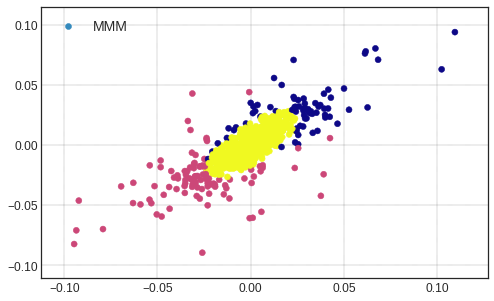}}}
\subfigure[]{{\includegraphics[width=5.0cm]{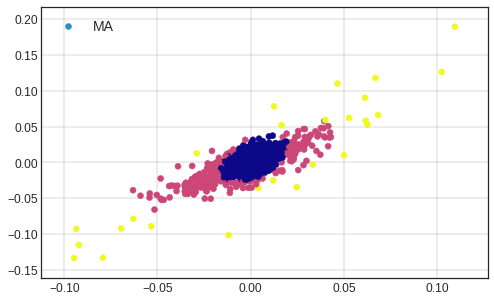}}}
\subfigure[]{{\includegraphics[width=5.0cm]{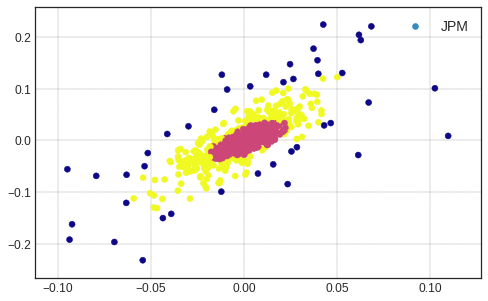}}}\\
\subfigure[]{{\includegraphics[width=5.0cm]{figs/corr_sp/JPM.png}}}
\subfigure[]{{\includegraphics[width=5.0cm]{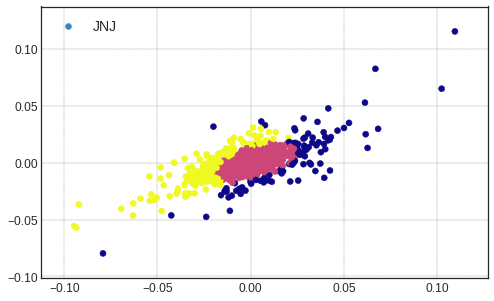}}}
\subfigure[]{{\includegraphics[width=5.0cm]{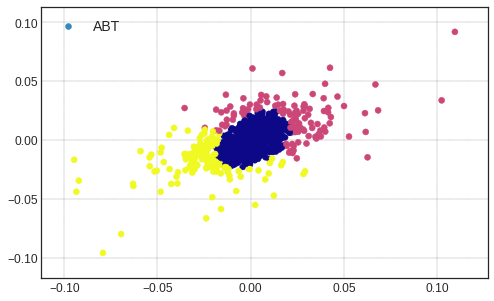}}}
\caption{Clustering stock returns against S\&P500 with Gmm $N_c = 3$}
\label{fig:corr_sp}
\end{figure*}

\begin{figure*}
\centering
\subfigure[]{{\includegraphics[width=5.0cm]{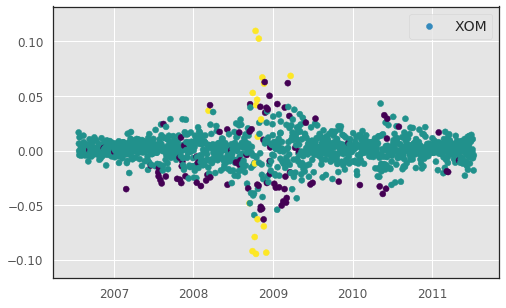}}}
\subfigure[]{{\includegraphics[width=5.0cm]{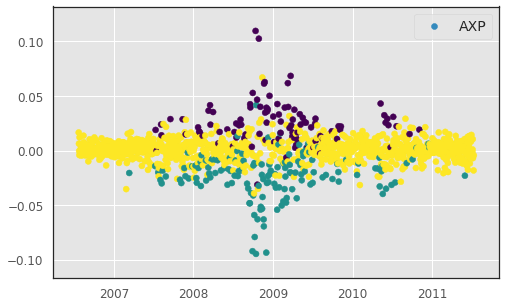}}}
\subfigure[]{{\includegraphics[width=5.0cm]{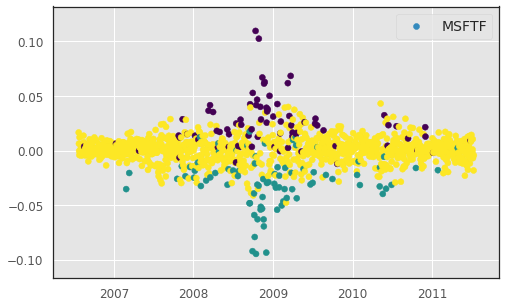}}}\\
\subfigure[]{{\includegraphics[width=5.0cm]{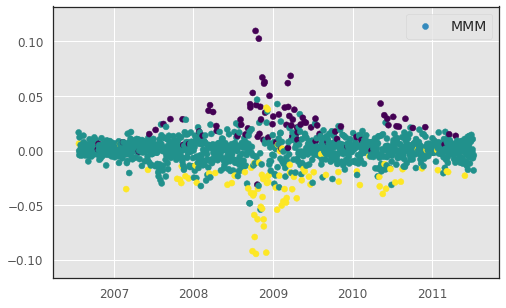}}}
\subfigure[]{{\includegraphics[width=5.0cm]{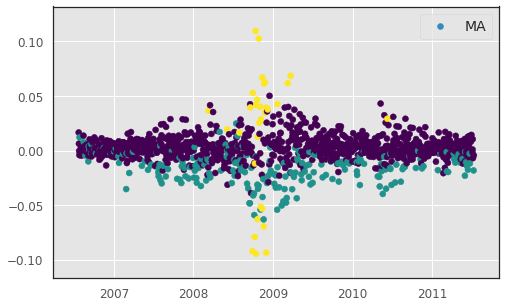}}}
\subfigure[]{{\includegraphics[width=5.0cm]{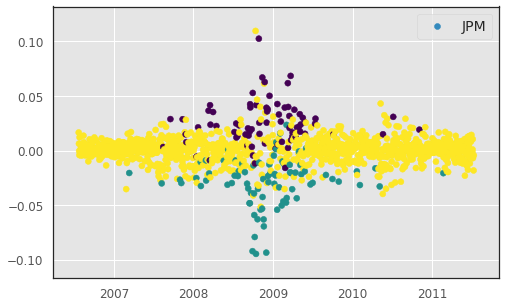}}}\\
\subfigure[]{{\includegraphics[width=5.0cm]{figs/clustering/JPM.png}}}
\subfigure[]{{\includegraphics[width=5.0cm]{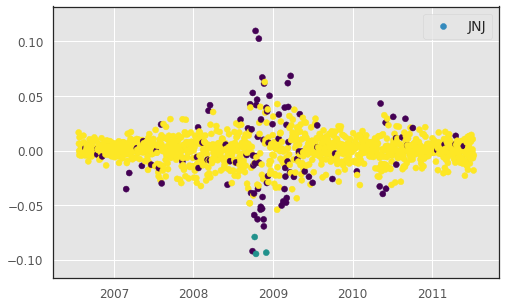}}}
\subfigure[]{{\includegraphics[width=5.0cm]{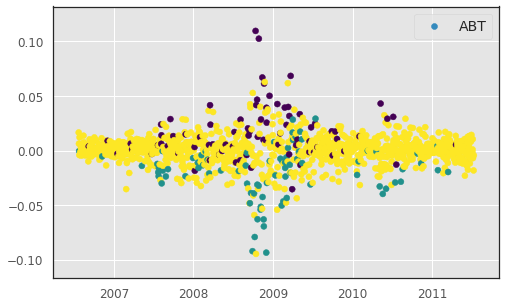}}}
\caption{Clustering stock returns during the time with $N_c = 3$}
\label{fig:clustering}
\end{figure*}

Figure \ref{fig:corr_sp} shows how the Gmm with $N_c = 3$ can model joint distribution of a stock against the market index. The horizontal axes are index of S\&P500, while the latter axes illustrate the individual stock returns. Every single cluster has a unique correlation structure. 
Since in higher dimensions correlation structures of stock returns become more complicated, Gmm can be trained under changing market conditions to simulate future returns. The graph for all stocks are shown in appendix in figures \ref{fig:corr_sp_a} and \ref{fig:corr_sp_b}. Figure \ref{fig:clustering} also shows clustering of time series returns with $N_c = 3$ components. All of stocks are shown in appendix in figures \ref{fig:clustering_a} and \ref{fig:clustering_b}. 

\subsection{Christofferson test and Quadratic Loss Function}

For checking the validity and accuracy of VaRs, we use the proposed method by \cite{christoffersen1998evaluating}. The VaR result should satisfy both unconditional coverage and independence properties of exceeding returns to be accepted. Two statistics are 
\begin{equation}
    LR_{uc} = -2 \ln{[(1-p)^{n-x}p^x]} + 2\ln{[(1 - \frac{x}{n})^{n-x}(\frac{x}{n})^x]} ,
\end{equation}
and
\begin{equation}
\begin{array}{rl}{L R_{i n d}=-2} & {\ln \left[\left(1-\hat{\pi}_{2}\right)^{n_{00}+n_{11}} \hat{\pi}_{2}^{n_{01}+n_{11}}\right]+2 \ln \left[\left(1-\hat{\pi}_{01}\right)^{n_{00}} \hat{\pi}_{01}\left(1-\hat{\pi}_{11}\right)^{n_{10}} \pi_{11}^{n_{11}}\right]} \end{array} ,
\end{equation}
where
$${\hat{\pi}_{01}=} {\frac{n_{01}}{n_{00}+n_{01}}, \quad \hat{\pi}_{11} =  \frac{n_{11}}{n_{10}+n_{11}}, \quad \hat{\pi}_{2}=\frac{n_{01}+n_{11}}{n_{00}+n_{10}+n_{01}+n_{11}}}.$$
Both statistics follow a $\chi^2_{df=2}$ distribution. Here, $n_{ij}$, ($i, j \in \{0,1\}$) are the number of days with $i$ exceedance and $j$ exceedance observed in the previous day. 

For a loss function $L$, we use a quadratic loss, proposed by \cite{lopez1999methods} and \cite{martens2009forecasting}, such that $L = \frac{\sum_{t=1}^{T} QL_t}{T}$, where
\begin{equation}
    QL_{t} = I_{r_{t} \leq VaR_{t}} (1 + (r_{t} - VaR_{t})^2)
\end{equation}
and $I$ is an indicator function.
% For checking the validity of ES, \cite{mcneil2005quantitative}, suggests a Zero Mean testing. According to this method, first the VaR should be valid (which we investigate it using Christofferson test), and second, the excess loss should be back-tested. By writing 
% \begin{equation}
%      ES_\alpha = VaR_\alpha + (ES_\alpha - VaR_\alpha),
% \end{equation} 
% the term $ES_\alpha - VaR_\alpha$ is excess loss and should be back-tested through the test statistic
% \begin{equation}
%     S = (L - ES_\alpha)1_{L>VaR_\alpha}, 
% \end{equation}
% For testing this hypothesis, we use a sample $t$ test, assuming $S$ samples are independent and identically distributed.

A VaR result is valid if the p-value for both unconditional statistics (UC) and independent statistics (ID) are greater than $0.01$. % The same confidence level is used for checking the validity of ES with its t-statistic. Note ES back-test is only applicable when the VaR test is passed.

\subsection{Calculating VaR and back-testing}

\subsubsection{Individual Stocks' VaR}

\begin{figure*}
\centering
\subfigure[]{{\includegraphics[width=5.0cm]{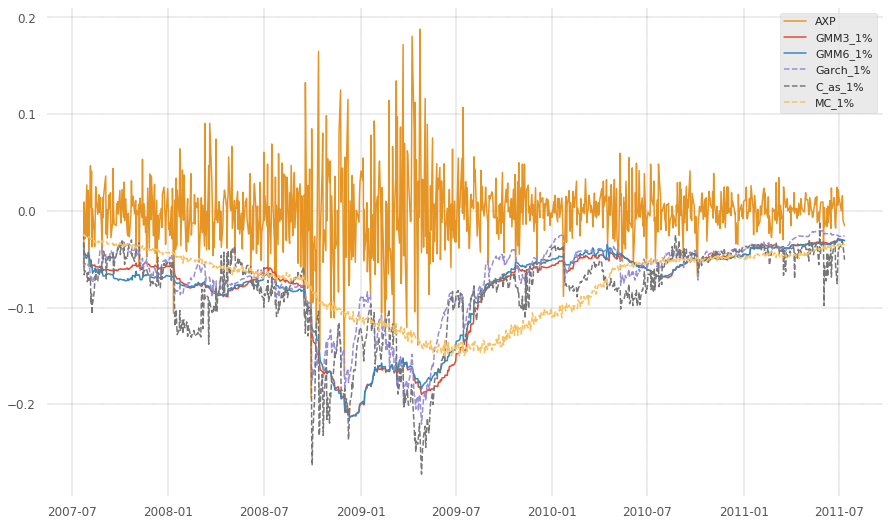}}}
\subfigure[]{{\includegraphics[width=5.0cm]{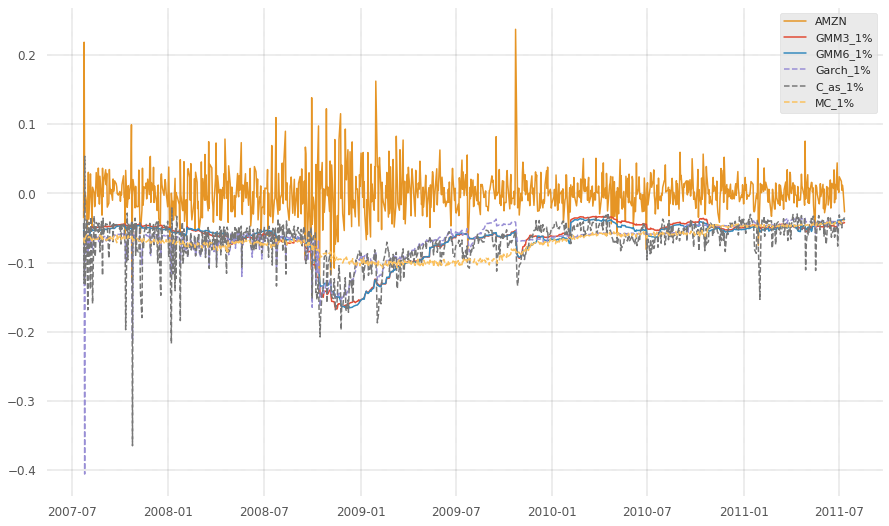}}}
\subfigure[]{{\includegraphics[width=5.0cm]{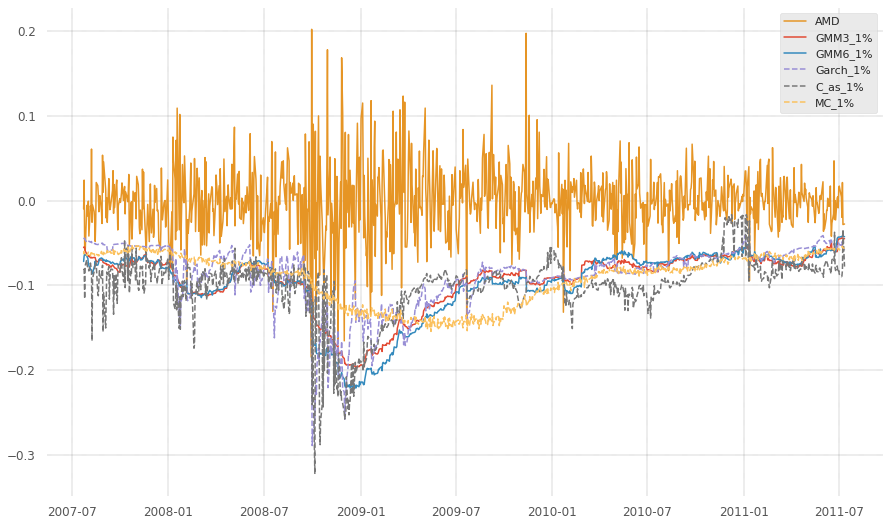}}}\\
\subfigure[]{{\includegraphics[width=5.0cm]{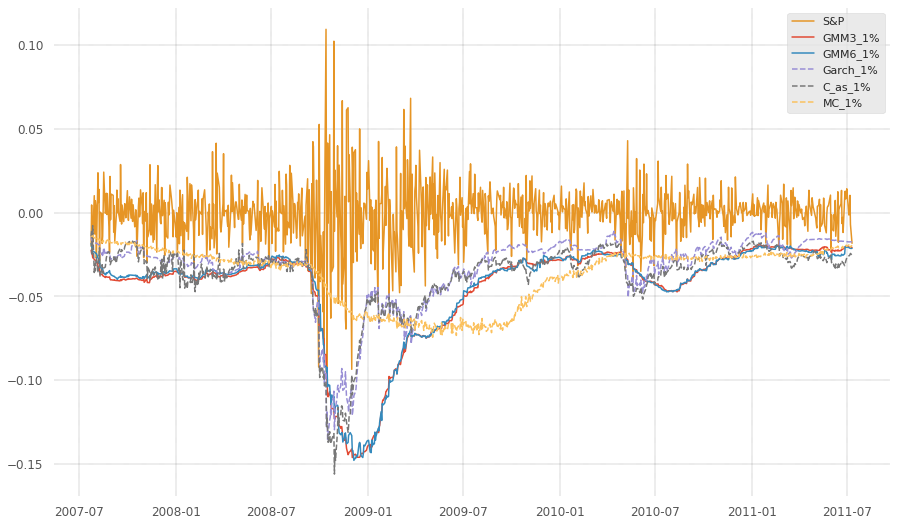}}}
\subfigure[]{{\includegraphics[width=5.0cm]{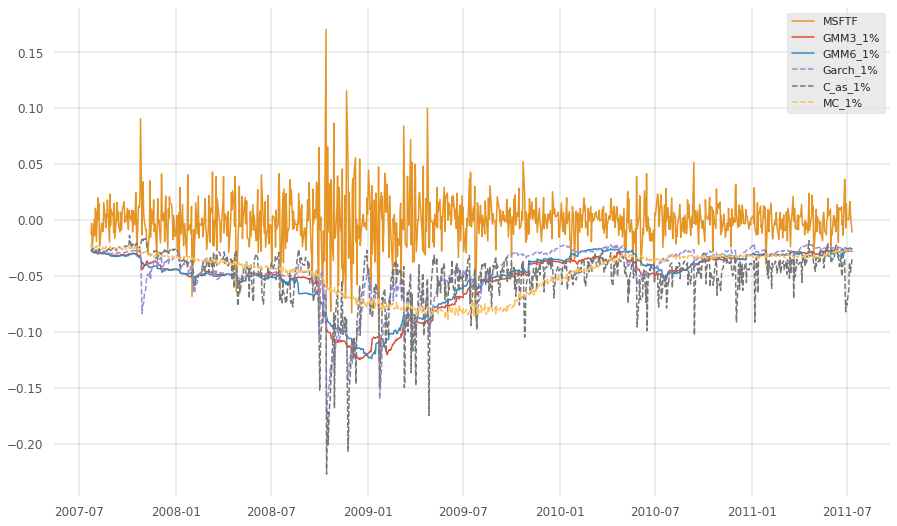}}}
\subfigure[]{{\includegraphics[width=5.0cm]{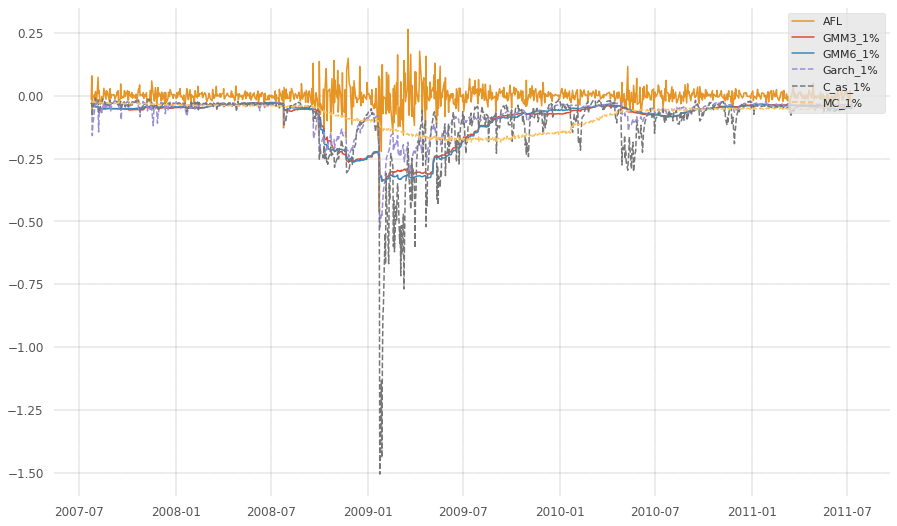}}}\\
\subfigure[]{{\includegraphics[width=5.0cm]{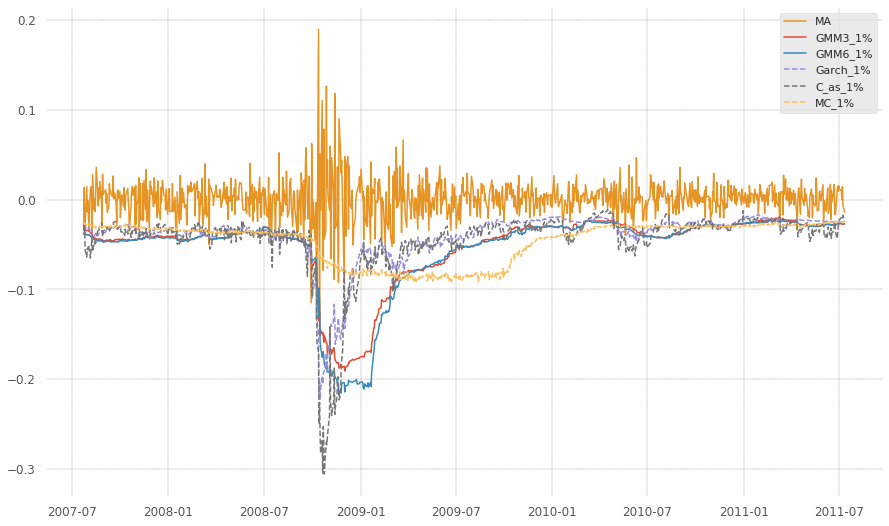}}}
\subfigure[]{{\includegraphics[width=5.0cm]{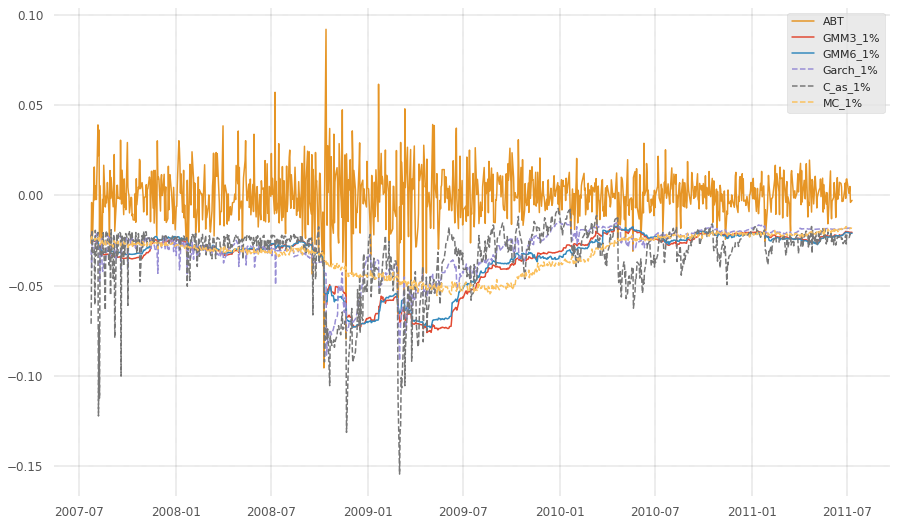}}}
\subfigure[]{{\includegraphics[width=5.0cm]{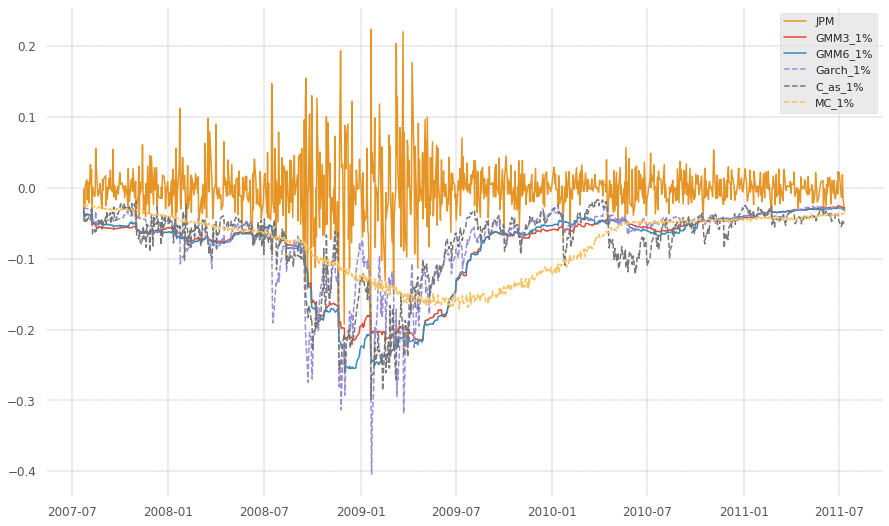}}}

\caption{Value at Risk 1 \% for individual stocks during a crisis period}
\label{fig:VAR1}
\end{figure*}

\begin{figure*}
\centering
\subfigure[]{{\includegraphics[width=5.0cm]{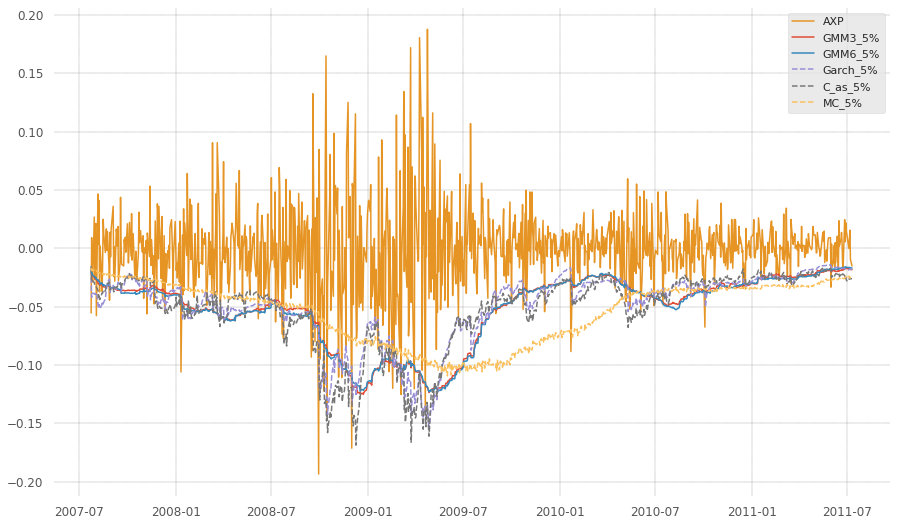}}}
\subfigure[]{{\includegraphics[width=5.0cm]{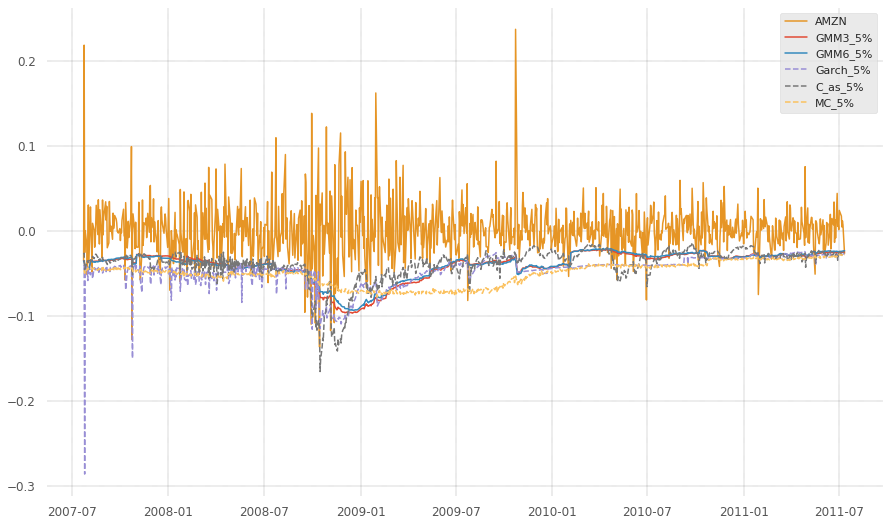}}}
\subfigure[]{{\includegraphics[width=5.0cm]{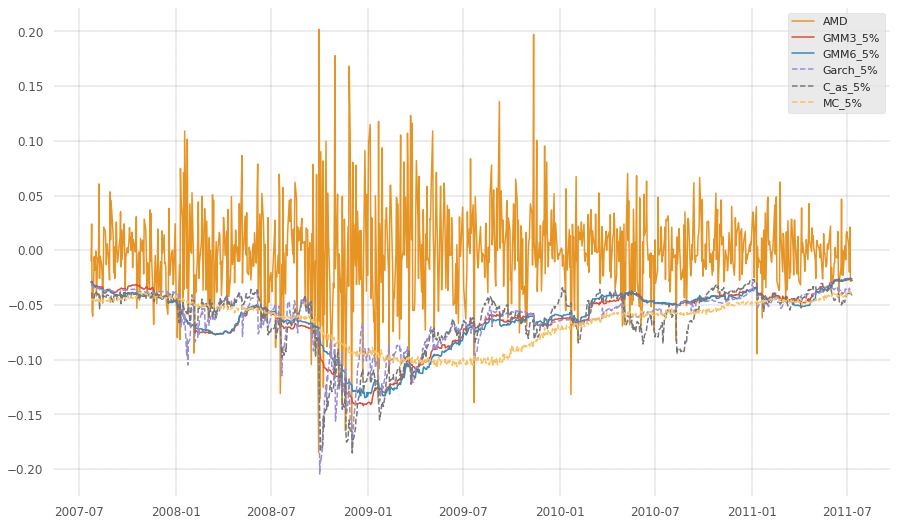}}}\\
\subfigure[]{{\includegraphics[width=5.0cm]{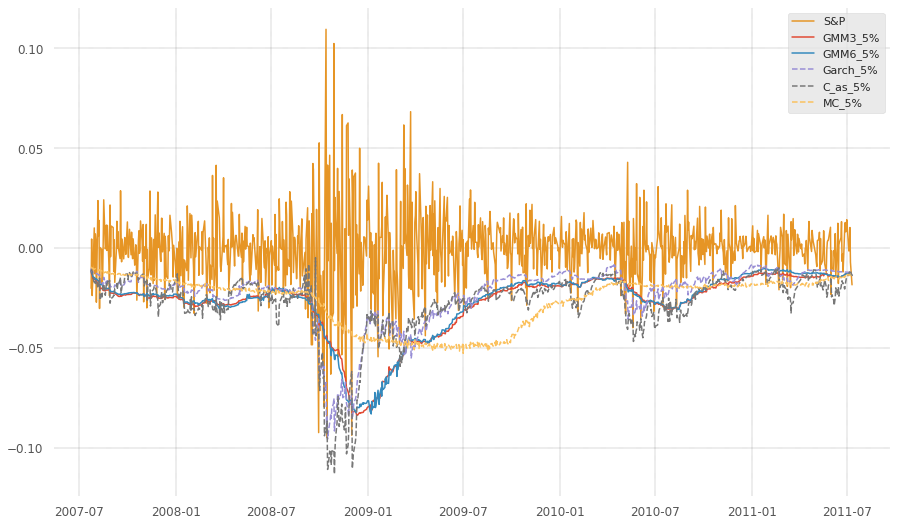}}}
\subfigure[]{{\includegraphics[width=5.0cm]{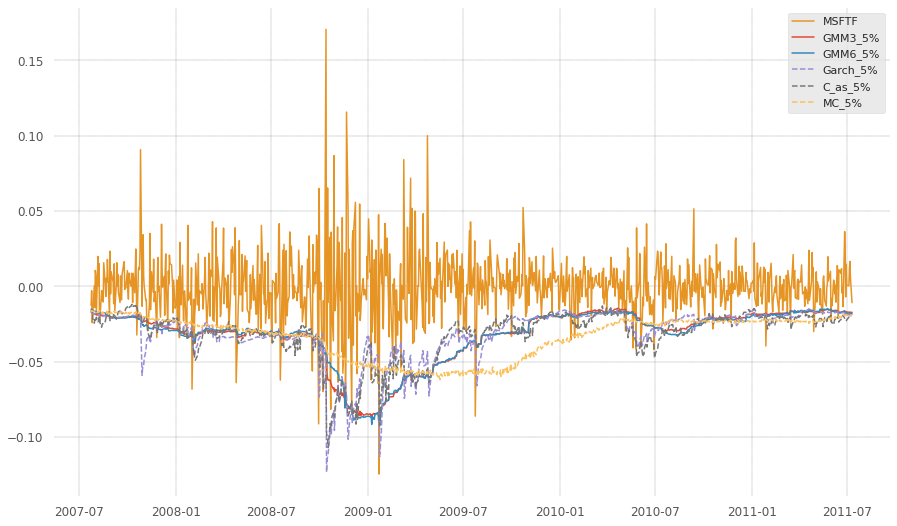}}}
\subfigure[]{{\includegraphics[width=5.0cm]{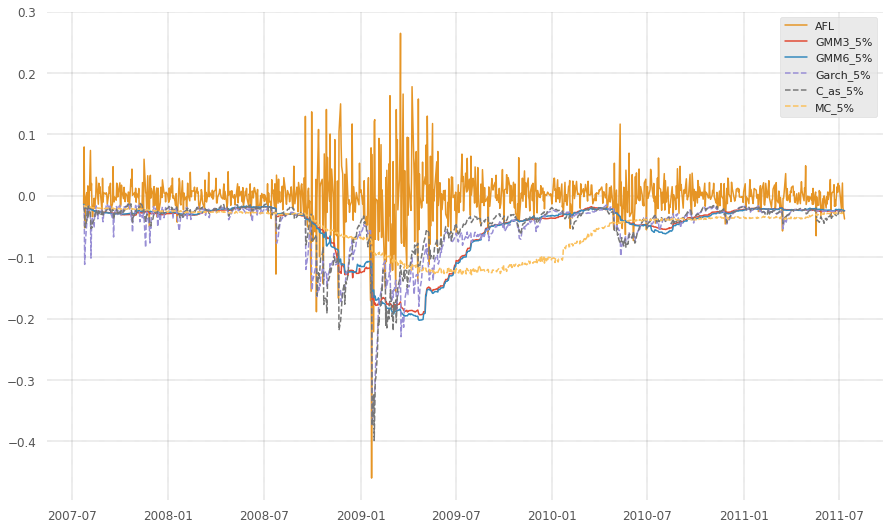}}}\\
\subfigure[]{{\includegraphics[width=5.0cm]{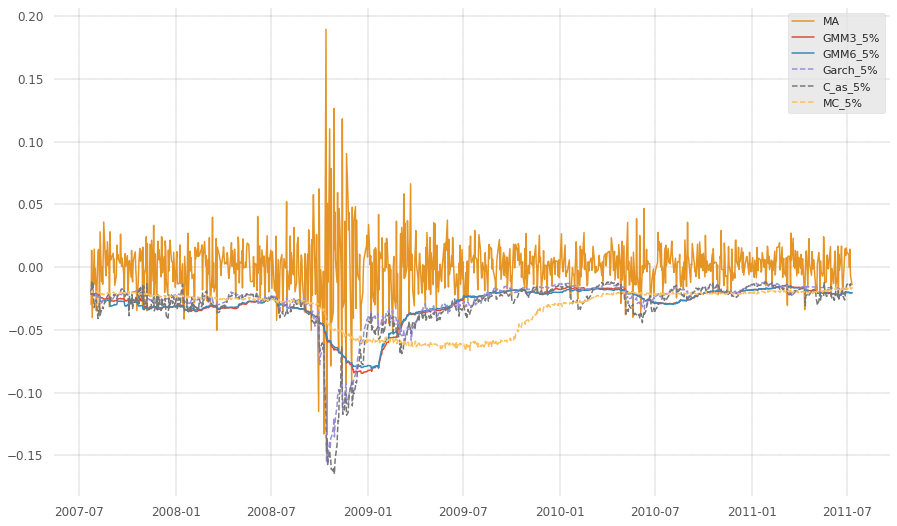}}}
\subfigure[]{{\includegraphics[width=5.0cm]{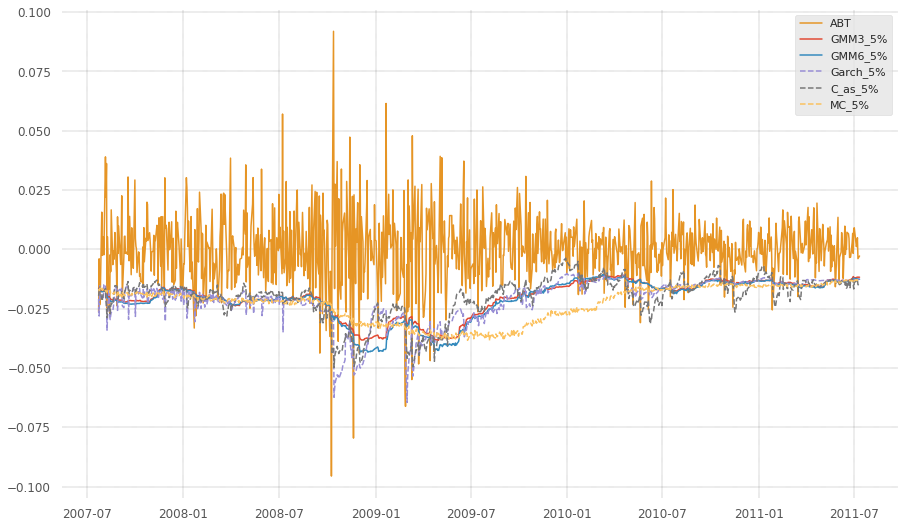}}}
\subfigure[]{{\includegraphics[width=5.0cm]{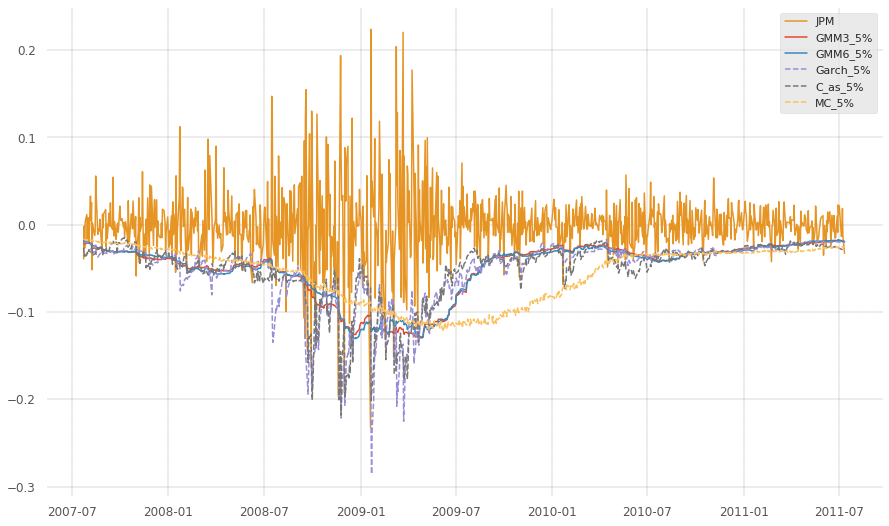}}}
\caption{Value at Risk 5 \% for individual stocks during a crisis period}
\label{fig:VAR5}
\end{figure*}

The VaR for 1000 days, starting from $7/24/2007$ for both individual stocks and the equally weighted portfolio, are calculated. The results of VaR for historical, parametric (variance-covariance or Normal), GBM Monte Carlo Simulation (MS), CaViaR adaptive (C-AD), CaViaR asymmetric (C-AS), CaViaR symmetric-VaR (C-SAV), CaViaR GARCH (C-GARCH) proposed by \cite{engle2004caviar}, and GARCH(1,1) are compared with Gmm ($N_c = 3, 4, 5,$ and $6$) in figure \ref{fig:VAR1} and \ref{fig:VAR5} for $1\%$ and $5 \%$ levels of significance. See figures \ref{fig:VAR1_a}, \ref{fig:VAR1_b}, \ref{fig:VAR5_a} and \ref{fig:VAR5_b} in appendix for all stocks. The calculated VaR are back-tested based on both unconditional and independence statistics of Christofferson back-testing, as well as a quadratic loss function. The decision for  rejecting a VaR output depends on the corresponding p-value compared to the significance level of $1\%$. The results are shown in tables \ref{tab_VaR1} and \ref{tab_VaR5} (1\% and 5\%) with their associated p-values. Not rejected VaRs are shown in green, while the rejected ones are in red. Where $\alpha = 1\%$ all of VaRs for different stocks and $N_c$s are not rejected. As table \ref{tab_VaR1} shows the number of failed VaRs for Historical Simulation (HS), Parametric Variance-Covariance VaR (Normal), Monte Carlo (MC), GARCH(1,1), CaViaR symmetric-VaR (C-SAV),  CaViaR asymmetric (C-AS),  CaViaR GARCH (C-G), CaViaR adaptive (C-AD), are $6, 12, 12, 10, 0, 2, 1$ and $5$, respectively. Table \ref{tab_lossVaR1} shows the quadratic loss $L$ for different VaRs. The Gmms-VaR are lower than other models for almost all stocks in quadratic loss. 
In table \ref{tab_VaR5}, the number of failed VaR cases for HS, Para, MC, GARCH(1,1), C-SAV, C-AS, C-GARCH and C-AD are $8, 4, 4, 4, 0, 1, 1 \text{ and } 5 $, while for the Gmms with different numbers of components, $N_c$, are $3, 1, 3$ and $2$. Figures \ref{fig:VAR1} and \ref{fig:VAR5} plot some of the individual stocks with HS VaR, Para VaR, MC VaR, GARCH(1,1) VaR C-SAV, C-AS, C-G, C-AD and Gmm VaR with $N_c = 3, 4, 5$ and $6$ for levels of 1\% and 5\%. Results of quadratic loss function $L$ for $VaR_{5\%}$ are also shown in table \ref{tab_lossVaR5}, suggesting a lower amount of losses for Gmms VaR, in many cases. 

We also apply christofferson's test on our model when $\sigma_{short}$ is changing in range $\{10, 20, ..., 90\}$. The rejected VaRs are shown by R and marked by red in tables \ref{tab_sigma_s3}, \ref{tab_sigma_s4}, \ref{tab_sigma_s5} and \ref{tab_sigma_s6}. Otherwise they are illustrated by NR (not rejected) and shown by green. We conclude that for small $\sigma_{short} \leq 30$ the VaRs are mostly rejected, while for greater amounts they have better results. Moreover, there is no significant differences between $\sigma_{short} = \{60, 70, 80\}$. We do not use $\sigma_{short} \geq 100$, because we want to apply the effect of recent days (closer than 100) on the model's output.

\subsubsection{Portfolio of Stocks' VaR} 

\begin{figure*}
\centering
\subfigure[]{{\includegraphics[width=7.0cm]{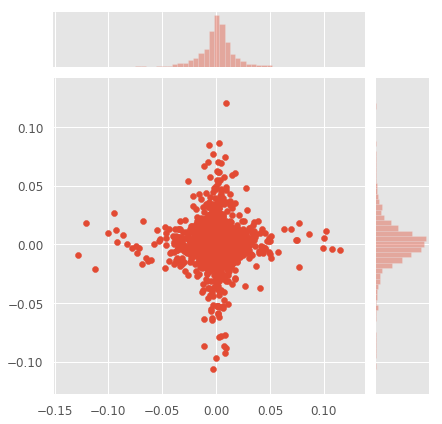}}}
\subfigure[]{{\includegraphics[width=7.0cm]{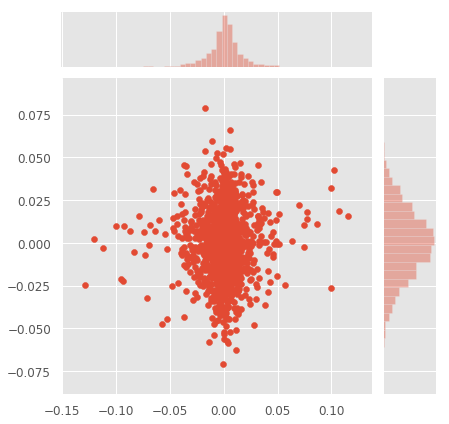}}}
\caption{Comparing Simulated portfolio returns between Gmm $N_c = 3$ (left) and Gaussian Copula (right)}
\label{fig:GC}
\end{figure*}

We consider a portfolio consisting of all the 14 stocks in our analysis and the S\&P500 index itself. Clearly, for a portfolio of stocks, the correlation structure between constituent assets is not negligible. As Gaussian Copulas (GC) are strong tools for modeling joint distribution of random variables, we compare our proposed model with the Gaussian Copula. Portfolio returns are simulated for 1000 times, and figure \ref{fig:GC} shows a comparison between GC and Gmm ($N_c = 3$) against real data. The vertical axes  represent real returns while the horizontal axes show the simulated ones. Clearly, during a crisis a GC is unable to capture the fat-tail distributions, in contrary to a Gmm. KS test is also employed for checking goodness of fit, and the corresponding p-values are shown in table \ref{tab_ksGC}, showing that simulated returns in Gmm conform to a real distribution, unlike those came from a GC.
% Table generated by Excel2LaTeX from sheet 'KS'
\begin{table}
% [width=1.0\linewidth,cols=5,pos=t!]
  \centering
  \caption{KS test results between real data and simulated one}
% \resizebox{\textwidth}{!}{
    \begin{tabular}{lccccc}
    % {\tblwidth}{@{}LCCCCC@{}}
    \toprule
      & \multicolumn{1}{p{4.215em}}{\textbf{GC}} & \multicolumn{1}{p{4.215em}}{\boldmath{}\textbf{GMM \newline{}$N_c = 3$}\unboldmath{}} & \multicolumn{1}{p{4.215em}}{\boldmath{}\textbf{GMM \newline{}$N_c = 4$}\unboldmath{}} & \multicolumn{1}{p{4.215em}}{\boldmath{}\textbf{GMM \newline{}$N_c = 5$}\unboldmath{}} & \multicolumn{1}{p{4.215em}}{\boldmath{}\textbf{GMM \newline{}$N_c = 6$}\unboldmath{}} \\
\midrule
statistic & 0.1214 & 0.0639 & 0.0575 & 0.0535 & 0.0615 \\
p-value & 1.86E-08 & 0.0120 & 0.0318 & 0.0554 & 0.0175 \\
 \\
& \textcolor[rgb]{ 1,  0,  0}{X} & \textcolor[rgb]{ 0,  .635,  .314}{ \checkmark} & \textcolor[rgb]{ 0,  .635,  .314}{ \checkmark} & \textcolor[rgb]{ 0,  .635,  .314}{ \checkmark} & \textcolor[rgb]{ 0,  .635,  .314}{ \checkmark} \\
    \bottomrule
    \end{tabular}%
% }
  \label{tab_ksGC}%
\vspace{1ex}\\
{\justifying \noindent Note: This table shows that if the simulated returns from different models are following the same distribution as real returns. GC denotes Gaussian Copula. The threshold for not rejecting the null hypothesis based on the p-value is considered 0.01.
\par}
\end{table}

\begin{figure*}
\centering
\subfigure[]{{\includegraphics[width=15.0cm, height=9cm]{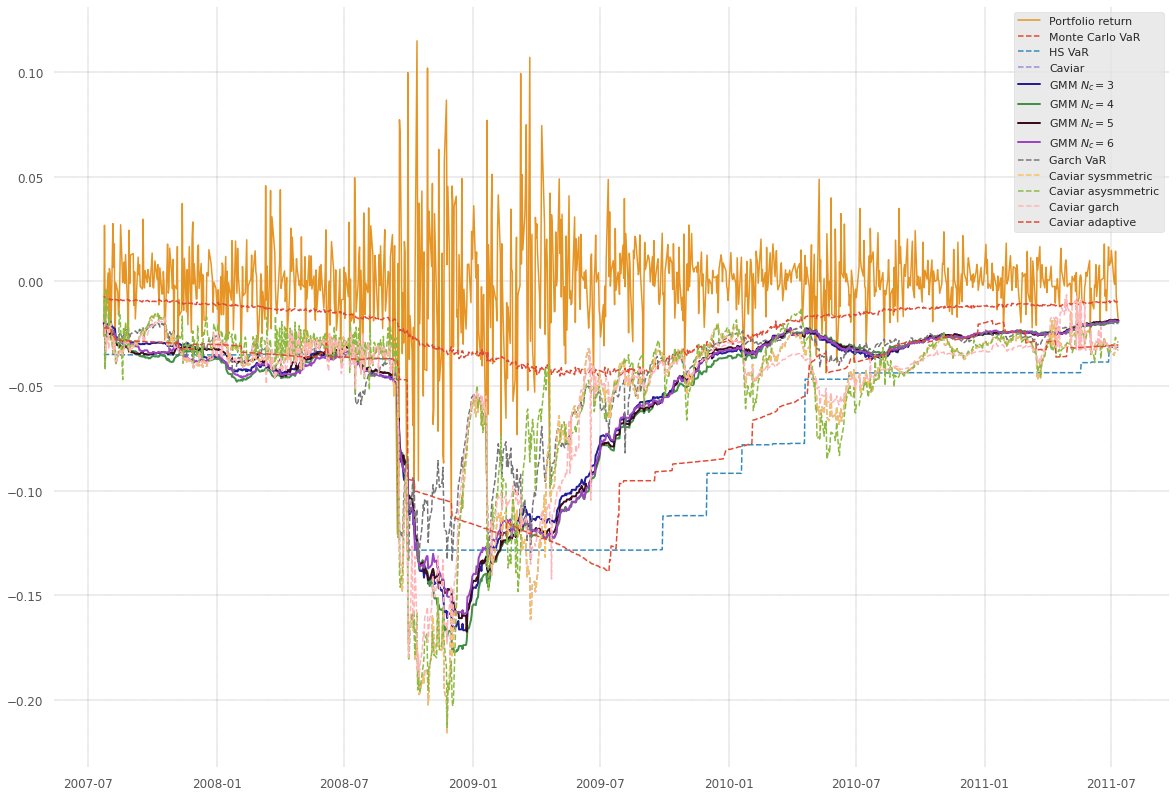}}}\\
\subfigure[]{{\includegraphics[width=15.0cm, height=9cm]{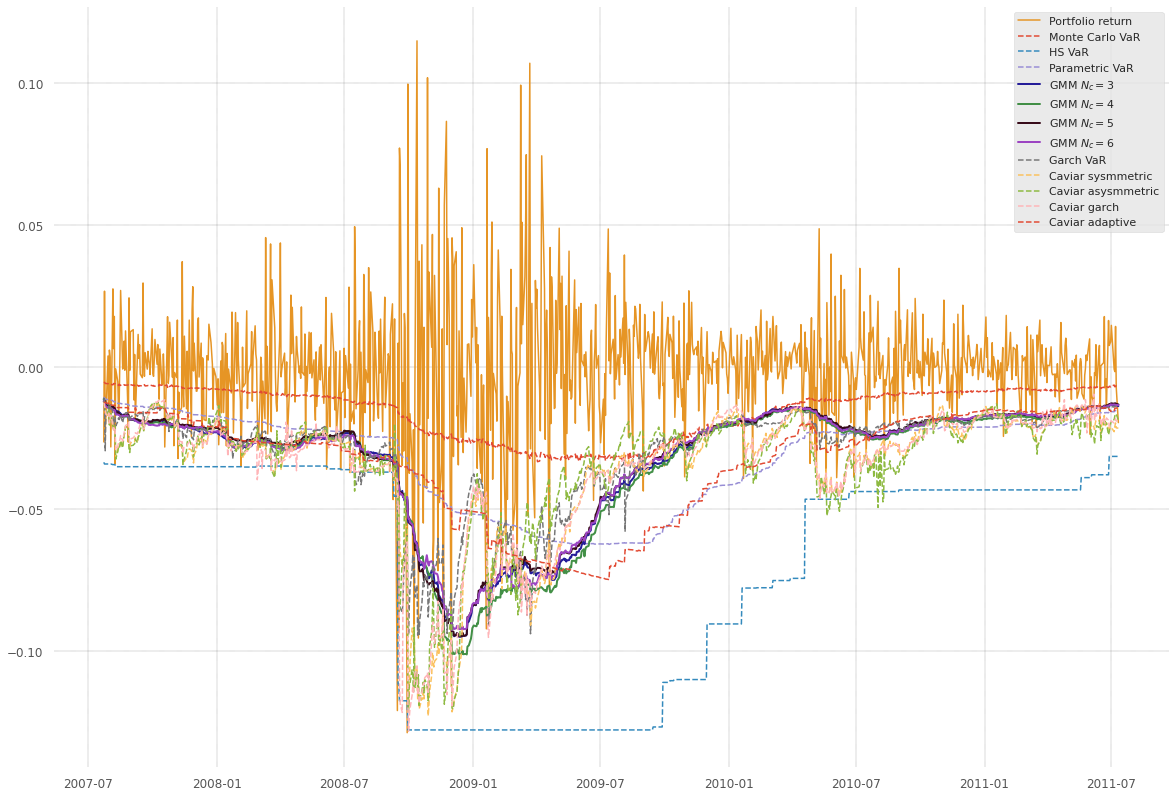}}}
\caption{Value at Risk 1\% (up) and 5\% (down) for a portfolio of stocks during a crisis period}
\label{fig:portfolio_var}
\end{figure*}

The VaRs are calculated with the same parameters we used in calculation of individual stocks, and shown in figure \ref{fig:portfolio_var}. The p-values for testing the VaR in unconditional coverage (UC), Independence (ID) and Conditional coverage (CC) tests are demonstrated in table \ref{tab_port_var}. When all these values are higher than the threshold $0.01$ the VaR is not rejected and shown by "NR", and otherwise, it is rejected and shown by "R". In the $5\%$ confidence level, all Gmms VaR and the GARCH(1,1) VaR are not rejected, whereas other VaRs are being rejected. In the $1\%$ level, HS VaR, Gmm $(N_c = 4)$ VaR and Gmm $(N_c = 6)$ VaR are not rejected and other VaRs are rejected. We rank the 9 VaR methods for each level, firstly according to whether they are rejected or not, and secondly based on their quadratic losses. As it can be seen, Gmm results are superior according to their UC, ID, and quadratic losses. The execution time for calculating VaRs in 1000 days are shown in table \ref{tab_times}. Although our Gmm model is based on Monte Carlo approach, and simulates 3000 returns for each day, its execution time is significantly lower than other models such as GARCH(1,1), GC, Parametric VaR and Monte Carlo simulation based on a GBM, as well as different CaViaR types. It does not necessitates of any correlation matrix, therefore the computation is fast, and Gmm can model the joint distributions using EM algorithm. 
\begin{table}
% [width=1.0\linewidth,cols=7,pos=t!]
  \centering
  \caption{Execution time for different VaR models}
    \begin{tabular}{lcccccc}
    % {\tblwidth}{@{}CCCCCCC@{}}
\toprule
\textbf{Model} & \multicolumn{1}{p{3.285em}}{\textbf{GARCH\newline{}(1,1)}} & \multicolumn{1}{p{4.215em}}{\boldmath{}\textbf{GMM \newline{}$N_c = 3$}\unboldmath{}} & \multicolumn{1}{p{4.215em}}{\boldmath{}\textbf{GMM \newline{}$N_c = 4$}\unboldmath{}} & \multicolumn{1}{p{4.215em}}{\boldmath{}\textbf{GMM \newline{}$N_c = 5$}\unboldmath{}} & \multicolumn{1}{p{4.215em}}{\boldmath{}\textbf{GMM \newline{}$N_c = 6$}\unboldmath{}} & \multicolumn{1}{p{2.93em}}{\textbf{GC}} \\
\midrule
\textbf{time (seconds)} & 260.60 & 26.94 & 29.54 & 31.75 & 33.93 & 604.60 \\
\midrule
\textbf{Model} & \multicolumn{1}{p{3.285em}}{\textbf{Para}} & \multicolumn{1}{p{4.215em}}{\textbf{MC}} & \multicolumn{1}{p{4.215em}}{\textbf{C-sy}} & \multicolumn{1}{p{4.215em}}{\textbf{C-as}} & \multicolumn{1}{p{4.215em}}{\textbf{C-G}} & \multicolumn{1}{p{2.93em}}{\textbf{C-ad}} \\
\midrule
\textbf{time (seconds)} & 247.00 & 2693.43 & 520.56 & 726.74 & 789.97 & 726.08 \\
\bottomrule

    \end{tabular}%
  \label{tab_times}%
     \vspace{1ex}
      {\raggedright Note: This table compares the computational time between different models. GC denotes the Gaussian Copula, Para is parametric (variance-covariance), MC is Monte Carlo Simulation, C-sy is CaViaR Symmetric, C-as is CaViaR Asymmetric, C-G is CaViaR GARCH, and C-ad is CaViaR Adaptive model.
      \par}
\end{table}%

On the other hand, for one single time parameter training $\boldsymbol{\theta}$ in a Gmm, by initializing with respect to k-means and previous parameters, the execution time for convergence is 0.0572 and 0.0305 seconds and the number of iterations is 20 and 2, respectively. This technique for parameters initialization lessens the computation cost as well as execution time. Overall, according to the algorithm speed, Gmm outperforms other models.
\begin{table}
% [width=1.0\linewidth,cols=16,pos=t!]
  \centering
  \caption{VaR 1 \% for individual stocks}
\resizebox{\textwidth}{!}{
    \begin{tabular}{lcccccccccccccccccc}
    % {\tblwidth}{@{}LCCCCCCCCCCCCCCCCCC@{}}
\toprule
      & \textbf{test} & \textbf{S\&P500} & \textbf{MMM} & \textbf{AXP} & \textbf{AMD} & \textbf{AIG} & \textbf{AFL} & \textbf{ABT} & \textbf{MSFTF} & \textbf{AAPL} & \textbf{AMZN} & \textbf{BAC} & \textbf{JPM} & \textbf{JNJ} & \textbf{XOM} & \textbf{MA} \\
\midrule
\multicolumn{1}{l}{\multirow{2}[1]{*}{\boldmath{}\textbf{GMM\newline{}$N_c = 3$}\unboldmath{}}} & UC    & \textcolor[rgb]{ 0,  .502,  0}{0.75} & \textcolor[rgb]{ 0,  .502,  0}{0.36} & \textcolor[rgb]{ 0,  .502,  0}{0.14} & \textcolor[rgb]{ 0,  .502,  0}{0.23} & \textcolor[rgb]{ 0,  .502,  0}{0.08} & \textcolor[rgb]{ 0,  .502,  0}{0.08} & \textcolor[rgb]{ 0,  .502,  0}{0.08} & \textcolor[rgb]{ 0,  .502,  0}{0.36} & \textcolor[rgb]{ 0,  .502,  0}{0.36} & \textcolor[rgb]{ 0,  .502,  0}{0.01} & \textcolor[rgb]{ 0,  .502,  0}{0.02} & \textcolor[rgb]{ 0,  .502,  0}{0.08} & \textcolor[rgb]{ 0,  .502,  0}{0.75} & \textcolor[rgb]{ 0,  .502,  0}{0.08} & \textcolor[rgb]{ 0,  .502,  0}{0.02} \\
      & ID    & \textcolor[rgb]{ 0,  .502,  0}{0.62} & \textcolor[rgb]{ 0,  .502,  0}{0.56} & \textcolor[rgb]{ 0,  .502,  0}{0.5} & \textcolor[rgb]{ 0,  .502,  0}{0.19} & \textcolor[rgb]{ 0,  .502,  0}{0.23} & \textcolor[rgb]{ 0,  .502,  0}{0.25} & \textcolor[rgb]{ 0,  .502,  0}{0.23} & \textcolor[rgb]{ 0,  .502,  0}{0.56} & \textcolor[rgb]{ 0,  .502,  0}{0.54} & \textcolor[rgb]{ 0,  .502,  0}{0.37} & \textcolor[rgb]{ 0,  .502,  0}{0.33} & \textcolor[rgb]{ 0,  .502,  0}{0.47} & \textcolor[rgb]{ 0,  .502,  0}{0.62} & \textcolor[rgb]{ 0,  .502,  0}{0.47} & \textcolor[rgb]{ 0,  .502,  0}{0.3} \\
\multicolumn{1}{l}{\multirow{2}[0]{*}{\boldmath{}\textbf{GMM\newline{}$N_c = 4$}\unboldmath{}}} & UC    & \textcolor[rgb]{ 0,  .502,  0}{0.23} & \textcolor[rgb]{ 0,  .502,  0}{0.02} & \textcolor[rgb]{ 0,  .502,  0}{0.08} & \textcolor[rgb]{ 0,  .502,  0}{0.36} & \textcolor[rgb]{ 0,  .502,  0}{0.04} & \textcolor[rgb]{ 0,  .502,  0}{0.04} & \textcolor[rgb]{ 0,  .502,  0}{0.36} & \textcolor[rgb]{ 0,  .502,  0}{0.36} & \textcolor[rgb]{ 0,  .502,  0}{0.08} & \textcolor[rgb]{ 0,  .502,  0}{0.01} & \textcolor[rgb]{ 0,  .502,  0}{0.01} & \textcolor[rgb]{ 0,  .502,  0}{0.04} & \textcolor[rgb]{ 0,  .502,  0}{0.54} & \textcolor[rgb]{ 0,  .502,  0}{0.08} & \textcolor[rgb]{ 0,  .502,  0}{0.01} \\
      & ID    & \textcolor[rgb]{ 0,  .502,  0}{0.53} & \textcolor[rgb]{ 0,  .502,  0}{0.04} & \textcolor[rgb]{ 0,  .502,  0}{0.23} & \textcolor[rgb]{ 0,  .502,  0}{0.56} & \textcolor[rgb]{ 0,  .502,  0}{0.29} & \textcolor[rgb]{ 0,  .502,  0}{0.29} & \textcolor[rgb]{ 0,  .502,  0}{0.14} & \textcolor[rgb]{ 0,  .502,  0}{0.56} & \textcolor[rgb]{ 0,  .502,  0}{0.23} & \textcolor[rgb]{ 0,  .502,  0}{0.37} & \textcolor[rgb]{ 0,  .502,  0}{0.39} & \textcolor[rgb]{ 0,  .502,  0}{0.44} & \textcolor[rgb]{ 0,  .502,  0}{0.59} & \textcolor[rgb]{ 0,  .502,  0}{0.25} & \textcolor[rgb]{ 0,  .502,  0}{0.34} \\
\multicolumn{1}{l}{\multirow{2}[0]{*}{\boldmath{}\textbf{GMM\newline{}$N_c = 5$}\unboldmath{}}} & UC    & \textcolor[rgb]{ 0,  .502,  0}{0.01} & \textcolor[rgb]{ 0,  .502,  0}{0.14} & \textcolor[rgb]{ 0,  .502,  0}{0.14} & \textcolor[rgb]{ 0,  .502,  0}{0.75} & \textcolor[rgb]{ 0,  .502,  0}{0.02} & \textcolor[rgb]{ 0,  .502,  0}{0.08} & \textcolor[rgb]{ 0,  .502,  0}{0.02} & \textcolor[rgb]{ 0,  .502,  0}{0.02} & \textcolor[rgb]{ 0,  .502,  0}{0.36} & \textcolor[rgb]{ 0,  .502,  0}{0.02} & \textcolor[rgb]{ 0,  .502,  0}{0.01} & \textcolor[rgb]{ 0,  .502,  0}{0.08} & \textcolor[rgb]{ 0,  .502,  0}{1} & \textcolor[rgb]{ 0,  .502,  0}{0.02} & \textcolor[rgb]{ 0,  .502,  0}{0.23} \\
      & ID    & \textcolor[rgb]{ 0,  .502,  0}{0.4} & \textcolor[rgb]{ 0,  .502,  0}{0.02} & \textcolor[rgb]{ 0,  .502,  0}{0.2} & \textcolor[rgb]{ 0,  .502,  0}{0.62} & \textcolor[rgb]{ 0,  .502,  0}{0.03} & \textcolor[rgb]{ 0,  .502,  0}{0.25} & \textcolor[rgb]{ 0,  .502,  0}{0.3} & \textcolor[rgb]{ 0,  .502,  0}{0.33} & \textcolor[rgb]{ 0,  .502,  0}{0.54} & \textcolor[rgb]{ 0,  .502,  0}{0.42} & \textcolor[rgb]{ 0,  .502,  0}{0.34} & \textcolor[rgb]{ 0,  .502,  0}{0.47} & \textcolor[rgb]{ 0,  .502,  0}{0.65} & \textcolor[rgb]{ 0,  .502,  0}{0.33} & \textcolor[rgb]{ 0,  .502,  0}{0.19} \\
\multicolumn{1}{l}{\multirow{2}[0]{*}{\boldmath{}\textbf{GMM\newline{}$N_c = 6$}\unboldmath{}}} & UC    & \textcolor[rgb]{ 0,  .502,  0}{0.14} & \textcolor[rgb]{ 0,  .502,  0}{1} & \textcolor[rgb]{ 0,  .502,  0}{0.14} & \textcolor[rgb]{ 0,  .502,  0}{0.23} & \textcolor[rgb]{ 0,  .502,  0}{0.01} & \textcolor[rgb]{ 0,  .502,  0}{0.02} & \textcolor[rgb]{ 0,  .502,  0}{0.08} & \textcolor[rgb]{ 0,  .502,  0}{0.14} & \textcolor[rgb]{ 0,  .502,  0}{0.54} & \textcolor[rgb]{ 0,  .502,  0}{0.02} & \textcolor[rgb]{ 0,  .502,  0}{0.02} & \textcolor[rgb]{ 0,  .502,  0}{0.08} & \textcolor[rgb]{ 0,  .502,  0}{0.75} & \textcolor[rgb]{ 0,  .502,  0}{0.14} & \textcolor[rgb]{ 0,  .502,  0}{0.14} \\
      & ID    & \textcolor[rgb]{ 0,  .502,  0}{0.5} & \textcolor[rgb]{ 0,  .502,  0}{0.65} & \textcolor[rgb]{ 0,  .502,  0}{0.2} & \textcolor[rgb]{ 0,  .502,  0}{0.53} & \textcolor[rgb]{ 0,  .502,  0}{0.04} & \textcolor[rgb]{ 0,  .502,  0}{0.33} & \textcolor[rgb]{ 0,  .502,  0}{0.23} & \textcolor[rgb]{ 0,  .502,  0}{0.5} & \textcolor[rgb]{ 0,  .502,  0}{0.57} & \textcolor[rgb]{ 0,  .502,  0}{0.42} & \textcolor[rgb]{ 0,  .502,  0}{0.3} & \textcolor[rgb]{ 0,  .502,  0}{0.47} & \textcolor[rgb]{ 0,  .502,  0}{0.62} & \textcolor[rgb]{ 0,  .502,  0}{0.22} & \textcolor[rgb]{ 0,  .502,  0}{0.22} \\
\multirow{2}[0]{*}{\textbf{HS}} & UC    & \textcolor[rgb]{ .753,  0,  0}{0} & \textcolor[rgb]{ 0,  .502,  0}{0.02} & \textcolor[rgb]{ .753,  0,  0}{0} & \textcolor[rgb]{ 0,  .502,  0}{0.04} & \textcolor[rgb]{ .753,  0,  0}{0} & \textcolor[rgb]{ 0,  .502,  0}{0.01} & \textcolor[rgb]{ 0,  .502,  0}{0.01} & \textcolor[rgb]{ .753,  0,  0}{0} & \textcolor[rgb]{ 0,  .502,  0}{0.14} & \textcolor[rgb]{ 0,  .502,  0}{0.01} & \textcolor[rgb]{ .753,  0,  0}{0} & \textcolor[rgb]{ 0,  .502,  0}{0.04} & \textcolor[rgb]{ .753,  0,  0}{0.01} & \textcolor[rgb]{ 0,  .502,  0}{0.08} & \textcolor[rgb]{ 0,  .502,  0}{0.14} \\
      & ID    & \textcolor[rgb]{ .753,  0,  0}{0.31} & \textcolor[rgb]{ 0,  .502,  0}{0.04} & \textcolor[rgb]{ .753,  0,  0}{0.33} & \textcolor[rgb]{ 0,  .502,  0}{0.29} & \textcolor[rgb]{ .753,  0,  0}{0.03} & \textcolor[rgb]{ 0,  .502,  0}{0.39} & \textcolor[rgb]{ 0,  .502,  0}{0.39} & \textcolor[rgb]{ .753,  0,  0}{0.34} & \textcolor[rgb]{ 0,  .502,  0}{0.5} & \textcolor[rgb]{ 0,  .502,  0}{0.37} & \textcolor[rgb]{ .753,  0,  0}{0.75} & \textcolor[rgb]{ 0,  .502,  0}{0.45} & \textcolor[rgb]{ .753,  0,  0}{0} & \textcolor[rgb]{ 0,  .502,  0}{0.25} & \textcolor[rgb]{ 0,  .502,  0}{0.22} \\
\multirow{2}[0]{*}{\textbf{Normal}} & UC    & \textcolor[rgb]{ .753,  0,  0}{0} & \textcolor[rgb]{ .753,  0,  0}{0} & \textcolor[rgb]{ .753,  0,  0}{0} & \textcolor[rgb]{ 0,  .502,  0}{0.01} & \textcolor[rgb]{ .753,  0,  0}{0} & \textcolor[rgb]{ .753,  0,  0}{0} & \textcolor[rgb]{ 0,  .502,  0}{0.04} & \textcolor[rgb]{ .753,  0,  0}{0} & \textcolor[rgb]{ .753,  0,  0}{0} & \textcolor[rgb]{ 0,  .502,  0}{0.14} & \textcolor[rgb]{ .753,  0,  0}{0} & \textcolor[rgb]{ .753,  0,  0}{0} & \textcolor[rgb]{ .753,  0,  0}{0} & \textcolor[rgb]{ .753,  0,  0}{0} & \textcolor[rgb]{ .753,  0,  0}{0} \\
      & ID    & \textcolor[rgb]{ .753,  0,  0}{0.29} & \textcolor[rgb]{ .753,  0,  0}{0.18} & \textcolor[rgb]{ .753,  0,  0}{0.22} & \textcolor[rgb]{ 0,  .502,  0}{0.41} & \textcolor[rgb]{ .753,  0,  0}{0} & \textcolor[rgb]{ .753,  0,  0}{0.71} & \textcolor[rgb]{ 0,  .502,  0}{0.27} & \textcolor[rgb]{ .753,  0,  0}{0.6} & \textcolor[rgb]{ .753,  0,  0}{0} & \textcolor[rgb]{ 0,  .502,  0}{0.02} & \textcolor[rgb]{ .753,  0,  0}{0.15} & \textcolor[rgb]{ .753,  0,  0}{0.58} & \textcolor[rgb]{ .753,  0,  0}{0.01} & \textcolor[rgb]{ .753,  0,  0}{0.12} & \textcolor[rgb]{ .753,  0,  0}{0.75} \\
\multirow{2}[0]{*}{\textbf{MC}} & UC    & \textcolor[rgb]{ .753,  0,  0}{0} & \textcolor[rgb]{ .753,  0,  0}{0} & \textcolor[rgb]{ .753,  0,  0}{0} & \textcolor[rgb]{ 0,  .502,  0}{0.04} & \textcolor[rgb]{ .753,  0,  0}{0} & \textcolor[rgb]{ .753,  0,  0}{0} & \textcolor[rgb]{ 0,  .502,  0}{0.02} & \textcolor[rgb]{ .753,  0,  0}{0} & \textcolor[rgb]{ .753,  0,  0}{0} & \textcolor[rgb]{ 0,  .502,  0}{0.14} & \textcolor[rgb]{ .753,  0,  0}{0} & \textcolor[rgb]{ .753,  0,  0}{0} & \textcolor[rgb]{ .753,  0,  0}{0} & \textcolor[rgb]{ .753,  0,  0}{0} & \textcolor[rgb]{ .753,  0,  0}{0} \\
      & ID    & \textcolor[rgb]{ .753,  0,  0}{0.33} & \textcolor[rgb]{ .753,  0,  0}{0.18} & \textcolor[rgb]{ .753,  0,  0}{0.18} & \textcolor[rgb]{ 0,  .502,  0}{0.44} & \textcolor[rgb]{ .753,  0,  0}{0} & \textcolor[rgb]{ .753,  0,  0}{0.71} & \textcolor[rgb]{ 0,  .502,  0}{0.3} & \textcolor[rgb]{ .753,  0,  0}{0.65} & \textcolor[rgb]{ .753,  0,  0}{0} & \textcolor[rgb]{ 0,  .502,  0}{0.02} & \textcolor[rgb]{ .753,  0,  0}{0.17} & \textcolor[rgb]{ .753,  0,  0}{0.58} & \textcolor[rgb]{ .753,  0,  0}{0} & \textcolor[rgb]{ .753,  0,  0}{0.16} & \textcolor[rgb]{ .753,  0,  0}{0.75} \\
\multirow{2}[0]{*}{\textbf{GARCH(1,1)}} & UC    & \textcolor[rgb]{ .753,  0,  0}{0} & \textcolor[rgb]{ .753,  0,  0}{0} & \textcolor[rgb]{ 0,  .502,  0}{0.01} & \textcolor[rgb]{ 0,  .502,  0}{0.02} & \textcolor[rgb]{ .753,  0,  0}{0} & \textcolor[rgb]{ 0,  .502,  0}{0.01} & \textcolor[rgb]{ .753,  0,  0}{0} & \textcolor[rgb]{ .753,  0,  0}{0} & \textcolor[rgb]{ 0,  .502,  0}{0.02} & \textcolor[rgb]{ 0,  .502,  0}{0.75} & \textcolor[rgb]{ .753,  0,  0}{0} & \textcolor[rgb]{ .753,  0,  0}{0} & \textcolor[rgb]{ .753,  0,  0}{0} & \textcolor[rgb]{ .753,  0,  0}{0} & \textcolor[rgb]{ .753,  0,  0}{0} \\
      & ID    & \textcolor[rgb]{ .753,  0,  0}{0.64} & \textcolor[rgb]{ .753,  0,  0}{0.04} & \textcolor[rgb]{ 0,  .502,  0}{0.39} & \textcolor[rgb]{ 0,  .502,  0}{0.33} & \textcolor[rgb]{ .753,  0,  0}{0.81} & \textcolor[rgb]{ 0,  .502,  0}{0.4} & \textcolor[rgb]{ .753,  0,  0}{0.46} & \textcolor[rgb]{ .753,  0,  0}{0.51} & \textcolor[rgb]{ 0,  .502,  0}{0.42} & \textcolor[rgb]{ 0,  .502,  0}{0.62} & \textcolor[rgb]{ .753,  0,  0}{0.32} & \textcolor[rgb]{ .753,  0,  0}{0.09} & \textcolor[rgb]{ .753,  0,  0}{0.28} & \textcolor[rgb]{ .753,  0,  0}{0.89} & \textcolor[rgb]{ .753,  0,  0}{0.14} \\
\multirow{2}[0]{*}{\textbf{C-SAV}} & UC    & \textcolor[rgb]{ 0,  .502,  0}{0.23} & \textcolor[rgb]{ 0,  .502,  0}{0.04} & \textcolor[rgb]{ 0,  .502,  0}{0.75} & \textcolor[rgb]{ 0,  .502,  0}{0.36} & \textcolor[rgb]{ 0,  .502,  0}{0.04} & \textcolor[rgb]{ 0,  .502,  0}{0.23} & \textcolor[rgb]{ 0,  .502,  0}{0.01} & \textcolor[rgb]{ 0,  .502,  0}{0.54} & \textcolor[rgb]{ 0,  .502,  0}{0.36} & \textcolor[rgb]{ 0,  .502,  0}{1} & \textcolor[rgb]{ 0,  .502,  0}{0.08} & \textcolor[rgb]{ 0,  .502,  0}{0.08} & \textcolor[rgb]{ 0,  .502,  0}{0.51} & \textcolor[rgb]{ 0,  .502,  0}{0.14} & \textcolor[rgb]{ 0,  .502,  0}{0.04} \\
      & ID    & \textcolor[rgb]{ 0,  .502,  0}{0.17} & \textcolor[rgb]{ 0,  .502,  0}{0.44} & \textcolor[rgb]{ 0,  .502,  0}{0.59} & \textcolor[rgb]{ 0,  .502,  0}{0.56} & \textcolor[rgb]{ 0,  .502,  0}{0.03} & \textcolor[rgb]{ 0,  .502,  0}{0.53} & \textcolor[rgb]{ 0,  .502,  0}{0.37} & \textcolor[rgb]{ 0,  .502,  0}{0.59} & \textcolor[rgb]{ 0,  .502,  0}{0.54} & \textcolor[rgb]{ 0,  .502,  0}{0.65} & \textcolor[rgb]{ 0,  .502,  0}{0.47} & \textcolor[rgb]{ 0,  .502,  0}{0.47} & \textcolor[rgb]{ 0,  .502,  0}{0.72} & \textcolor[rgb]{ 0,  .502,  0}{0.5} & \textcolor[rgb]{ 0,  .502,  0}{0.29} \\
\multirow{2}[0]{*}{\textbf{C-AS}} & UC    & \textcolor[rgb]{ .753,  0,  0}{0} & \textcolor[rgb]{ 0,  .502,  0}{0.01} & \textcolor[rgb]{ 0,  .502,  0}{0.54} & \textcolor[rgb]{ 0,  .502,  0}{0.23} & \textcolor[rgb]{ 0,  .502,  0}{0.23} & \textcolor[rgb]{ .753,  0,  0}{0} & \textcolor[rgb]{ 0,  .502,  0}{0.01} & \textcolor[rgb]{ 0,  .502,  0}{0.02} & \textcolor[rgb]{ 0,  .502,  0}{0.36} & \textcolor[rgb]{ 0,  .502,  0}{0.75} & \textcolor[rgb]{ 0,  .502,  0}{0.54} & \textcolor[rgb]{ 0,  .502,  0}{0.02} & \textcolor[rgb]{ 0,  .502,  0}{0.75} & \textcolor[rgb]{ 0,  .502,  0}{0.08} & \textcolor[rgb]{ 0,  .502,  0}{0.02} \\
      & ID    & \textcolor[rgb]{ .753,  0,  0}{0.33} & \textcolor[rgb]{ 0,  .502,  0}{0.41} & \textcolor[rgb]{ 0,  .502,  0}{0.57} & \textcolor[rgb]{ 0,  .502,  0}{0.53} & \textcolor[rgb]{ 0,  .502,  0}{0.19} & \textcolor[rgb]{ .753,  0,  0}{0.07} & \textcolor[rgb]{ 0,  .502,  0}{0.37} & \textcolor[rgb]{ 0,  .502,  0}{0.42} & \textcolor[rgb]{ 0,  .502,  0}{0.54} & \textcolor[rgb]{ 0,  .502,  0}{0.62} & \textcolor[rgb]{ 0,  .502,  0}{0.59} & \textcolor[rgb]{ 0,  .502,  0}{0.03} & \textcolor[rgb]{ 0,  .502,  0}{0.62} & \textcolor[rgb]{ 0,  .502,  0}{0.23} & \textcolor[rgb]{ 0,  .502,  0}{0.42} \\
\multirow{2}[0]{*}{\textbf{C-GARCH}} & UC    & \textcolor[rgb]{ 0,  .502,  0}{0.02} & \textcolor[rgb]{ 0,  .502,  0}{0.02} & \textcolor[rgb]{ 0,  .502,  0}{1} & \textcolor[rgb]{ 0,  .502,  0}{0.75} & \textcolor[rgb]{ 0,  .502,  0}{0.02} & \textcolor[rgb]{ 0,  .502,  0}{0.23} & \textcolor[rgb]{ 0,  .502,  0}{0.36} & \textcolor[rgb]{ 0,  .502,  0}{0.04} & \textcolor[rgb]{ 0,  .502,  0}{1} & \textcolor[rgb]{ 0,  .502,  0}{0.36} & \textcolor[rgb]{ 0,  .502,  0}{0.23} & \textcolor[rgb]{ 0,  .502,  0}{0.01} & \textcolor[rgb]{ 0,  .502,  0}{0.54} & \textcolor[rgb]{ 0,  .502,  0}{0.14} & \textcolor[rgb]{ .753,  0,  0}{0} \\
      & ID    & \textcolor[rgb]{ 0,  .502,  0}{0.42} & \textcolor[rgb]{ 0,  .502,  0}{0.33} & \textcolor[rgb]{ 0,  .502,  0}{0.61} & \textcolor[rgb]{ 0,  .502,  0}{0.62} & \textcolor[rgb]{ 0,  .502,  0}{0.04} & \textcolor[rgb]{ 0,  .502,  0}{0.53} & \textcolor[rgb]{ 0,  .502,  0}{0.56} & \textcolor[rgb]{ 0,  .502,  0}{0.44} & \textcolor[rgb]{ 0,  .502,  0}{0.61} & \textcolor[rgb]{ 0,  .502,  0}{0.56} & \textcolor[rgb]{ 0,  .502,  0}{0.53} & \textcolor[rgb]{ 0,  .502,  0}{0.34} & \textcolor[rgb]{ 0,  .502,  0}{0.59} & \textcolor[rgb]{ 0,  .502,  0}{0.5} & \textcolor[rgb]{ .753,  0,  0}{0.3} \\
\multirow{2}[1]{*}{\textbf{C-AD}} & UC    & \textcolor[rgb]{ .753,  0,  0}{0} & \textcolor[rgb]{ 0,  .502,  0}{0.36} & \textcolor[rgb]{ .753,  0,  0}{0} & \textcolor[rgb]{ 0,  .502,  0}{0.08} & \textcolor[rgb]{ .753,  0,  0}{0} & \textcolor[rgb]{ 0,  .502,  0}{0.08} & \textcolor[rgb]{ 0,  .502,  0}{0.14} & \textcolor[rgb]{ 0,  .502,  0}{0.04} & \textcolor[rgb]{ 0,  .502,  0}{0.23} & \textcolor[rgb]{ 0,  .502,  0}{0.08} & \textcolor[rgb]{ .753,  0,  0}{0} & \textcolor[rgb]{ .753,  0,  0}{0} & \textcolor[rgb]{ 0,  .502,  0}{0.14} & \textcolor[rgb]{ 0,  .502,  0}{0.04} & \textcolor[rgb]{ 0,  .502,  0}{0.02} \\
      & ID    & \textcolor[rgb]{ .753,  0,  0}{0.33} & \textcolor[rgb]{ 0,  .502,  0}{0.56} & \textcolor[rgb]{ .753,  0,  0}{0.35} & \textcolor[rgb]{ 0,  .502,  0}{0.47} & \textcolor[rgb]{ .753,  0,  0}{0.06} & \textcolor[rgb]{ 0,  .502,  0}{0.47} & \textcolor[rgb]{ 0,  .502,  0}{0.2} & \textcolor[rgb]{ 0,  .502,  0}{0.44} & \textcolor[rgb]{ 0,  .502,  0}{0.17} & \textcolor[rgb]{ 0,  .502,  0}{0.25} & \textcolor[rgb]{ .753,  0,  0}{0.51} & \textcolor[rgb]{ .753,  0,  0}{0.16} & \textcolor[rgb]{ 0,  .502,  0}{0.02} & \textcolor[rgb]{ 0,  .502,  0}{0.29} & \textcolor[rgb]{ 0,  .502,  0}{0.3} \\
\bottomrule
    \end{tabular}%
}
  \label{tab_VaR1}%
       \vspace{1ex}\\
      {\justifying \noindent Note: This table compares the performance of VaR 1\% between different models based on christofferson unconditional coverage (UC), Independence (ID) test. The p-values more than 0.01, are considered as "good" VaRs and shown with green. VaRs with at least one p-value less than 0.01 are shown by red, and are not satisfying christofferson's criteria. HS shows historical simulation VaR, GC denotes the Gaussian Copula VaR, Normal is variance-covariance VaR, MC is Monte Carlo Simulation VaR, C-SAV is CaViaR Symmetric VaR, C-AS is CaViaR Asymmetric VaR, C-GARCH is CaViaR GARCH VaR, and C-AD is CaViaR Adaptive model VaR.
      \par}
\end{table}%

\begin{table}
% [width=1.0\linewidth,cols=16,pos=t!]
  \centering
  \caption{VaR 5 \% for individual stocks}
\resizebox{\textwidth}{!}{
    \begin{tabular}{lcccccccccccccccccc}
    % {\tblwidth}{@{}LCCCCCCCCCCCCCCCCCC@{}}
\toprule
      & \textbf{test} & \textbf{S\&P500} & \textbf{MMM} & \textbf{AXP} & \textbf{AMD} & \textbf{AIG} & \textbf{AFL} & \textbf{ABT} & \textbf{MSFTF} & \textbf{AAPL} & \textbf{AMZN} & \textbf{BAC} & \textbf{JPM} & \textbf{JNJ} & \textbf{XOM} & \textbf{MA} \\
\midrule
\multicolumn{1}{l}{\multirow{2}[1]{*}{\boldmath{}\textbf{GMM\newline{}$N_c = 3$}\unboldmath{}}} & UC    & \textcolor[rgb]{ 0,  .502,  0}{0.32} & \textcolor[rgb]{ 0,  .502,  0}{0.07} & \textcolor[rgb]{ 0,  .502,  0}{0.09} & \textcolor[rgb]{ .753,  0,  0}{0.16} & \textcolor[rgb]{ 0,  .502,  0}{0.02} & \textcolor[rgb]{ 0,  .502,  0}{0.66} & \textcolor[rgb]{ 0,  .502,  0}{0.25} & \textcolor[rgb]{ 0,  .502,  0}{0.88} & \textcolor[rgb]{ .753,  0,  0}{0.47} & \textcolor[rgb]{ 0,  .502,  0}{0.39} & \textcolor[rgb]{ 0,  .502,  0}{0.01} & \textcolor[rgb]{ 0,  .502,  0}{0.56} & \textcolor[rgb]{ .753,  0,  0}{0.16} & \textcolor[rgb]{ 0,  .502,  0}{0.25} & \textcolor[rgb]{ 0,  .502,  0}{0.32} \\
      & ID    & \textcolor[rgb]{ 0,  .502,  0}{0.67} & \textcolor[rgb]{ 0,  .502,  0}{0.32} & \textcolor[rgb]{ 0,  .502,  0}{0.77} & \textcolor[rgb]{ .753,  0,  0}{0} & \textcolor[rgb]{ 0,  .502,  0}{0.47} & \textcolor[rgb]{ 0,  .502,  0}{0.45} & \textcolor[rgb]{ 0,  .502,  0}{0.39} & \textcolor[rgb]{ 0,  .502,  0}{0.16} & \textcolor[rgb]{ .753,  0,  0}{0} & \textcolor[rgb]{ 0,  .502,  0}{0.3} & \textcolor[rgb]{ 0,  .502,  0}{0.05} & \textcolor[rgb]{ 0,  .502,  0}{0.52} & \textcolor[rgb]{ .753,  0,  0}{0} & \textcolor[rgb]{ 0,  .502,  0}{0.39} & \textcolor[rgb]{ 0,  .502,  0}{0.05} \\
\multicolumn{1}{l}{\multirow{2}[0]{*}{\boldmath{}\textbf{GMM\newline{}$N_c = 4$}\unboldmath{}}} & UC    & \textcolor[rgb]{ 0,  .502,  0}{0.32} & \textcolor[rgb]{ 0,  .502,  0}{0.05} & \textcolor[rgb]{ 0,  .502,  0}{0.05} & \textcolor[rgb]{ 0,  .502,  0}{0.16} & \textcolor[rgb]{ 0,  .502,  0}{0.04} & \textcolor[rgb]{ 0,  .502,  0}{0.32} & \textcolor[rgb]{ 0,  .502,  0}{0.2} & \textcolor[rgb]{ 0,  .502,  0}{0.47} & \textcolor[rgb]{ 0,  .502,  0}{0.66} & \textcolor[rgb]{ 0,  .502,  0}{0.56} & \textcolor[rgb]{ 0,  .502,  0}{0.03} & \textcolor[rgb]{ 0,  .502,  0}{0.66} & \textcolor[rgb]{ .753,  0,  0}{0.32} & \textcolor[rgb]{ 0,  .502,  0}{0.09} & \textcolor[rgb]{ 0,  .502,  0}{0.39} \\
      & ID    & \textcolor[rgb]{ 0,  .502,  0}{0.43} & \textcolor[rgb]{ 0,  .502,  0}{0.36} & \textcolor[rgb]{ 0,  .502,  0}{0.64} & \textcolor[rgb]{ 0,  .502,  0}{0.01} & \textcolor[rgb]{ 0,  .502,  0}{0.7} & \textcolor[rgb]{ 0,  .502,  0}{0.32} & \textcolor[rgb]{ 0,  .502,  0}{0.43} & \textcolor[rgb]{ 0,  .502,  0}{0.04} & \textcolor[rgb]{ 0,  .502,  0}{0.01} & \textcolor[rgb]{ 0,  .502,  0}{0.23} & \textcolor[rgb]{ 0,  .502,  0}{0.03} & \textcolor[rgb]{ 0,  .502,  0}{0.48} & \textcolor[rgb]{ .753,  0,  0}{0} & \textcolor[rgb]{ 0,  .502,  0}{1} & \textcolor[rgb]{ 0,  .502,  0}{0.04} \\
\multicolumn{1}{l}{\multirow{2}[0]{*}{\boldmath{}\textbf{GMM\newline{}$N_c = 5$}\unboldmath{}}} & UC    & \textcolor[rgb]{ 0,  .502,  0}{0.09} & \textcolor[rgb]{ 0,  .502,  0}{0.12} & \textcolor[rgb]{ 0,  .502,  0}{0.05} & \textcolor[rgb]{ 0,  .502,  0}{0.05} & \textcolor[rgb]{ 0,  .502,  0}{0.01} & \textcolor[rgb]{ 0,  .502,  0}{0.32} & \textcolor[rgb]{ 0,  .502,  0}{0.32} & \textcolor[rgb]{ .753,  0,  0}{0.39} & \textcolor[rgb]{ .753,  0,  0}{0.99} & \textcolor[rgb]{ 0,  .502,  0}{0.39} & \textcolor[rgb]{ 0,  .502,  0}{0.03} & \textcolor[rgb]{ 0,  .502,  0}{0.39} & \textcolor[rgb]{ .753,  0,  0}{0.2} & \textcolor[rgb]{ 0,  .502,  0}{0.12} & \textcolor[rgb]{ 0,  .502,  0}{0.09} \\
      & ID    & \textcolor[rgb]{ 0,  .502,  0}{0.93} & \textcolor[rgb]{ 0,  .502,  0}{1} & \textcolor[rgb]{ 0,  .502,  0}{0.64} & \textcolor[rgb]{ 0,  .502,  0}{0.01} & \textcolor[rgb]{ 0,  .502,  0}{0.3} & \textcolor[rgb]{ 0,  .502,  0}{0.32} & \textcolor[rgb]{ 0,  .502,  0}{0.34} & \textcolor[rgb]{ .753,  0,  0}{0} & \textcolor[rgb]{ .753,  0,  0}{0} & \textcolor[rgb]{ 0,  .502,  0}{0.29} & \textcolor[rgb]{ 0,  .502,  0}{0.09} & \textcolor[rgb]{ 0,  .502,  0}{0.62} & \textcolor[rgb]{ .753,  0,  0}{0} & \textcolor[rgb]{ 0,  .502,  0}{1} & \textcolor[rgb]{ 0,  .502,  0}{0.12} \\
\multicolumn{1}{l}{\multirow{2}[0]{*}{\boldmath{}\textbf{GMM\newline{}$N_c = 6$}\unboldmath{}}} & UC    & \textcolor[rgb]{ 0,  .502,  0}{0.07} & \textcolor[rgb]{ 0,  .502,  0}{0.16} & \textcolor[rgb]{ 0,  .502,  0}{0.12} & \textcolor[rgb]{ 0,  .502,  0}{0.16} & \textcolor[rgb]{ 0,  .502,  0}{0.02} & \textcolor[rgb]{ 0,  .502,  0}{0.32} & \textcolor[rgb]{ 0,  .502,  0}{0.39} & \textcolor[rgb]{ 0,  .502,  0}{0.66} & \textcolor[rgb]{ .753,  0,  0}{0.47} & \textcolor[rgb]{ 0,  .502,  0}{0.47} & \textcolor[rgb]{ 0,  .502,  0}{0.07} & \textcolor[rgb]{ 0,  .502,  0}{0.39} & \textcolor[rgb]{ .753,  0,  0}{0.25} & \textcolor[rgb]{ 0,  .502,  0}{0.25} & \textcolor[rgb]{ 0,  .502,  0}{0.47} \\
      & ID    & \textcolor[rgb]{ 0,  .502,  0}{0.99} & \textcolor[rgb]{ 0,  .502,  0}{1} & \textcolor[rgb]{ 0,  .502,  0}{0.85} & \textcolor[rgb]{ 0,  .502,  0}{0.01} & \textcolor[rgb]{ 0,  .502,  0}{0.04} & \textcolor[rgb]{ 0,  .502,  0}{0.32} & \textcolor[rgb]{ 0,  .502,  0}{0.31} & \textcolor[rgb]{ 0,  .502,  0}{0.02} & \textcolor[rgb]{ .753,  0,  0}{0} & \textcolor[rgb]{ 0,  .502,  0}{0.26} & \textcolor[rgb]{ 0,  .502,  0}{0.02} & \textcolor[rgb]{ 0,  .502,  0}{0.62} & \textcolor[rgb]{ .753,  0,  0}{0} & \textcolor[rgb]{ 0,  .502,  0}{0.39} & \textcolor[rgb]{ 0,  .502,  0}{0.03} \\
\multirow{2}[0]{*}{\textbf{HS}} & UC    & \textcolor[rgb]{ 0,  .502,  0}{0.07} & \textcolor[rgb]{ .753,  0,  0}{0} & \textcolor[rgb]{ .753,  0,  0}{0} & \textcolor[rgb]{ 0,  .502,  0}{0.04} & \textcolor[rgb]{ .753,  0,  0}{0} & \textcolor[rgb]{ 0,  .502,  0}{0.01} & \textcolor[rgb]{ 0,  .502,  0}{0.32} & \textcolor[rgb]{ .753,  0,  0}{0.03} & \textcolor[rgb]{ .753,  0,  0}{0.88} & \textcolor[rgb]{ 0,  .502,  0}{0.2} & \textcolor[rgb]{ .753,  0,  0}{0} & \textcolor[rgb]{ .753,  0,  0}{0} & \textcolor[rgb]{ .753,  0,  0}{0.07} & \textcolor[rgb]{ 0,  .502,  0}{0.2} & \textcolor[rgb]{ 0,  .502,  0}{0.09} \\
      & ID    & \textcolor[rgb]{ 0,  .502,  0}{0.14} & \textcolor[rgb]{ .753,  0,  0}{0.22} & \textcolor[rgb]{ .753,  0,  0}{1} & \textcolor[rgb]{ 0,  .502,  0}{0.01} & \textcolor[rgb]{ .753,  0,  0}{0.02} & \textcolor[rgb]{ 0,  .502,  0}{0.06} & \textcolor[rgb]{ 0,  .502,  0}{0.14} & \textcolor[rgb]{ .753,  0,  0}{0} & \textcolor[rgb]{ .753,  0,  0}{0} & \textcolor[rgb]{ 0,  .502,  0}{0.41} & \textcolor[rgb]{ .753,  0,  0}{0} & \textcolor[rgb]{ .753,  0,  0}{0.87} & \textcolor[rgb]{ .753,  0,  0}{0} & \textcolor[rgb]{ 0,  .502,  0}{0.19} & \textcolor[rgb]{ 0,  .502,  0}{0.12} \\
\multirow{2}[0]{*}{\textbf{Normal}} & UC    & \textcolor[rgb]{ 0,  .502,  0}{0.01} & \textcolor[rgb]{ 0,  .502,  0}{0.25} & \textcolor[rgb]{ .753,  0,  0}{0} & \textcolor[rgb]{ 0,  .502,  0}{0.09} & \textcolor[rgb]{ 0,  .502,  0}{0.01} & \textcolor[rgb]{ 0,  .502,  0}{0.07} & \textcolor[rgb]{ 0,  .502,  0}{0.78} & \textcolor[rgb]{ .753,  0,  0}{0.2} & \textcolor[rgb]{ 0,  .502,  0}{0.89} & \textcolor[rgb]{ 0,  .502,  0}{0.01} & \textcolor[rgb]{ .753,  0,  0}{0} & \textcolor[rgb]{ 0,  .502,  0}{0.2} & \textcolor[rgb]{ .753,  0,  0}{0.56} & \textcolor[rgb]{ 0,  .502,  0}{0.25} & \textcolor[rgb]{ 0,  .502,  0}{0.09} \\
      & ID    & \textcolor[rgb]{ 0,  .502,  0}{0.96} & \textcolor[rgb]{ 0,  .502,  0}{0.72} & \textcolor[rgb]{ .753,  0,  0}{1} & \textcolor[rgb]{ 0,  .502,  0}{0.04} & \textcolor[rgb]{ 0,  .502,  0}{0.02} & \textcolor[rgb]{ 0,  .502,  0}{0.29} & \textcolor[rgb]{ 0,  .502,  0}{0.1} & \textcolor[rgb]{ .753,  0,  0}{0} & \textcolor[rgb]{ 0,  .502,  0}{0.03} & \textcolor[rgb]{ 0,  .502,  0}{0.42} & \textcolor[rgb]{ .753,  0,  0}{0.03} & \textcolor[rgb]{ 0,  .502,  0}{0.19} & \textcolor[rgb]{ .753,  0,  0}{0} & \textcolor[rgb]{ 0,  .502,  0}{0.06} & \textcolor[rgb]{ 0,  .502,  0}{0.12} \\
\multirow{2}[0]{*}{\textbf{MC}} & UC    & \textcolor[rgb]{ 0,  .502,  0}{0.01} & \textcolor[rgb]{ 0,  .502,  0}{0.2} & \textcolor[rgb]{ .753,  0,  0}{0} & \textcolor[rgb]{ 0,  .502,  0}{0.09} & \textcolor[rgb]{ 0,  .502,  0}{0.02} & \textcolor[rgb]{ 0,  .502,  0}{0.12} & \textcolor[rgb]{ 0,  .502,  0}{0.89} & \textcolor[rgb]{ .753,  0,  0}{0.12} & \textcolor[rgb]{ 0,  .502,  0}{0.99} & \textcolor[rgb]{ 0,  .502,  0}{0.01} & \textcolor[rgb]{ .753,  0,  0}{0} & \textcolor[rgb]{ 0,  .502,  0}{0.16} & \textcolor[rgb]{ .753,  0,  0}{0.77} & \textcolor[rgb]{ 0,  .502,  0}{0.25} & \textcolor[rgb]{ 0,  .502,  0}{0.04} \\
      & ID    & \textcolor[rgb]{ 0,  .502,  0}{0.96} & \textcolor[rgb]{ 0,  .502,  0}{0.77} & \textcolor[rgb]{ .753,  0,  0}{1} & \textcolor[rgb]{ 0,  .502,  0}{0.04} & \textcolor[rgb]{ 0,  .502,  0}{0.01} & \textcolor[rgb]{ 0,  .502,  0}{0.46} & \textcolor[rgb]{ 0,  .502,  0}{0.12} & \textcolor[rgb]{ .753,  0,  0}{0} & \textcolor[rgb]{ 0,  .502,  0}{0.04} & \textcolor[rgb]{ 0,  .502,  0}{0.42} & \textcolor[rgb]{ .753,  0,  0}{0.01} & \textcolor[rgb]{ 0,  .502,  0}{0.21} & \textcolor[rgb]{ .753,  0,  0}{0} & \textcolor[rgb]{ 0,  .502,  0}{0.06} & \textcolor[rgb]{ 0,  .502,  0}{0.07} \\
\multirow{2}[0]{*}{\textbf{GARCH(1,1)}} & UC    & \textcolor[rgb]{ .753,  0,  0}{0} & \textcolor[rgb]{ 0,  .502,  0}{0.09} & \textcolor[rgb]{ 0,  .502,  0}{0.07} & \textcolor[rgb]{ 0,  .502,  0}{0.25} & \textcolor[rgb]{ 0,  .502,  0}{0.32} & \textcolor[rgb]{ 0,  .502,  0}{0.56} & \textcolor[rgb]{ 0,  .502,  0}{0.99} & \textcolor[rgb]{ 0,  .502,  0}{0.47} & \textcolor[rgb]{ .753,  0,  0}{0.56} & \textcolor[rgb]{ 0,  .502,  0}{0.01} & \textcolor[rgb]{ 0,  .502,  0}{0.32} & \textcolor[rgb]{ 0,  .502,  0}{0.47} & \textcolor[rgb]{ 0,  .502,  0}{0.89} & \textcolor[rgb]{ .753,  0,  0}{0} & \textcolor[rgb]{ .753,  0,  0}{0} \\
      & ID    & \textcolor[rgb]{ .753,  0,  0}{0.03} & \textcolor[rgb]{ 0,  .502,  0}{0.27} & \textcolor[rgb]{ 0,  .502,  0}{0.65} & \textcolor[rgb]{ 0,  .502,  0}{0.16} & \textcolor[rgb]{ 0,  .502,  0}{0.67} & \textcolor[rgb]{ 0,  .502,  0}{0.72} & \textcolor[rgb]{ 0,  .502,  0}{0.36} & \textcolor[rgb]{ 0,  .502,  0}{0.11} & \textcolor[rgb]{ .753,  0,  0}{0} & \textcolor[rgb]{ 0,  .502,  0}{0.88} & \textcolor[rgb]{ 0,  .502,  0}{0.34} & \textcolor[rgb]{ 0,  .502,  0}{0.57} & \textcolor[rgb]{ 0,  .502,  0}{0.32} & \textcolor[rgb]{ .753,  0,  0}{0.61} & \textcolor[rgb]{ .753,  0,  0}{0.39} \\
\multirow{2}[0]{*}{\textbf{C-SAV}} & UC    & \textcolor[rgb]{ 0,  .502,  0}{0.39} & \textcolor[rgb]{ 0,  .502,  0}{0.05} & \textcolor[rgb]{ 0,  .502,  0}{0.66} & \textcolor[rgb]{ 0,  .502,  0}{0.01} & \textcolor[rgb]{ 0,  .502,  0}{0.39} & \textcolor[rgb]{ 0,  .502,  0}{0.25} & \textcolor[rgb]{ 0,  .502,  0}{0.03} & \textcolor[rgb]{ 0,  .502,  0}{0.47} & \textcolor[rgb]{ 0,  .502,  0}{0.39} & \textcolor[rgb]{ 0,  .502,  0}{0.56} & \textcolor[rgb]{ 0,  .502,  0}{0.66} & \textcolor[rgb]{ 0,  .502,  0}{0.66} & \textcolor[rgb]{ 0,  .502,  0}{0.32} & \textcolor[rgb]{ 0,  .502,  0}{0.07} & \textcolor[rgb]{ 0,  .502,  0}{0.56} \\
      & ID    & \textcolor[rgb]{ 0,  .502,  0}{0.16} & \textcolor[rgb]{ 0,  .502,  0}{0.96} & \textcolor[rgb]{ 0,  .502,  0}{0.02} & \textcolor[rgb]{ 0,  .502,  0}{0.12} & \textcolor[rgb]{ 0,  .502,  0}{0.62} & \textcolor[rgb]{ 0,  .502,  0}{0.62} & \textcolor[rgb]{ 0,  .502,  0}{1} & \textcolor[rgb]{ 0,  .502,  0}{0.16} & \textcolor[rgb]{ 0,  .502,  0}{0.12} & \textcolor[rgb]{ 0,  .502,  0}{0.09} & \textcolor[rgb]{ 0,  .502,  0}{0.07} & \textcolor[rgb]{ 0,  .502,  0}{0.59} & \textcolor[rgb]{ 0,  .502,  0}{0.14} & \textcolor[rgb]{ 0,  .502,  0}{0.7} & \textcolor[rgb]{ 0,  .502,  0}{0.55} \\
\multirow{2}[0]{*}{\textbf{C-AS}} & UC    & \textcolor[rgb]{ 0,  .502,  0}{0.66} & \textcolor[rgb]{ .753,  0,  0}{0} & \textcolor[rgb]{ 0,  .502,  0}{0.77} & \textcolor[rgb]{ 0,  .502,  0}{0.09} & \textcolor[rgb]{ 0,  .502,  0}{0.47} & \textcolor[rgb]{ 0,  .502,  0}{0.32} & \textcolor[rgb]{ 0,  .502,  0}{0.02} & \textcolor[rgb]{ 0,  .502,  0}{0.88} & \textcolor[rgb]{ 0,  .502,  0}{0.66} & \textcolor[rgb]{ 0,  .502,  0}{0.05} & \textcolor[rgb]{ 0,  .502,  0}{0.12} & \textcolor[rgb]{ 0,  .502,  0}{0.32} & \textcolor[rgb]{ 0,  .502,  0}{0.04} & \textcolor[rgb]{ 0,  .502,  0}{0.09} & \textcolor[rgb]{ 0,  .502,  0}{0.32} \\
      & ID    & \textcolor[rgb]{ 0,  .502,  0}{0.19} & \textcolor[rgb]{ .753,  0,  0}{0.25} & \textcolor[rgb]{ 0,  .502,  0}{0.02} & \textcolor[rgb]{ 0,  .502,  0}{0.27} & \textcolor[rgb]{ 0,  .502,  0}{0.99} & \textcolor[rgb]{ 0,  .502,  0}{0.32} & \textcolor[rgb]{ 0,  .502,  0}{1} & \textcolor[rgb]{ 0,  .502,  0}{0.24} & \textcolor[rgb]{ 0,  .502,  0}{0.07} & \textcolor[rgb]{ 0,  .502,  0}{0.16} & \textcolor[rgb]{ 0,  .502,  0}{1} & \textcolor[rgb]{ 0,  .502,  0}{0.88} & \textcolor[rgb]{ 0,  .502,  0}{0.38} & \textcolor[rgb]{ 0,  .502,  0}{0.77} & \textcolor[rgb]{ 0,  .502,  0}{0.74} \\
\multirow{2}[0]{*}{\textbf{C-GARCH}} & UC    & \textcolor[rgb]{ 0,  .502,  0}{0.56} & \textcolor[rgb]{ 0,  .502,  0}{0.03} & \textcolor[rgb]{ 0,  .502,  0}{0.66} & \textcolor[rgb]{ .753,  0,  0}{0} & \textcolor[rgb]{ 0,  .502,  0}{0.56} & \textcolor[rgb]{ 0,  .502,  0}{0.05} & \textcolor[rgb]{ 0,  .502,  0}{0.18} & \textcolor[rgb]{ 0,  .502,  0}{0.39} & \textcolor[rgb]{ 0,  .502,  0}{0.38} & \textcolor[rgb]{ 0,  .502,  0}{0.16} & \textcolor[rgb]{ 0,  .502,  0}{0.2} & \textcolor[rgb]{ 0,  .502,  0}{0.56} & \textcolor[rgb]{ 0,  .502,  0}{0.12} & \textcolor[rgb]{ 0,  .502,  0}{0.16} & \textcolor[rgb]{ 0,  .502,  0}{0.47} \\
      & ID    & \textcolor[rgb]{ 0,  .502,  0}{0.04} & \textcolor[rgb]{ 0,  .502,  0}{0.85} & \textcolor[rgb]{ 0,  .502,  0}{0.22} & \textcolor[rgb]{ .753,  0,  0}{0.31} & \textcolor[rgb]{ 0,  .502,  0}{0.52} & \textcolor[rgb]{ 0,  .502,  0}{0.69} & \textcolor[rgb]{ 0,  .502,  0}{0.1} & \textcolor[rgb]{ 0,  .502,  0}{0.3} & \textcolor[rgb]{ 0,  .502,  0}{0.46} & \textcolor[rgb]{ 0,  .502,  0}{0.21} & \textcolor[rgb]{ 0,  .502,  0}{0.77} & \textcolor[rgb]{ 0,  .502,  0}{0.96} & \textcolor[rgb]{ 0,  .502,  0}{0.04} & \textcolor[rgb]{ 0,  .502,  0}{1} & \textcolor[rgb]{ 0,  .502,  0}{1} \\
\multirow{2}[1]{*}{\textbf{C-AD}} & UC    & \textcolor[rgb]{ 0,  .502,  0}{0.01} & \textcolor[rgb]{ 0,  .502,  0}{0.03} & \textcolor[rgb]{ 0,  .502,  0}{0.01} & \textcolor[rgb]{ 0,  .502,  0}{0.05} & \textcolor[rgb]{ .753,  0,  0}{0} & \textcolor[rgb]{ .753,  0,  0}{0} & \textcolor[rgb]{ 0,  .502,  0}{0.39} & \textcolor[rgb]{ 0,  .502,  0}{0.01} & \textcolor[rgb]{ .753,  0,  0}{0.89} & \textcolor[rgb]{ 0,  .502,  0}{0.56} & \textcolor[rgb]{ .753,  0,  0}{0} & \textcolor[rgb]{ .753,  0,  0}{0} & \textcolor[rgb]{ 0,  .502,  0}{0.05} & \textcolor[rgb]{ 0,  .502,  0}{0.03} & \textcolor[rgb]{ 0,  .502,  0}{0.12} \\
      & ID    & \textcolor[rgb]{ 0,  .502,  0}{0.75} & \textcolor[rgb]{ 0,  .502,  0}{0.45} & \textcolor[rgb]{ 0,  .502,  0}{1} & \textcolor[rgb]{ 0,  .502,  0}{0.02} & \textcolor[rgb]{ .753,  0,  0}{0.53} & \textcolor[rgb]{ .753,  0,  0}{0.08} & \textcolor[rgb]{ 0,  .502,  0}{0.31} & \textcolor[rgb]{ 0,  .502,  0}{0.01} & \textcolor[rgb]{ .753,  0,  0}{0} & \textcolor[rgb]{ 0,  .502,  0}{0.09} & \textcolor[rgb]{ .753,  0,  0}{0.2} & \textcolor[rgb]{ .753,  0,  0}{0.24} & \textcolor[rgb]{ 0,  .502,  0}{0.01} & \textcolor[rgb]{ 0,  .502,  0}{0.45} & \textcolor[rgb]{ 0,  .502,  0}{0.25} \\
\bottomrule
    \end{tabular}%
}
  \label{tab_VaR5}%
         \vspace{1ex}\\
      {\justifying \noindent Note: This table compares the performance of VaR 5\% between different models based on christofferson unconditional coverage (UC), Independence (ID) test. The p-values more than 0.01, are considered as "good" VaRs and shown with green. VaRs with at least one p-value less than 0.01 are shown by red, and are not satisfying christofferson's criteria. HS shows historical simulation VaR, GC denotes the Gaussian Copula VaR, Normal is variance-covariance VaR, MC is Monte Carlo Simulation VaR, C-SAV is CaViaR Symmetric VaR, C-AS is CaViaR Asymmetric VaR, C-GARCH is CaViaR GARCH VaR, and C-AD is CaViaR Adaptive model VaR.
      \par}
\end{table}%

\begin{table}
% [width=1\linewidth,cols=7,pos=t!]
  \centering
  \caption{Losses for VaR 1\%}
\resizebox{\textwidth}{!}{
    \begin{tabular}{lcccccc}
    % {\tblwidth}{@{}LCCCCCC@{}}
\toprule
      & \multicolumn{1}{p{5.215em}}{\boldmath{}\textbf{GMM\newline{}$N_c = 3$}\unboldmath{}} & \multicolumn{1}{p{5.215em}}{\boldmath{}\textbf{GMM\newline{}$N_c = 4$}\unboldmath{}} & \multicolumn{1}{p{5.215em}}{\boldmath{}\textbf{GMM\newline{}$N_c = 5$}\unboldmath{}} & \multicolumn{1}{p{5.215em}}{\boldmath{}\textbf{GMM\newline{}$N_c = 6$}\unboldmath{}} & \textbf{HS} & \textbf{Para} \\
\midrule
\textbf{S\&P500} & \textcolor[rgb]{ 0,  .502,  0}{0.01100} & \textcolor[rgb]{ 0,  .502,  0}{0.01400} & \textcolor[rgb]{ 0,  .502,  0}{0.01900} & \textcolor[rgb]{ 0,  .502,  0}{0.01500} & \textcolor[rgb]{ .753,  0,  0}{0.02301} & \textcolor[rgb]{ .753,  0,  0}{0.04001} \\
\textbf{MMM} & \textcolor[rgb]{ 0,  .502,  0}{0.01301} & \textcolor[rgb]{ 0,  .502,  0}{0.01801} & \textcolor[rgb]{ 0,  .502,  0}{0.01501} & \textcolor[rgb]{ 0,  .502,  0}{0.01001} & \textcolor[rgb]{ 0,  .502,  0}{0.01801} & \textcolor[rgb]{ .753,  0,  0}{0.02601} \\
\textbf{AXP} & \textcolor[rgb]{ 0,  .502,  0}{0.01501} & \textcolor[rgb]{ 0,  .502,  0}{0.01601} & \textcolor[rgb]{ 0,  .502,  0}{0.01501} & \textcolor[rgb]{ 0,  .502,  0}{0.01501} & \textcolor[rgb]{ .753,  0,  0}{0.02202} & \textcolor[rgb]{ .753,  0,  0}{0.02803} \\
\textbf{AMD} & \textcolor[rgb]{ 0,  .502,  0}{0.01401} & \textcolor[rgb]{ 0,  .502,  0}{0.01301} & \textcolor[rgb]{ 0,  .502,  0}{0.01101} & \textcolor[rgb]{ 0,  .502,  0}{0.01401} & \textcolor[rgb]{ 0,  .502,  0}{0.01702} & \textcolor[rgb]{ 0,  .502,  0}{0.02002} \\
\textbf{AIG} & \textcolor[rgb]{ 0,  .502,  0}{0.01645} & \textcolor[rgb]{ 0,  .502,  0}{0.01847} & \textcolor[rgb]{ 0,  .502,  0}{0.01849} & \textcolor[rgb]{ 0,  .502,  0}{0.01944} & \textcolor[rgb]{ .753,  0,  0}{0.02684} & \textcolor[rgb]{ .753,  0,  0}{0.04001} \\
\textbf{AFL} & \textcolor[rgb]{ 0,  .502,  0}{0.01608} & \textcolor[rgb]{ 0,  .502,  0}{0.01709} & \textcolor[rgb]{ 0,  .502,  0}{0.01608} & \textcolor[rgb]{ 0,  .502,  0}{0.01808} & \textcolor[rgb]{ 0,  .502,  0}{0.01913} & \textcolor[rgb]{ .753,  0,  0}{0.02620} \\
\textbf{ABT} & \textcolor[rgb]{ 0,  .502,  0}{0.01600} & \textcolor[rgb]{ 0,  .502,  0}{0.01300} & \textcolor[rgb]{ 0,  .502,  0}{0.01800} & \textcolor[rgb]{ 0,  .502,  0}{0.01600} & \textcolor[rgb]{ 0,  .502,  0}{0.02001} & \textcolor[rgb]{ 0,  .502,  0}{0.01701} \\
\textbf{MSFTF} & \textcolor[rgb]{ 0,  .502,  0}{0.01300} & \textcolor[rgb]{ 0,  .502,  0}{0.01300} & \textcolor[rgb]{ 0,  .502,  0}{0.01800} & \textcolor[rgb]{ 0,  .502,  0}{0.01500} & \textcolor[rgb]{ .753,  0,  0}{0.02101} & \textcolor[rgb]{ .753,  0,  0}{0.02401} \\
\textbf{AAPL} & \textcolor[rgb]{ 0,  .502,  0}{0.01302} & \textcolor[rgb]{ 0,  .502,  0}{0.01602} & \textcolor[rgb]{ 0,  .502,  0}{0.01302} & \textcolor[rgb]{ 0,  .502,  0}{0.01202} & \textcolor[rgb]{ 0,  .502,  0}{0.01502} & \textcolor[rgb]{ .753,  0,  0}{0.02403} \\
\textbf{AMZN} & \textcolor[rgb]{ 0,  .502,  0}{0.02001} & \textcolor[rgb]{ 0,  .502,  0}{0.02001} & \textcolor[rgb]{ 0,  .502,  0}{0.01801} & \textcolor[rgb]{ 0,  .502,  0}{0.01801} & \textcolor[rgb]{ 0,  .502,  0}{0.01901} & \textcolor[rgb]{ 0,  .502,  0}{0.01501} \\
\textbf{BAC} & \textcolor[rgb]{ 0,  .502,  0}{0.01804} & \textcolor[rgb]{ 0,  .502,  0}{0.02004} & \textcolor[rgb]{ 0,  .502,  0}{0.01903} & \textcolor[rgb]{ 0,  .502,  0}{0.01803} & \textcolor[rgb]{ .753,  0,  0}{0.02711} & \textcolor[rgb]{ .753,  0,  0}{0.03514} \\
\textbf{JPM} & \textcolor[rgb]{ 0,  .502,  0}{0.01701} & \textcolor[rgb]{ 0,  .502,  0}{0.01801} & \textcolor[rgb]{ 0,  .502,  0}{0.01701} & \textcolor[rgb]{ 0,  .502,  0}{0.01701} & \textcolor[rgb]{ 0,  .502,  0}{0.01703} & \textcolor[rgb]{ .753,  0,  0}{0.02404} \\
\textbf{JNJ} & \textcolor[rgb]{ 0,  .502,  0}{0.01100} & \textcolor[rgb]{ 0,  .502,  0}{0.01200} & \textcolor[rgb]{ 0,  .502,  0}{0.01000} & \textcolor[rgb]{ 0,  .502,  0}{0.01100} & \textcolor[rgb]{ .753,  0,  0}{0.01900} & \textcolor[rgb]{ .753,  0,  0}{0.02301} \\
\textbf{XOM} & \textcolor[rgb]{ 0,  .502,  0}{0.01601} & \textcolor[rgb]{ 0,  .502,  0}{0.01601} & \textcolor[rgb]{ 0,  .502,  0}{0.01801} & \textcolor[rgb]{ 0,  .502,  0}{0.01501} & \textcolor[rgb]{ 0,  .502,  0}{0.01602} & \textcolor[rgb]{ .753,  0,  0}{0.02402} \\
\textbf{MA} & \textcolor[rgb]{ 0,  .502,  0}{0.01801} & \textcolor[rgb]{ 0,  .502,  0}{0.01901} & \textcolor[rgb]{ 0,  .502,  0}{0.01401} & \textcolor[rgb]{ 0,  .502,  0}{0.01501} & \textcolor[rgb]{ 0,  .502,  0}{0.01502} & \textcolor[rgb]{ .753,  0,  0}{0.02702} \\
\midrule
      & \textbf{MC} & \multicolumn{1}{p{5.215em}}{\textbf{GARCH\newline{}(1,1)}} & \textbf{C-SAV} & \textbf{C-AS} & \textbf{C-G} & \textbf{C-AD} \\
\midrule
\textbf{S\&P500} & \textcolor[rgb]{ .753,  0,  0}{0.04101} & \textcolor[rgb]{ .753,  0,  0}{0.04000} & \textcolor[rgb]{ 0,  .502,  0}{0.01400} & \textcolor[rgb]{ .753,  0,  0}{0.02200} & \textcolor[rgb]{ 0,  .502,  0}{0.01800} & \textcolor[rgb]{ .753,  0,  0}{0.02201} \\
\textbf{MMM} & \textcolor[rgb]{ .753,  0,  0}{0.02601} & \textcolor[rgb]{ .753,  0,  0}{0.02701} & \textcolor[rgb]{ 0,  .502,  0}{0.01701} & \textcolor[rgb]{ 0,  .502,  0}{0.02001} & \textcolor[rgb]{ 0,  .502,  0}{0.01801} & \textcolor[rgb]{ 0,  .502,  0}{0.01301} \\
\textbf{AXP} & \textcolor[rgb]{ .753,  0,  0}{0.03003} & \textcolor[rgb]{ 0,  .502,  0}{0.02002} & \textcolor[rgb]{ 0,  .502,  0}{0.01101} & \textcolor[rgb]{ 0,  .502,  0}{0.01202} & \textcolor[rgb]{ 0,  .502,  0}{0.01001} & \textcolor[rgb]{ .753,  0,  0}{0.02102} \\
\textbf{AMD} & \textcolor[rgb]{ 0,  .502,  0}{0.01702} & \textcolor[rgb]{ 0,  .502,  0}{0.01802} & \textcolor[rgb]{ 0,  .502,  0}{0.01302} & \textcolor[rgb]{ 0,  .502,  0}{0.01402} & \textcolor[rgb]{ 0,  .502,  0}{0.15000} & \textcolor[rgb]{ 0,  .502,  0}{0.01601} \\
\textbf{AIG} & \textcolor[rgb]{ .753,  0,  0}{0.03901} & \textcolor[rgb]{ .753,  0,  0}{0.02850} & \textcolor[rgb]{ 0,  .502,  0}{0.01837} & \textcolor[rgb]{ 0,  .502,  0}{0.01422} & \textcolor[rgb]{ 0,  .502,  0}{0.01907} & \textcolor[rgb]{ .753,  0,  0}{0.03074} \\
\textbf{AFL} & \textcolor[rgb]{ .753,  0,  0}{0.02620} & \textcolor[rgb]{ 0,  .502,  0}{0.02111} & \textcolor[rgb]{ 0,  .502,  0}{0.01408} & \textcolor[rgb]{ .753,  0,  0}{0.02110} & \textcolor[rgb]{ 0,  .502,  0}{0.01409} & \textcolor[rgb]{ 0,  .502,  0}{0.01612} \\
\textbf{ABT} & \textcolor[rgb]{ 0,  .502,  0}{0.01801} & \textcolor[rgb]{ .753,  0,  0}{0.02100} & \textcolor[rgb]{ 0,  .502,  0}{0.02000} & \textcolor[rgb]{ 0,  .502,  0}{0.02000} & \textcolor[rgb]{ 0,  .502,  0}{0.02000} & \textcolor[rgb]{ 0,  .502,  0}{0.01501} \\
\textbf{MSFTF} & \textcolor[rgb]{ .753,  0,  0}{0.02501} & \textcolor[rgb]{ .753,  0,  0}{0.02201} & \textcolor[rgb]{ 0,  .502,  0}{0.01200} & \textcolor[rgb]{ 0,  .502,  0}{0.01800} & \textcolor[rgb]{ 0,  .502,  0}{0.01701} & \textcolor[rgb]{ 0,  .502,  0}{0.01700} \\
\textbf{AAPL} & \textcolor[rgb]{ .753,  0,  0}{0.02402} & \textcolor[rgb]{ 0,  .502,  0}{0.01802} & \textcolor[rgb]{ 0,  .502,  0}{0.01302} & \textcolor[rgb]{ 0,  .502,  0}{0.01302} & \textcolor[rgb]{ 0,  .502,  0}{0.01002} & \textcolor[rgb]{ 0,  .502,  0}{0.01402} \\
\textbf{AMZN} & \textcolor[rgb]{ 0,  .502,  0}{0.01501} & \textcolor[rgb]{ 0,  .502,  0}{0.01101} & \textcolor[rgb]{ 0,  .502,  0}{0.01001} & \textcolor[rgb]{ 0,  .502,  0}{0.01102} & \textcolor[rgb]{ 0,  .502,  0}{0.01301} & \textcolor[rgb]{ 0,  .502,  0}{0.01601} \\
\textbf{BAC} & \textcolor[rgb]{ .753,  0,  0}{0.03613} & \textcolor[rgb]{ .753,  0,  0}{0.02204} & \textcolor[rgb]{ 0,  .502,  0}{0.01603} & \textcolor[rgb]{ 0,  .502,  0}{0.01203} & \textcolor[rgb]{ 0,  .502,  0}{0.01403} & \textcolor[rgb]{ .753,  0,  0}{0.02210} \\
\textbf{JPM} & \textcolor[rgb]{ .753,  0,  0}{0.02404} & \textcolor[rgb]{ .753,  0,  0}{0.02302} & \textcolor[rgb]{ 0,  .502,  0}{0.01601} & \textcolor[rgb]{ 0,  .502,  0}{0.01802} & \textcolor[rgb]{ 0,  .502,  0}{0.01902} & \textcolor[rgb]{ .753,  0,  0}{0.02602} \\
\textbf{JNJ} & \textcolor[rgb]{ .753,  0,  0}{0.02500} & \textcolor[rgb]{ .753,  0,  0}{0.02400} & \textcolor[rgb]{ 0,  .502,  0}{0.00800} & \textcolor[rgb]{ 0,  .502,  0}{0.01100} & \textcolor[rgb]{ 0,  .502,  0}{0.01200} & \textcolor[rgb]{ 0,  .502,  0}{0.01500} \\
\textbf{XOM} & \textcolor[rgb]{ .753,  0,  0}{0.02602} & \textcolor[rgb]{ .753,  0,  0}{0.02901} & \textcolor[rgb]{ 0,  .502,  0}{0.01501} & \textcolor[rgb]{ 0,  .502,  0}{0.01601} & \textcolor[rgb]{ 0,  .502,  0}{0.01500} & \textcolor[rgb]{ 0,  .502,  0}{0.01701} \\
\textbf{MA} & \textcolor[rgb]{ .753,  0,  0}{0.02702} & \textcolor[rgb]{ .753,  0,  0}{0.02501} & \textcolor[rgb]{ 0,  .502,  0}{0.01701} & \textcolor[rgb]{ 0,  .502,  0}{0.01801} & \textcolor[rgb]{ .753,  0,  0}{0.02301} & \textcolor[rgb]{ 0,  .502,  0}{0.01802} \\
\bottomrule

    \end{tabular}%
}
  \label{tab_lossVaR1}%
         \vspace{1ex}\\
      {\justifying \noindent Note: This table compares the performance of VaR 1\% between different models based on Quadratic loss function. The color red shows that if the hypothesis that VaR is satisfying the christofferson's criteria is rejected. HS shows historical simulation VaR, GC denotes the Gaussian Copula VaR, Para is parametric or variance-covariance VaR, MC is Monte Carlo Simulation VaR, C-SAV is CaViaR Symmetric VaR, C-AS is CaViaR Asymmetric VaR, C-GARCH is CaViaR GARCH VaR, and C-AD is CaViaR Adaptive model VaR.
      \par}
\end{table}%

\begin{table}
% [width=1.0\linewidth,cols=7,pos=t!]
  \centering
  \caption{Losses for VaR 5\%}
\resizebox{\textwidth}{!}{
    \begin{tabular}{lccccccc}
    % {\tblwidth}{@{}LCCCCCCC@{}}
\toprule
      & \multicolumn{1}{p{5.215em}}{\boldmath{}\textbf{GMM\newline{}$N_c = 3$}\unboldmath{}} & \multicolumn{1}{p{5.215em}}{\boldmath{}\textbf{GMM\newline{}$N_c = 4$}\unboldmath{}} & \multicolumn{1}{p{5.215em}}{\boldmath{}\textbf{GMM\newline{}$N_c = 5$}\unboldmath{}} & \multicolumn{1}{p{5.215em}}{\boldmath{}\textbf{GMM\newline{}$N_c = 6$}\unboldmath{}} & \textbf{HS} & \textbf{Para} \\
\midrule
\textbf{S\&P500} & \textcolor[rgb]{ 0,  .502,  0}{0.05801} & \textcolor[rgb]{ 0,  .502,  0}{0.05801} & \textcolor[rgb]{ 0,  .502,  0}{0.06301} & \textcolor[rgb]{ 0,  .502,  0}{0.06401} & \textcolor[rgb]{ 0,  .502,  0}{0.06402} & \textcolor[rgb]{ 0,  .502,  0}{0.07102} \\
\textbf{MMM} & \textcolor[rgb]{ 0,  .502,  0}{0.06302} & \textcolor[rgb]{ 0,  .502,  0}{0.06402} & \textcolor[rgb]{ 0,  .502,  0}{0.06102} & \textcolor[rgb]{ 0,  .502,  0}{0.06002} & \textcolor[rgb]{ .753,  0,  0}{0.07202} & \textcolor[rgb]{ 0,  .502,  0}{0.05802} \\
\textbf{AXP} & \textcolor[rgb]{ 0,  .502,  0}{0.06205} & \textcolor[rgb]{ 0,  .502,  0}{0.06405} & \textcolor[rgb]{ 0,  .502,  0}{0.06405} & \textcolor[rgb]{ 0,  .502,  0}{0.06105} & \textcolor[rgb]{ .753,  0,  0}{0.07608} & \textcolor[rgb]{ .753,  0,  0}{0.07407} \\
\textbf{AMD} & \textcolor[rgb]{ .753,  0,  0}{0.06105} & \textcolor[rgb]{ 0,  .502,  0}{0.06106} & \textcolor[rgb]{ 0,  .502,  0}{0.06506} & \textcolor[rgb]{ 0,  .502,  0}{0.06106} & \textcolor[rgb]{ 0,  .502,  0}{0.06507} & \textcolor[rgb]{ 0,  .502,  0}{0.06207} \\
\textbf{AIG} & \textcolor[rgb]{ 0,  .502,  0}{0.06922} & \textcolor[rgb]{ 0,  .502,  0}{0.06723} & \textcolor[rgb]{ 0,  .502,  0}{0.07128} & \textcolor[rgb]{ 0,  .502,  0}{0.06925} & \textcolor[rgb]{ .753,  0,  0}{0.08447} & \textcolor[rgb]{ 0,  .502,  0}{0.06922} \\
\textbf{AFL} & \textcolor[rgb]{ 0,  .502,  0}{0.05419} & \textcolor[rgb]{ 0,  .502,  0}{0.05821} & \textcolor[rgb]{ 0,  .502,  0}{0.05820} & \textcolor[rgb]{ 0,  .502,  0}{0.05820} & \textcolor[rgb]{ 0,  .502,  0}{0.07031} & \textcolor[rgb]{ 0,  .502,  0}{0.06427} \\
\textbf{ABT} & \textcolor[rgb]{ 0,  .502,  0}{0.05801} & \textcolor[rgb]{ 0,  .502,  0}{0.05901} & \textcolor[rgb]{ 0,  .502,  0}{0.05701} & \textcolor[rgb]{ 0,  .502,  0}{0.05601} & \textcolor[rgb]{ 0,  .502,  0}{0.05702} & \textcolor[rgb]{ 0,  .502,  0}{0.04801} \\
\textbf{MSFTF} & \textcolor[rgb]{ 0,  .502,  0}{0.05102} & \textcolor[rgb]{ 0,  .502,  0}{0.05502} & \textcolor[rgb]{ .753,  0,  0}{0.05602} & \textcolor[rgb]{ 0,  .502,  0}{0.05302} & \textcolor[rgb]{ .753,  0,  0}{0.06603} & \textcolor[rgb]{ .753,  0,  0}{0.05902} \\
\textbf{AAPL} & \textcolor[rgb]{ .753,  0,  0}{0.05504} & \textcolor[rgb]{ 0,  .502,  0}{0.05304} & \textcolor[rgb]{ .753,  0,  0}{0.05004} & \textcolor[rgb]{ .753,  0,  0}{0.05504} & \textcolor[rgb]{ .753,  0,  0}{0.05105} & \textcolor[rgb]{ 0,  .502,  0}{0.04905} \\
\textbf{AMZN} & \textcolor[rgb]{ 0,  .502,  0}{0.05704} & \textcolor[rgb]{ 0,  .502,  0}{0.05504} & \textcolor[rgb]{ 0,  .502,  0}{0.05704} & \textcolor[rgb]{ 0,  .502,  0}{0.05605} & \textcolor[rgb]{ 0,  .502,  0}{0.06006} & \textcolor[rgb]{ 0,  .502,  0}{0.03304} \\
\textbf{BAC} & \textcolor[rgb]{ 0,  .502,  0}{0.06917} & \textcolor[rgb]{ 0,  .502,  0}{0.06717} & \textcolor[rgb]{ 0,  .502,  0}{0.06717} & \textcolor[rgb]{ 0,  .502,  0}{0.06416} & \textcolor[rgb]{ .753,  0,  0}{0.08831} & \textcolor[rgb]{ .753,  0,  0}{0.07625} \\
\textbf{JPM} & \textcolor[rgb]{ 0,  .502,  0}{0.05507} & \textcolor[rgb]{ 0,  .502,  0}{0.05407} & \textcolor[rgb]{ 0,  .502,  0}{0.05706} & \textcolor[rgb]{ 0,  .502,  0}{0.05707} & \textcolor[rgb]{ .753,  0,  0}{0.07612} & \textcolor[rgb]{ 0,  .502,  0}{0.06009} \\
\textbf{JNJ} & \textcolor[rgb]{ .753,  0,  0}{0.06001} & \textcolor[rgb]{ .753,  0,  0}{0.05701} & \textcolor[rgb]{ .753,  0,  0}{0.05901} & \textcolor[rgb]{ .753,  0,  0}{0.05801} & \textcolor[rgb]{ .753,  0,  0}{0.06301} & \textcolor[rgb]{ .753,  0,  0}{0.05401} \\
\textbf{XOM} & \textcolor[rgb]{ 0,  .502,  0}{0.05803} & \textcolor[rgb]{ 0,  .502,  0}{0.06203} & \textcolor[rgb]{ 0,  .502,  0}{0.06103} & \textcolor[rgb]{ 0,  .502,  0}{0.05803} & \textcolor[rgb]{ 0,  .502,  0}{0.05904} & \textcolor[rgb]{ 0,  .502,  0}{0.05804} \\
\textbf{MA} & \textcolor[rgb]{ 0,  .502,  0}{0.05703} & \textcolor[rgb]{ 0,  .502,  0}{0.05603} & \textcolor[rgb]{ 0,  .502,  0}{0.06203} & \textcolor[rgb]{ 0,  .502,  0}{0.05503} & \textcolor[rgb]{ 0,  .502,  0}{0.06205} & \textcolor[rgb]{ 0,  .502,  0}{0.06204} \\
\midrule
      & \textbf{MC} & \multicolumn{1}{p{5.215em}}{\textbf{GARCH\newline{}(1,1)}} & \textbf{C-SAV} & \textbf{C-AS} & \textbf{C-G} & \textbf{C-AD} \\
\midrule
\textbf{S\&P500} & \textcolor[rgb]{ 0,  .502,  0}{0.07102} & \textcolor[rgb]{ .753,  0,  0}{0.08201} & \textcolor[rgb]{ 0,  .502,  0}{0.05601} & \textcolor[rgb]{ 0,  .502,  0}{0.05401} & \textcolor[rgb]{ 0,  .502,  0}{0.06810} & \textcolor[rgb]{ 0,  .502,  0}{0.06902} \\
\textbf{MMM} & \textcolor[rgb]{ 0,  .502,  0}{0.05902} & \textcolor[rgb]{ 0,  .502,  0}{0.06202} & \textcolor[rgb]{ 0,  .502,  0}{0.06401} & \textcolor[rgb]{ .753,  0,  0}{0.07402} & \textcolor[rgb]{ 0,  .502,  0}{0.06602} & \textcolor[rgb]{ 0,  .502,  0}{0.06602} \\
\textbf{AXP} & \textcolor[rgb]{ .753,  0,  0}{0.07407} & \textcolor[rgb]{ 0,  .502,  0}{0.06305} & \textcolor[rgb]{ 0,  .502,  0}{0.05304} & \textcolor[rgb]{ 0,  .502,  0}{0.05204} & \textcolor[rgb]{ 0,  .502,  0}{0.05304} & \textcolor[rgb]{ 0,  .502,  0}{0.06907} \\
\textbf{AMD} & \textcolor[rgb]{ 0,  .502,  0}{0.06207} & \textcolor[rgb]{ 0,  .502,  0}{0.05806} & \textcolor[rgb]{ 0,  .502,  0}{0.06805} & \textcolor[rgb]{ 0,  .502,  0}{0.06205} & \textcolor[rgb]{ .753,  0,  0}{0.07506} & \textcolor[rgb]{ .753,  0,  0}{0.06407} \\
\textbf{AIG} & \textcolor[rgb]{ 0,  .502,  0}{0.06821} & \textcolor[rgb]{ 0,  .502,  0}{0.05874} & \textcolor[rgb]{ 0,  .502,  0}{0.05770} & \textcolor[rgb]{ 0,  .502,  0}{0.05671} & \textcolor[rgb]{ 0,  .502,  0}{0.05565} & \textcolor[rgb]{ .753,  0,  0}{0.09841} \\
\textbf{AFL} & \textcolor[rgb]{ 0,  .502,  0}{0.06227} & \textcolor[rgb]{ 0,  .502,  0}{0.05517} & \textcolor[rgb]{ 0,  .502,  0}{0.05921} & \textcolor[rgb]{ 0,  .502,  0}{0.05818} & \textcolor[rgb]{ 0,  .502,  0}{0.06519} & \textcolor[rgb]{ 0,  .502,  0}{0.07730} \\
\textbf{ABT} & \textcolor[rgb]{ 0,  .502,  0}{0.04901} & \textcolor[rgb]{ 0,  .502,  0}{0.05001} & \textcolor[rgb]{ 0,  .502,  0}{0.06601} & \textcolor[rgb]{ 0,  .502,  0}{0.06701} & \textcolor[rgb]{ 0,  .502,  0}{0.06010} & \textcolor[rgb]{ 0,  .502,  0}{0.05601} \\
\textbf{MSFTF} & \textcolor[rgb]{ .753,  0,  0}{0.06102} & \textcolor[rgb]{ 0,  .502,  0}{0.05502} & \textcolor[rgb]{ 0,  .502,  0}{0.05502} & \textcolor[rgb]{ 0,  .502,  0}{0.05102} & \textcolor[rgb]{ 0,  .502,  0}{0.05602} & \textcolor[rgb]{ 0,  .502,  0}{0.06802} \\
\textbf{AAPL} & \textcolor[rgb]{ 0,  .502,  0}{0.05005} & \textcolor[rgb]{ .753,  0,  0}{0.05404} & \textcolor[rgb]{ 0,  .502,  0}{0.05604} & \textcolor[rgb]{ 0,  .502,  0}{0.05303} & \textcolor[rgb]{ 0,  .502,  0}{0.50100} & \textcolor[rgb]{ .753,  0,  0}{0.04905} \\
\textbf{AMZN} & \textcolor[rgb]{ 0,  .502,  0}{0.03304} & \textcolor[rgb]{ 0,  .502,  0}{0.03402} & \textcolor[rgb]{ 0,  .502,  0}{0.05503} & \textcolor[rgb]{ 0,  .502,  0}{0.06503} & \textcolor[rgb]{ 0,  .502,  0}{0.06104} & \textcolor[rgb]{ 0,  .502,  0}{0.05505} \\
\textbf{BAC} & \textcolor[rgb]{ .753,  0,  0}{0.07725} & \textcolor[rgb]{ 0,  .502,  0}{0.05813} & \textcolor[rgb]{ 0,  .502,  0}{0.05309} & \textcolor[rgb]{ 0,  .502,  0}{0.06114} & \textcolor[rgb]{ 0,  .502,  0}{0.06011} & \textcolor[rgb]{ .753,  0,  0}{0.09329} \\
\textbf{JPM} & \textcolor[rgb]{ 0,  .502,  0}{0.06109} & \textcolor[rgb]{ 0,  .502,  0}{0.05605} & \textcolor[rgb]{ 0,  .502,  0}{0.05406} & \textcolor[rgb]{ 0,  .502,  0}{0.05807} & \textcolor[rgb]{ 0,  .502,  0}{0.05507} & \textcolor[rgb]{ .753,  0,  0}{0.07411} \\
\textbf{JNJ} & \textcolor[rgb]{ .753,  0,  0}{0.05201} & \textcolor[rgb]{ 0,  .502,  0}{0.04901} & \textcolor[rgb]{ 0,  .502,  0}{0.05701} & \textcolor[rgb]{ 0,  .502,  0}{0.06501} & \textcolor[rgb]{ 0,  .502,  0}{0.06101} & \textcolor[rgb]{ 0,  .502,  0}{0.06401} \\
\textbf{XOM} & \textcolor[rgb]{ 0,  .502,  0}{0.05804} & \textcolor[rgb]{ .753,  0,  0}{0.06902} & \textcolor[rgb]{ 0,  .502,  0}{0.06302} & \textcolor[rgb]{ 0,  .502,  0}{0.06204} & \textcolor[rgb]{ 0,  .502,  0}{0.06002} & \textcolor[rgb]{ 0,  .502,  0}{0.06604} \\
\textbf{MA} & \textcolor[rgb]{ 0,  .502,  0}{0.06504} & \textcolor[rgb]{ .753,  0,  0}{0.08202} & \textcolor[rgb]{ 0,  .502,  0}{0.05401} & \textcolor[rgb]{ 0,  .502,  0}{0.05702} & \textcolor[rgb]{ 0,  .502,  0}{0.05502} & \textcolor[rgb]{ 0,  .502,  0}{0.06104} \\
\bottomrule
    \end{tabular}%
}
  \label{tab_lossVaR5}%
           \vspace{1ex}\\
      {\justifying \noindent Note: This table compares the performance of VaR 5\% between different models based on Quadratic loss function. The color red shows that if the hypothesis that VaR is satisfying the christoffeson's criteria is rejected. HS shows historical simulation VaR, GC denotes the Gaussian Copula VaR, Normal is variance-covariance VaR, MC is Monte Carlo Simulation VaR, C-SAV is CaViaR Symmetric VaR, C-AS is CaViaR Asymmetric VaR, C-GARCH is CaViaR GARCH VaR, and C-AD is CaViaR Adaptive model VaR.
      \par}
\end{table}%

\begin{table}
% [width=1.0\linewidth,cols=7,pos=t!]
  \centering
  \caption{The back-testing results for VaRs of a portfolio}
%   \resizebox{\textwidth}{!}{
    \begin{tabular}{lcccccc}
    % {\tblwidth}{@{}LCCCCCC@{}} 
\toprule
\textbf{rank} & \textbf{Model (VaR 1\%)} & \textbf{UC} & \textbf{ID} & \textbf{CC} & \textbf{result} & \textbf{quadratic loss} \\
\midrule
1     & \textbf{HS} & 0.17  & 0.79  & 0.17  & NR    & 0.0060 \\
2     & \boldmath{}\textbf{GMM $N_c = 4$}\unboldmath{} & 0.01  & 0.4   & 0.01  & NR    & 0.0190 \\
3     & \boldmath{}\textbf{GMM $N_c = 6$}\unboldmath{} & 0.01  & 0.37  & 0.01  & NR    & 0.0210 \\
4     & \textbf{C-GARCH} & 0     & 0.42  & 0     & R     & 0.0180 \\
5     & \textbf{C-ad} & 0     & 0.47  & 0     & R     & 0.0160 \\
6     & \boldmath{}\textbf{GMM $N_c = 3$}\unboldmath{} & 0     & 0.34  & 0     & R     & 0.0220 \\
7     & \boldmath{}\textbf{GMM $N_c = 5$}\unboldmath{} & 0     & 0.3   & 0     & R     & 0.0240 \\
8     & \textbf{C-sav} & 0     & 0.29  & 0     & R     & 0.0240 \\
9     & \textbf{C-as} & 0     & 0.25  & 0     & R     & 0.0260 \\
10    & \textbf{Garch} & 0     & 0.18  & 0     & R     & 0.0300 \\
11    & \textbf{GC} & 0     & 0     & 0     & R     & 0.0360 \\
12    & \textbf{Para} & 0     & 0.79  & 0     & R     & 0.0400 \\
13    & \textbf{MC} & 0     & 0.39  & 0     & R     & 0.1471 \\
\midrule
\textbf{rank} & \textbf{Model (VaR 5\%)} & \textbf{UC} & \textbf{ID} & \textbf{CC} & \textbf{result} & \textbf{quadratic loss} \\
\midrule
1     & \textbf{C-sav} & 0.56  & 0.63  & 0.56  & NR    & 0.0540 \\
2     & \textbf{C-as} & 0.39  & 0.14  & 0.39  & NR    & 0.0570 \\
3     & \textbf{C-GARCH} & 0.56  & 0.55  & 0.56  & NR    & 0.0550 \\
4     & \boldmath{}\textbf{GMM $N_c = 4$}\unboldmath{} & 0.12  & 0.3   & 0.12  & NR    & 0.0620 \\
5     & \boldmath{}\textbf{GMM $N_c = 6$}\unboldmath{} & 0.05  & 0.54  & 0.05  & NR    & 0.0650 \\
6     & \boldmath{}\textbf{GMM $N_c = 3$}\unboldmath{} & 0.03  & 0.46  & 0.03  & NR    & 0.0670 \\
7     & \textbf{Garch} & 0.01  & 0.15  & 0.01  & NR    & 0.0690 \\
8     & \boldmath{}\textbf{GMM $N_c = 5$}\unboldmath{} & 0.01  & 0.15  & 0.01  & NR    & 0.0690 \\
9     & \textbf{HS} & 0     & 0.79  & 0     & R     & 0.0060 \\
10    & \textbf{GC} & 0     & 0.68  & 0     & R     & 0.0720 \\
11    & \textbf{C-ad} & 0     & 0.82  & 0     & R     & 0.0750 \\
12    & \textbf{Para} & 0     & 0.33  & 0     & R     & 0.0790 \\
13    & \textbf{MC} & 0     & 0.39  & 0     & R     & 0.1951 \\
\bottomrule
    \end{tabular}%
% }
  \label{tab_port_var}%
           \vspace{1ex}\\
      {\justifying \noindent Note: This table compares the performance of portfolio VaRs different models based on both Christoffersen's test and Quadratic loss function. If Christoffersen's test rejects goodness of a VaR it is shown by R, while otherwise it is shown by NR (Not Rejected). P-values for unconditional test (UC), Independence test (ID), and Condition Coverage (CC) are shown. HS shows historical simulation VaR,  GC denotes the Gaussian Copula VaR, Normal is variance-covariance VaR, MC is Monte Carlo Simulation VaR, C-SAV is CaViaR Symmetric VaR, C-AS is CaViaR Asymmetric VaR, C-GARCH is CaViaR GARCH VaR, and C-AD is CaViaR Adaptive model VaR.
      \par}
\end{table}%

\section{Conclusion and Future Works}

A Gmm-based Monte Carlo approach for simulating stock prices with applications in value-at-risk has been introduced. Our model can be an alternative to GBM and Copula approaches. Our results were mostly approved by Christoffersen's test and using a quadratic loss function in a crisis period. While the traditional methods are time-consuming and fail the validity tests, our portfolio's VaR algorithm is by far less time consuming and its results were mostly approved by Christoffersen's test. The power of Gmm's simulation algorithm can be seen when estimating the risk of a portfolio of stocks. Gmm can produce the portfolio structure, with the same joint distribution and preserves non-linear correlation between individual stocks quite well. 

Since our model's predictive ability is approved by back-testing routines, using other forecasting approaches like GARCH models is not necessary. However, a combination with these models might also provide accurate and robust results. Another direction to do further research is to apply our technique for valuation of derivatives. Predicting the trend of market is also another application of our model which might be used in algorithmic trading machines. Without a shadow of a doubt, the algorithm speed is vital for professional traders working in financial institutions.  While being highly accurate, our model's computational cost is quite low. This is particularly interesting that as our results show, three number of Gmm components is sufficient for capturing complex market conditions, even during a crisis.

This method can be applied for simulation of not only stocks, but many other financial instruments such as plain bonds or bonds with embedded options. Luckily, Gmm is capable of capturing complex data structure, specifically the tail of return distributions. Another advantage of using Gmm for calculating VaR is that there is no assumption for a linear correlation between different assets.

\clearpage

\setcitestyle{numbers} % set the citation style to ``numbers''.
\bibliographystyle{plainnat}
\bibliography{refs}

\begin{thebibliography}{44}
\providecommand{\natexlab}[1]{#1}
\providecommand{\url}[1]{\texttt{#1}}
\expandafter\ifx\csname urlstyle\endcsname\relax
  \providecommand{\doi}[1]{doi: #1}\else
  \providecommand{\doi}{doi: \begingroup \urlstyle{rm}\Url}\fi

\bibitem[Abidin and Jaffar(2014)]{abidin2014forecasting}
Siti Nazifah~Zainol Abidin and Maheran~Mohd Jaffar.
\newblock Forecasting share prices of small size companies in bursa malaysia
  using geometric brownian motion.
\newblock \emph{Applied Mathematics \& Information Sciences}, 8\penalty0
  (1):\penalty0 107, 2014.

\bibitem[A{\"\i}t-Sahalia and Xiu(2016)]{ait2016increased}
Yacine A{\"\i}t-Sahalia and Dacheng Xiu.
\newblock Increased correlation among asset classes: Are volatility or jumps to
  blame, or both?
\newblock \emph{Journal of Econometrics}, 194\penalty0 (2):\penalty0 205--219,
  2016.

\bibitem[Bishop(2006)]{bishop2006pattern}
Christopher~M Bishop.
\newblock \emph{Pattern recognition and machine learning}.
\newblock springer, 2006.

\bibitem[Black and Scholes(1973)]{black1973pricing}
Fischer Black and Myron Scholes.
\newblock The pricing of options and corporate liabilities.
\newblock \emph{Journal of political economy}, 81\penalty0 (3):\penalty0
  637--654, 1973.

\bibitem[Britten-Jones and Schaefer(1999)]{britten1999non}
Mark Britten-Jones and Stephen~M Schaefer.
\newblock Non-linear value-at-risk.
\newblock \emph{Review of Finance}, 2\penalty0 (2):\penalty0 161--187, 1999.

\bibitem[Chen et~al.(2012)Chen, Gerlach, Hwang, and McAleer]{Chen2012}
Cathy~W.S. Chen, Richard Gerlach, Bruce~B.K. Hwang, and Michael McAleer.
\newblock {Forecasting Value-at-Risk using nonlinear regression quantiles and
  the intra-day range}.
\newblock \emph{International Journal of Forecasting}, 28\penalty0
  (3):\penalty0 557--574, 2012.
\newblock ISSN 01692070.

\bibitem[Chen and Hong(2007)]{chen2007monte}
Nan Chen and L~Jeff Hong.
\newblock Monte carlo simulation in financial engineering.
\newblock In \emph{2007 Winter Simulation Conference}, pages 919--931. IEEE,
  2007.

\bibitem[Chin et~al.(1999)Chin, Weigend, and Zimmermann]{chin1999computing}
Elion Chin, Andreas~S Weigend, and Heinz Zimmermann.
\newblock Computing portfolio risk using gaussian mixtures and independent
  component analysis.
\newblock In \emph{Proceedings of the IEEE/IAFE 1999 Conference on
  Computational Intelligence for Financial Engineering (CIFEr)(IEEE Cat. No.
  99TH8408)}, pages 74--117. IEEE, 1999.

\bibitem[Christoffersen(1998)]{christoffersen1998evaluating}
Peter~F Christoffersen.
\newblock Evaluating interval forecasts.
\newblock \emph{International economic review}, pages 841--862, 1998.

\bibitem[Danielsson et~al.(2016)Danielsson, James, Valenzuela, and
  Zer]{Danielsson2016}
Jon Danielsson, Kevin~R. James, Marcela Valenzuela, and Ilknur Zer.
\newblock {Model risk of risk models}.
\newblock \emph{Journal of Financial Stability}, 23:\penalty0 79--91, 2016.

\bibitem[Date and Bustreo(2016)]{date2016measuring}
Paresh Date and Roberto Bustreo.
\newblock Measuring the risk of a non-linear portfolio with fat-tailed risk
  factors through a probability conserving transformation.
\newblock \emph{IMA Journal of Management Mathematics}, 27\penalty0
  (2):\penalty0 157--180, 2016.

\bibitem[Dowd(2007)]{dowd2007measuring}
Kevin Dowd.
\newblock \emph{Measuring market risk}.
\newblock John Wiley \& Sons, 2007.

\bibitem[Duffie and Pan(2001)]{duffie2001analytical}
Darrell Duffie and Jun Pan.
\newblock Analytical value-at-risk with jumps and credit risk.
\newblock \emph{Finance and Stochastics}, 5\penalty0 (2):\penalty0 155--180,
  2001.

\bibitem[Engle and Manganelli(2004)]{engle2004caviar}
Robert~F Engle and Simone Manganelli.
\newblock Caviar: Conditional autoregressive value at risk by regression
  quantiles.
\newblock \emph{Journal of Business \& Economic Statistics}, 22\penalty0
  (4):\penalty0 367--381, 2004.

\bibitem[Engle and Sheppard(2001)]{engle2001theoretical}
Robert~F Engle and Kevin Sheppard.
\newblock Theoretical and empirical properties of dynamic conditional
  correlation multivariate garch.
\newblock Technical report, National Bureau of Economic Research, 2001.

\bibitem[Fama(1965)]{fama1965behavior}
Eugene~F Fama.
\newblock The behavior of stock-market prices.
\newblock \emph{The journal of Business}, 38\penalty0 (1):\penalty0 34--105,
  1965.

\bibitem[Frank et~al.(2009)]{frank2009linkages}
Nathaniel Frank et~al.
\newblock Linkages between asset classes during the financial crisis,
  accounting for market microstructure noise and non-synchronous trading.
\newblock Technical report, Economics Group, Nuffield College, University of
  Oxford, 2009.

\bibitem[Fuertes and Olmo(2013)]{Fuertes2013}
Ana~Maria Fuertes and Jose Olmo.
\newblock {Optimally harnessing inter-day and intra-day information for daily
  value-at-risk prediction}.
\newblock \emph{International Journal of Forecasting}, 29\penalty0
  (1):\penalty0 28--42, 2013.

\bibitem[Furi{\'o} and Climent(2013)]{furio2013extreme}
Dolores Furi{\'o} and Francisco~J Climent.
\newblock Extreme value theory versus traditional garch approaches applied to
  financial data: a comparative evaluation.
\newblock \emph{Quantitative Finance}, 13\penalty0 (1):\penalty0 45--63, 2013.

\bibitem[Glasserman(2013)]{glasserman2013monte}
Paul Glasserman.
\newblock \emph{Monte Carlo methods in financial engineering}, volume~53.
\newblock Springer Science \& Business Media, 2013.

\bibitem[Glasserman et~al.(2002)Glasserman, Heidelberger, and
  Shahabuddin]{glasserman2002portfolio}
Paul Glasserman, Philip Heidelberger, and Perwez Shahabuddin.
\newblock Portfolio value-at-risk with heavy-tailed risk factors.
\newblock \emph{Mathematical Finance}, 12\penalty0 (3):\penalty0 239--269,
  2002.

\bibitem[Haugen and Haugen(2001)]{haugen2001modern}
Robert~A Haugen and Robert~A Haugen.
\newblock \emph{Modern investment theory}, volume~5.
\newblock Prentice Hall Upper Saddle River, NJ, 2001.

\bibitem[Jorion(2000)]{jorion2000value}
Philippe Jorion.
\newblock Value at risk.
\newblock 2000.

\bibitem[Klaassen and van Eeghen(2009)]{klaassen2009economic}
Pieter Klaassen and Idzard van Eeghen.
\newblock \emph{Economic Capital: How It Works, and What Every Manager Needs to
  Know}.
\newblock Elsevier, 2009.

\bibitem[Kon(1984)]{kon1984models}
Stanley~J Kon.
\newblock Models of stock returns—a comparison.
\newblock \emph{The Journal of Finance}, 39\penalty0 (1):\penalty0 147--165,
  1984.

\bibitem[Lopez et~al.(1999)]{lopez1999methods}
Jose~A Lopez et~al.
\newblock Methods for evaluating value-at-risk estimates.
\newblock \emph{Economic review}, 2:\penalty0 3--17, 1999.

\bibitem[Martens et~al.(2009)Martens, Van~Dijk, and
  De~Pooter]{martens2009forecasting}
Martin Martens, Dick Van~Dijk, and Michiel De~Pooter.
\newblock Forecasting s\&p 500 volatility: Long memory, level shifts, leverage
  effects, day-of-the-week seasonality, and macroeconomic announcements.
\newblock \emph{International Journal of forecasting}, 25\penalty0
  (2):\penalty0 282--303, 2009.

\bibitem[Meng and Taylor(2018)]{Meng2018}
Xiaochun Meng and James~W. Taylor.
\newblock {An approximate long-memory range-based approach for value at risk
  estimation}.
\newblock \emph{International Journal of Forecasting}, 34\penalty0
  (3):\penalty0 377--388, 2018.
\newblock ISSN 01692070.

\bibitem[Merton et~al.(1973)]{merton1973intertemporal}
Robert~C Merton et~al.
\newblock An intertemporal capital asset pricing model.
\newblock \emph{Econometrica}, 41\penalty0 (5):\penalty0 867--887, 1973.

\bibitem[Morgan et~al.(1996)]{morgan1996riskmetrics}
JP~Morgan et~al.
\newblock Riskmetrics technical document.
\newblock 1996.

\bibitem[Morone(2008)]{morone2008financial}
Andrea Morone.
\newblock Financial markets in the laboratory: an experimental analysis of some
  stylized facts.
\newblock \emph{Quantitative Finance}, 8\penalty0 (5):\penalty0 513--532, 2008.

\bibitem[Peng et~al.(2019)Peng, Hu, Chen, Zeng, and Yang]{peng2019modeling}
Wei Peng, Shichao Hu, Wang Chen, Yu-feng Zeng, and Lu~Yang.
\newblock Modeling the joint dynamic value at risk of the volatility index, oil
  price, and exchange rate.
\newblock \emph{International Review of Economics \& Finance}, 59:\penalty0
  137--149, 2019.

\bibitem[Reddy and Clinton(2016)]{reddy2016simulating}
Krishna Reddy and Vaughan Clinton.
\newblock Simulating stock prices using geometric brownian motion: Evidence
  from australian companies.
\newblock \emph{Australasian Accounting, Business and Finance Journal},
  10\penalty0 (3):\penalty0 23--47, 2016.

\bibitem[Rollett and Manohar(2004)]{rollett2004monte}
AD~Rollett and P~Manohar.
\newblock The monte carlo method continuum scale simulation of engineering
  materials, 2004.

\bibitem[Rossignolo et~al.(2012)Rossignolo, Fethi, and Shaban]{Rossignolo2012}
Adrian~F. Rossignolo, Meryem~Duygun Fethi, and Mohamed Shaban.
\newblock {Value-at-Risk models and Basel capital charges. Evidence from
  Emerging and Frontier stock markets}.
\newblock \emph{Journal of Financial Stability}, 8\penalty0 (4):\penalty0
  303--319, 2012.

\bibitem[Sengupta(2004)]{sengupta2004financial}
Chandan Sengupta.
\newblock \emph{Financial modeling using excel and VBA}, volume 152.
\newblock John Wiley \& Sons, 2004.

\bibitem[Stoyanov et~al.(2011)Stoyanov, Rachev, Racheva-Yotova, and
  Fabozzi]{stoyanov2011fat}
Stoyan~V Stoyanov, Svetlozar~T Rachev, Boryana Racheva-Yotova, and Frank~J
  Fabozzi.
\newblock Fat-tailed models for risk estimation.
\newblock \emph{The Journal of Portfolio Management}, 37\penalty0 (2):\penalty0
  107--117, 2011.

\bibitem[Sui(2003)]{sui2003value}
Sen Sui.
\newblock \emph{Value-at-Risk Analysis of Portfolio Return Model Using
  Independent Component Analysis and Gaussian Mixture Mode.}
\newblock PhD thesis, The Chinese University of Hong Kong, 2003.

\bibitem[Takahashi et~al.(2019)Takahashi, Chen, and
  Tanaka-Ishii]{takahashi2019modeling}
Shuntaro Takahashi, Yu~Chen, and Kumiko Tanaka-Ishii.
\newblock Modeling financial time-series with generative adversarial networks.
\newblock \emph{Physica A: Statistical Mechanics and its Applications},
  527:\penalty0 121261, 2019.

\bibitem[Tan(2005)]{tan2005modeling}
K~Tan.
\newblock Modeling returns distribution based on radical normal distributions.
\newblock \emph{Journal of the society for studies on industrial economies},
  46\penalty0 (3):\penalty0 449--467, 2005.

\bibitem[Tan and Tokinaga(2007)]{tan2007approximation}
K~Tan and S~Tokinaga.
\newblock An approximation of returns distribution based upon ga optimized
  mixture distribution and its applications.
\newblock In \emph{Proceedings of the Fourth International Conference on
  Computational Intelligence, Robotics and Autonomous Systems}, pages 307--312,
  2007.

\bibitem[Wang et~al.(2018)Wang, Shen, and Liu]{wang2018conditional}
Zhiwen Wang, Chen Shen, and Feng Liu.
\newblock A conditional model of wind power forecast errors and its application
  in scenario generation.
\newblock \emph{Applied energy}, 212:\penalty0 771--785, 2018.

\bibitem[Wilson(1999)]{wilson1999value}
T~Wilson.
\newblock Value at risk in risk management and analysis. vol. 1. c. alexander,
  editor, 1999.

\bibitem[Wong and Farooq(2019)]{wong2019combined}
Melvin Wong and Bilal Farooq.
\newblock A combined entropy and utility based generative model for large scale
  multiple discrete-continuous travel behaviour data.
\newblock \emph{arXiv preprint arXiv:1901.06415}, 2019.

\end{thebibliography}

\clearpage

\appendix
\section*{Appendix}
\section{Appendix} \label{sec:appendix}

\begin{figure*}
\centering
\subfigure{{\includegraphics[width=7.33cm]{figs/Nc3_dist/SP.png}}}
\subfigure{{\includegraphics[width=7.33cm]{figs/Nc3_dist/MMM.png}}}\\
\subfigure{{\includegraphics[width=7.33cm]{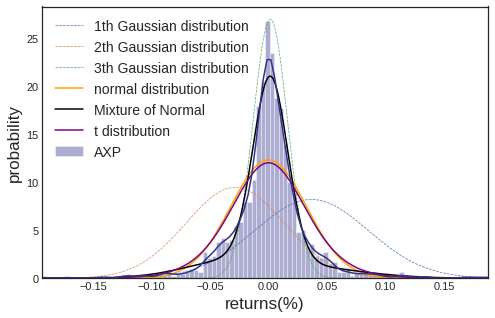}}}
\subfigure{{\includegraphics[width=7.33cm]{figs/Nc3_dist/AMD.png} }}\\
\subfigure{{\includegraphics[width=7.33cm]{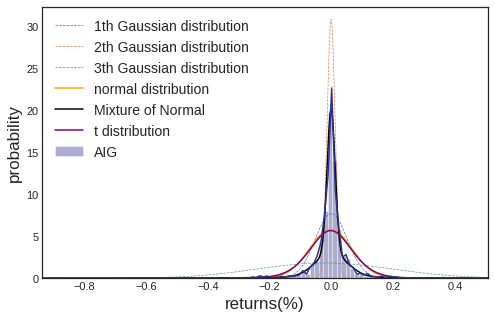} }}
\subfigure{{\includegraphics[width=7.33cm]{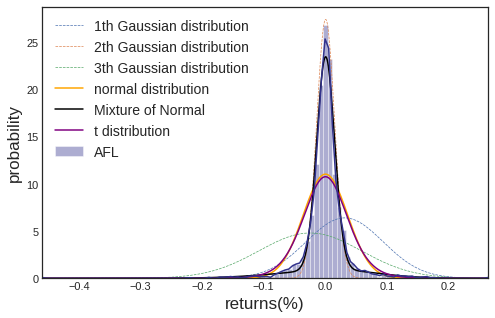} }}\\
\subfigure{{\includegraphics[width=7.33cm]{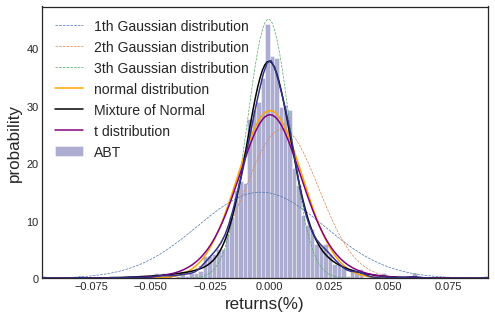}}}
\subfigure{{\includegraphics[width=7.33cm]{figs/Nc3_dist/MSFTF.png} }}\\

\caption{Mixture of Normal distribution for individual stocks during a crisis period}
\label{fig:Nc3_a}
\end{figure*}

\begin{figure*}
\centering
\subfigure{{\includegraphics[width=7.33cm]{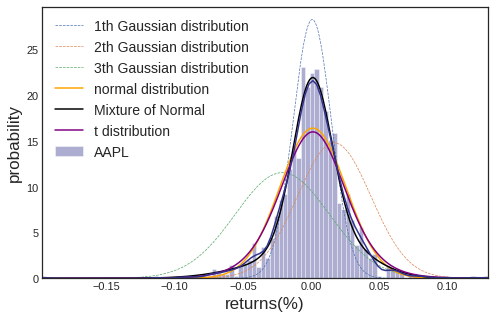}}}
\subfigure{{\includegraphics[width=7.33cm]{figs/Nc3_dist/AMZN.png}}}\\
\subfigure{{\includegraphics[width=7.33cm]{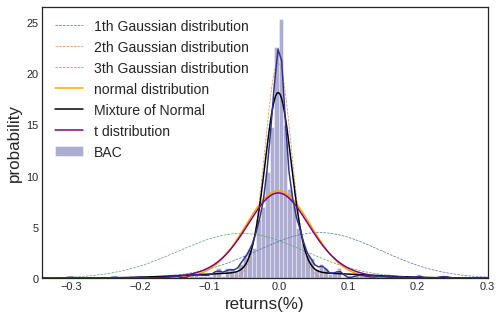}}}
\subfigure{{\includegraphics[width=7.33cm]{figs/Nc3_dist/JPM.png} }}\\
\subfigure{{\includegraphics[width=7.33cm]{figs/Nc3_dist/JNJ.png} }}
\subfigure{{\includegraphics[width=7.33cm]{figs/Nc3_dist/XOM.png} }}\\
\subfigure{{\includegraphics[width=7.33cm]{figs/Nc3_dist/MA.png}}}

\caption{Mixture of Normal distribution for individual stocks during a crisis period}
\label{fig:Nc3_b}
\end{figure*}

\begin{figure*}
\centering
\subfigure{{\includegraphics[width=7.32cm]{figs/corr_sp/MMM.png}}}
\subfigure{{\includegraphics[width=7.32cm]{figs/corr_sp/AXP.png}}}\\
\subfigure{{\includegraphics[width=7.32cm]{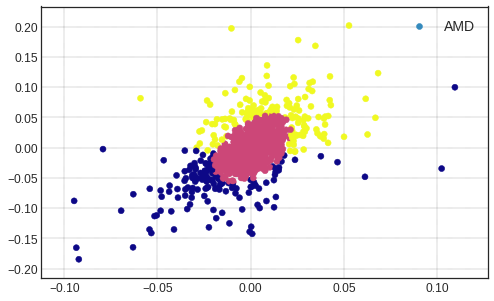}}}
\subfigure{{\includegraphics[width=7.32cm]{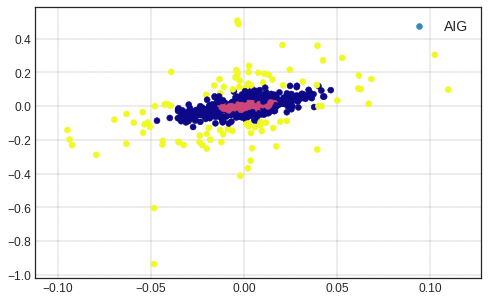} }}\\
\subfigure{{\includegraphics[width=7.32cm]{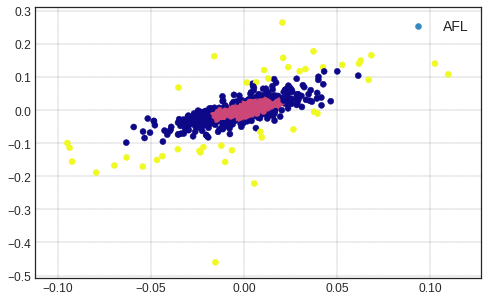} }}
\subfigure{{\includegraphics[width=7.32cm]{figs/corr_sp/ABT.png} }}\\
\subfigure{{\includegraphics[width=7.32cm]{figs/corr_sp/MSFTF.png}}}
\subfigure{{\includegraphics[width=7.32cm]{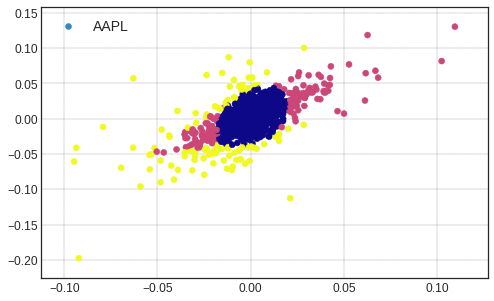} }}\\

\caption{Clustering stock returns against S\&P500 with Gmm $N_c = 3$}
\label{fig:corr_sp_a}
\end{figure*}

\begin{figure*}
\centering
\subfigure{{\includegraphics[width=7.32cm]{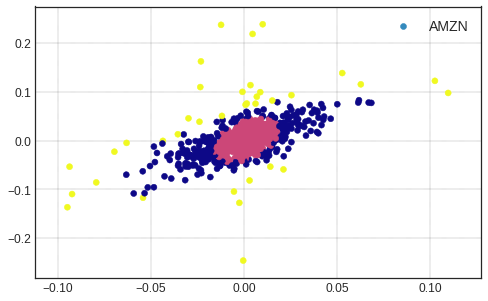}}}
\subfigure{{\includegraphics[width=7.32cm]{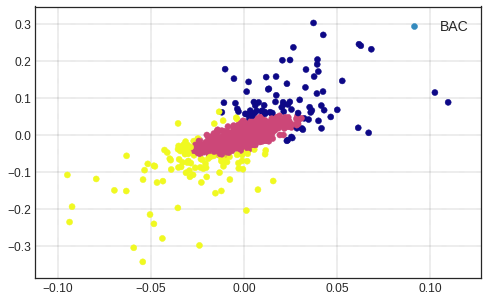}}}\\
\subfigure{{\includegraphics[width=7.32cm]{figs/corr_sp/JPM.png}}}
\subfigure{{\includegraphics[width=7.32cm]{figs/corr_sp/JNJ.png} }}\\
\subfigure{{\includegraphics[width=7.32cm]{figs/corr_sp/XOM.png} }}
\subfigure{{\includegraphics[width=7.32cm]{figs/corr_sp/MA.png} }}\\

\caption{Clustering stock returns against S\&P500 with Gmm $N_c = 3$}
\label{fig:corr_sp_b}
\end{figure*}

\begin{figure*}
\centering
\subfigure{{\includegraphics[width=7.31cm]{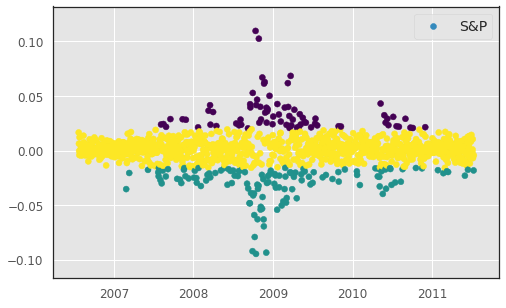}}}
\subfigure{{\includegraphics[width=7.31cm]{figs/clustering/MMM.png}}}\\
\subfigure{{\includegraphics[width=7.31cm]{figs/clustering/AXP.png}}}
\subfigure{{\includegraphics[width=7.31cm]{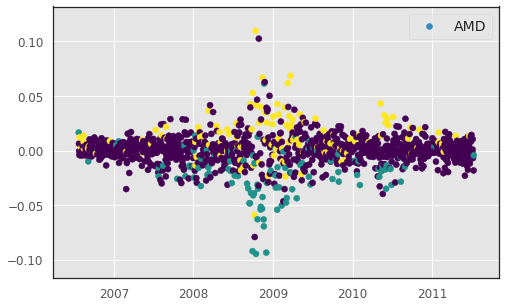} }}\\
\subfigure{{\includegraphics[width=7.31cm]{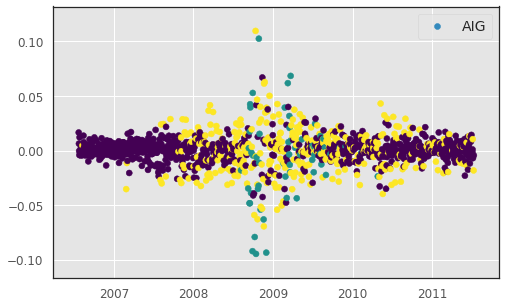} }}
\subfigure{{\includegraphics[width=7.31cm]{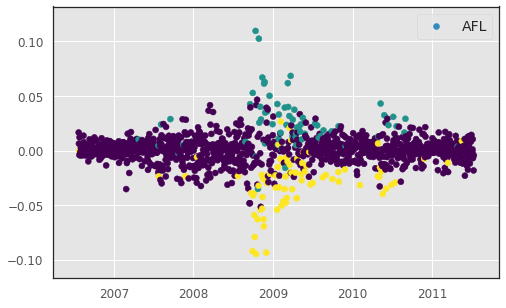} }}\\
\subfigure{{\includegraphics[width=7.31cm]{figs/clustering/ABT.png}}}
\subfigure{{\includegraphics[width=7.31cm]{figs/clustering/MSFTF.png} }}\\

\caption{Clustering stock returns during the time with $N_c = 3$}
\label{fig:clustering_a}
\end{figure*}

\begin{figure*}
\centering
\subfigure{{\includegraphics[width=7.31cm]{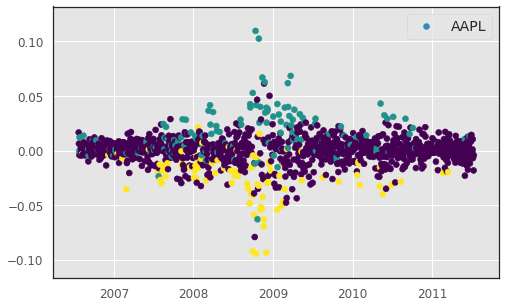}}}
\subfigure{{\includegraphics[width=7.31cm]{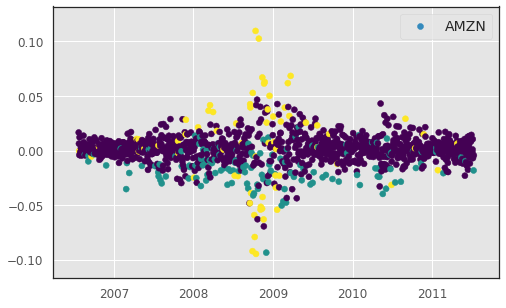} }}\\
\subfigure{{\includegraphics[width=7.31cm]{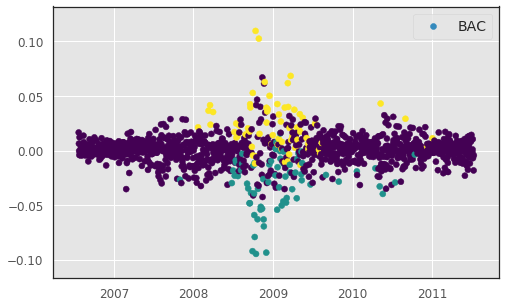}}}
\subfigure{{\includegraphics[width=7.31cm]{figs/clustering/JPM.png} }}\\
\subfigure{{\includegraphics[width=7.31cm]{figs/clustering/JNJ.png}}}
\subfigure{{\includegraphics[width=7.31cm]{figs/clustering/XOM.png} }}\\
\subfigure{{\includegraphics[width=7.31cm]{figs/clustering/MA.png}}}
\caption{Clustering stock returns during the time with $N_c = 3$}
\label{fig:clustering_b}
\end{figure*}

\begin{figure*}
\centering
\subfigure{{\includegraphics[width=7.31cm]{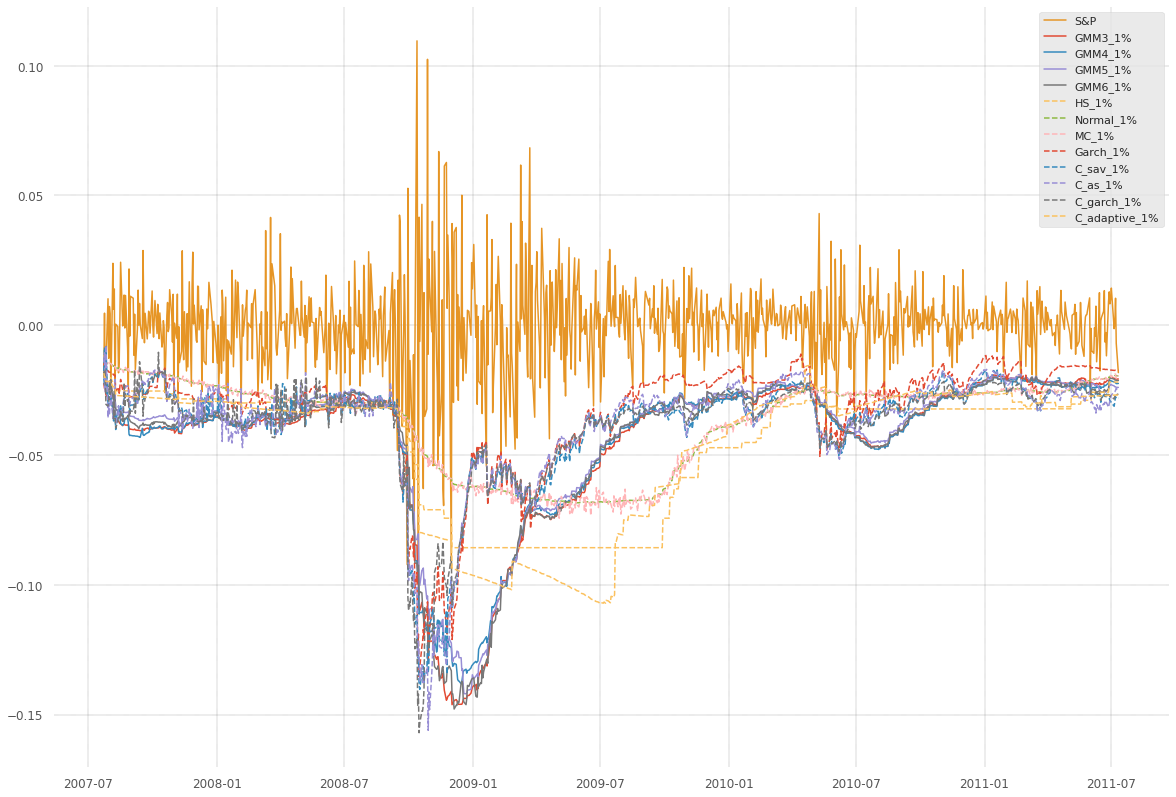}}}
\subfigure{{\includegraphics[width=7.31cm]{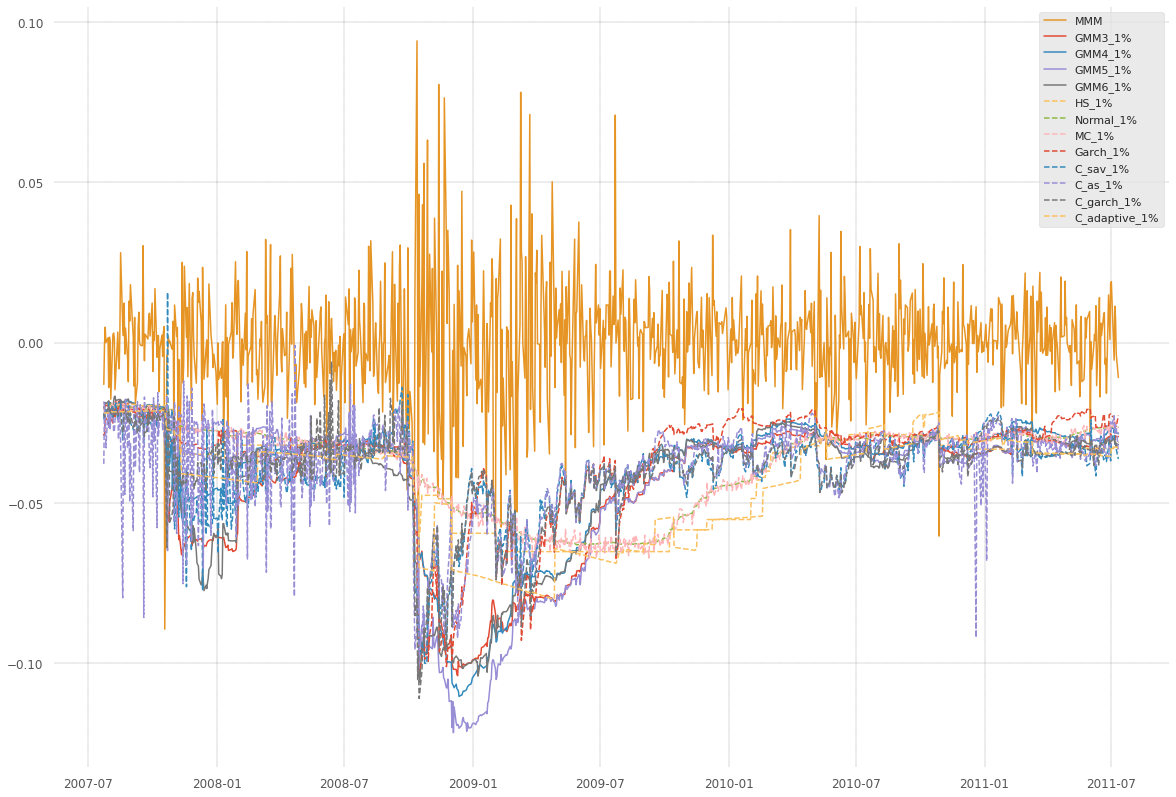}}}\\
\subfigure{{\includegraphics[width=7.31cm]{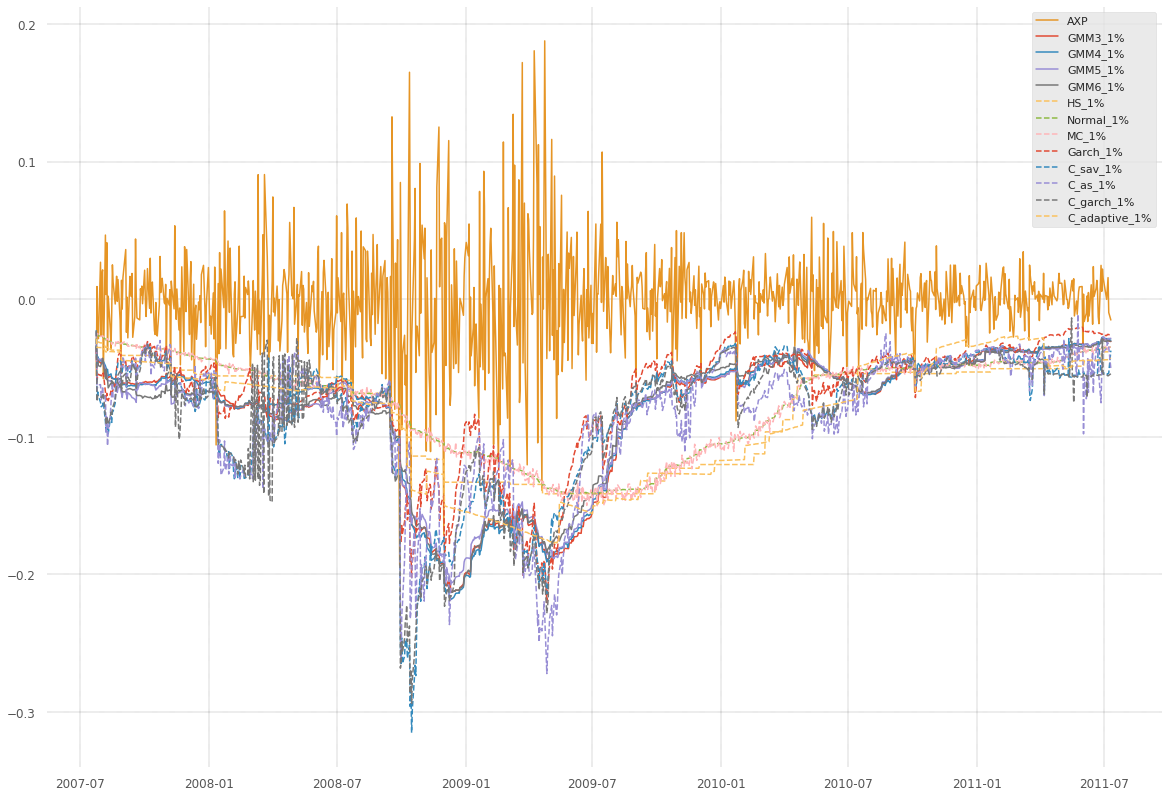}}}
\subfigure{{\includegraphics[width=7.31cm]{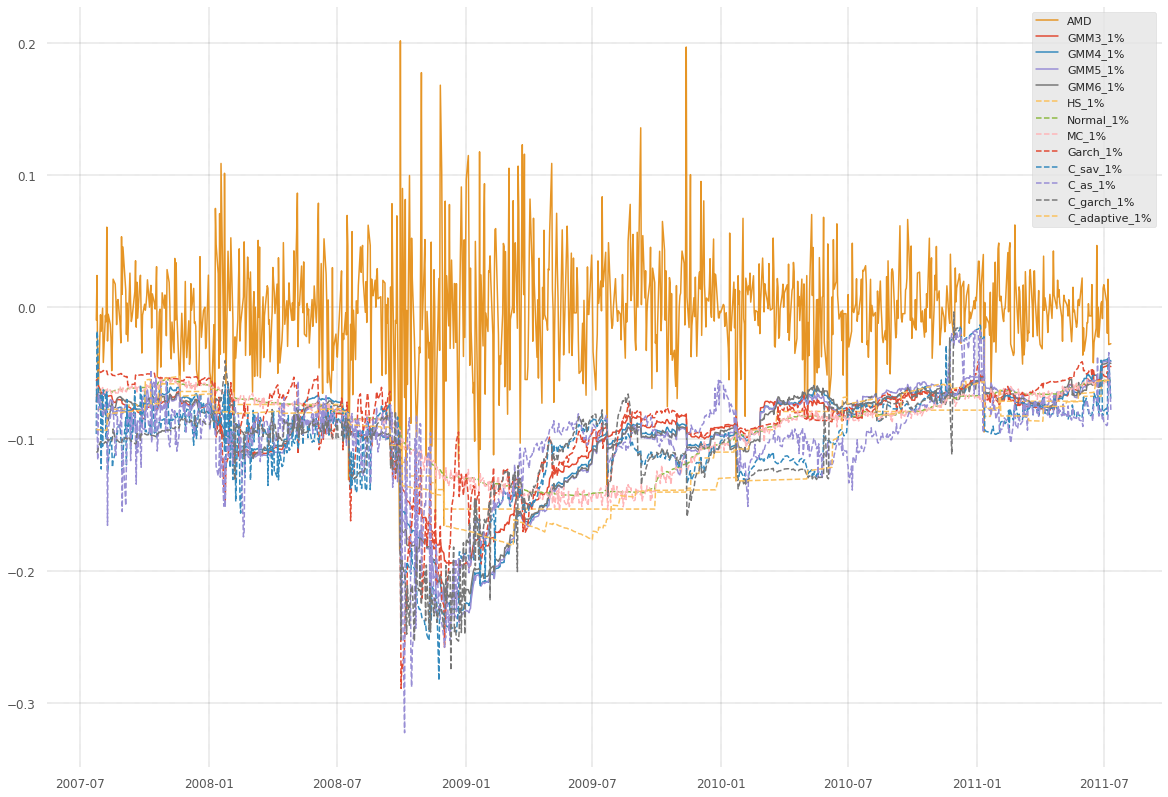}}}\\
\subfigure{{\includegraphics[width=7.31cm]{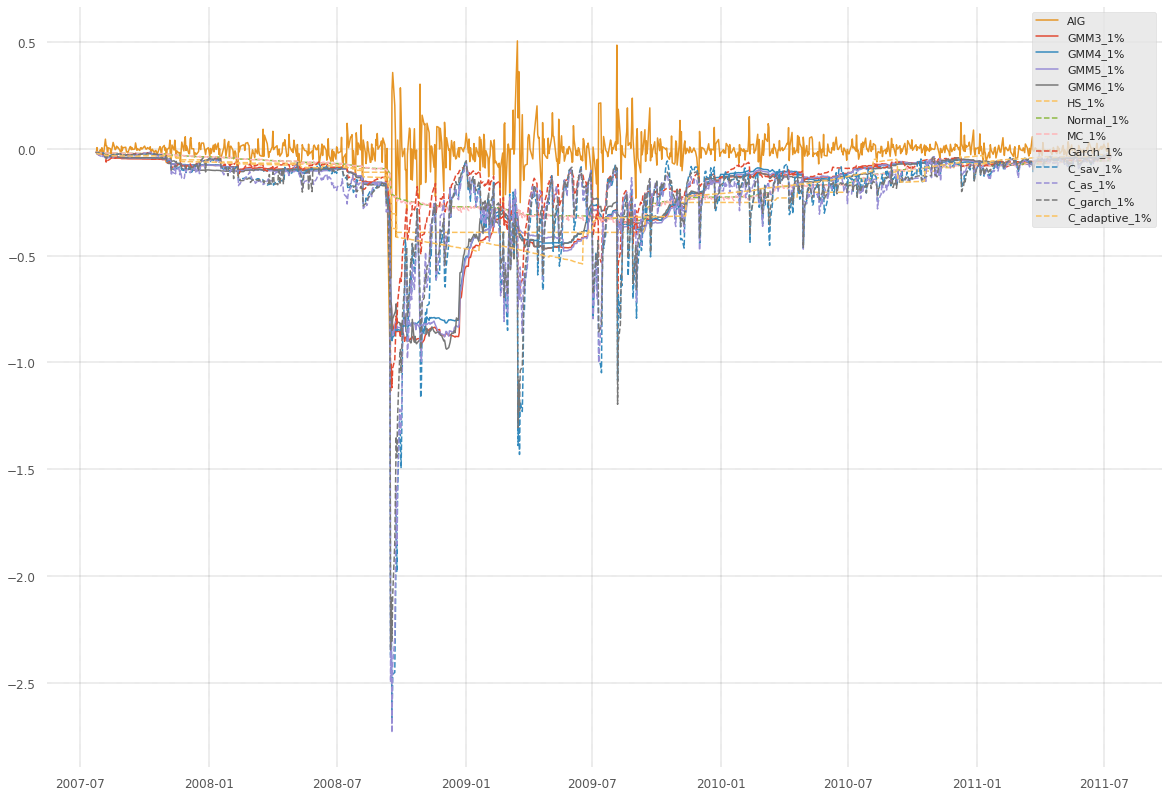}}}
\subfigure{{\includegraphics[width=7.31cm]{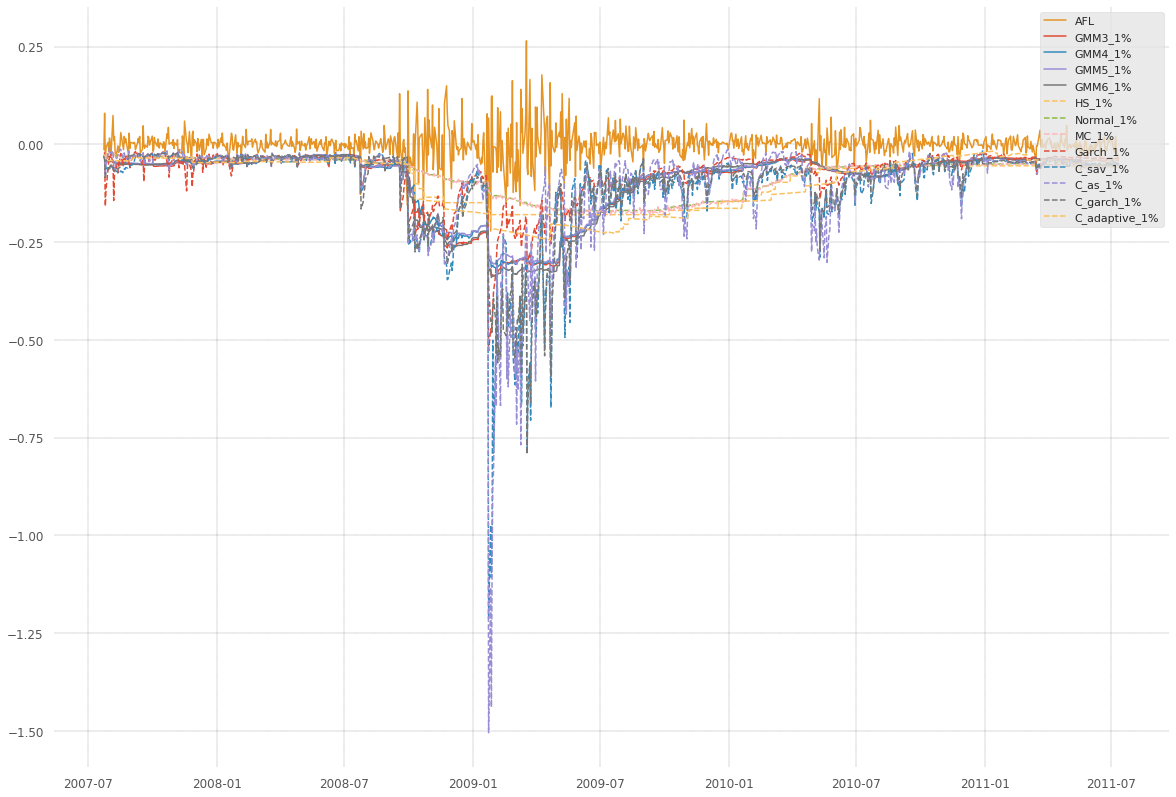}}}\\
\subfigure{{\includegraphics[width=7.31cm]{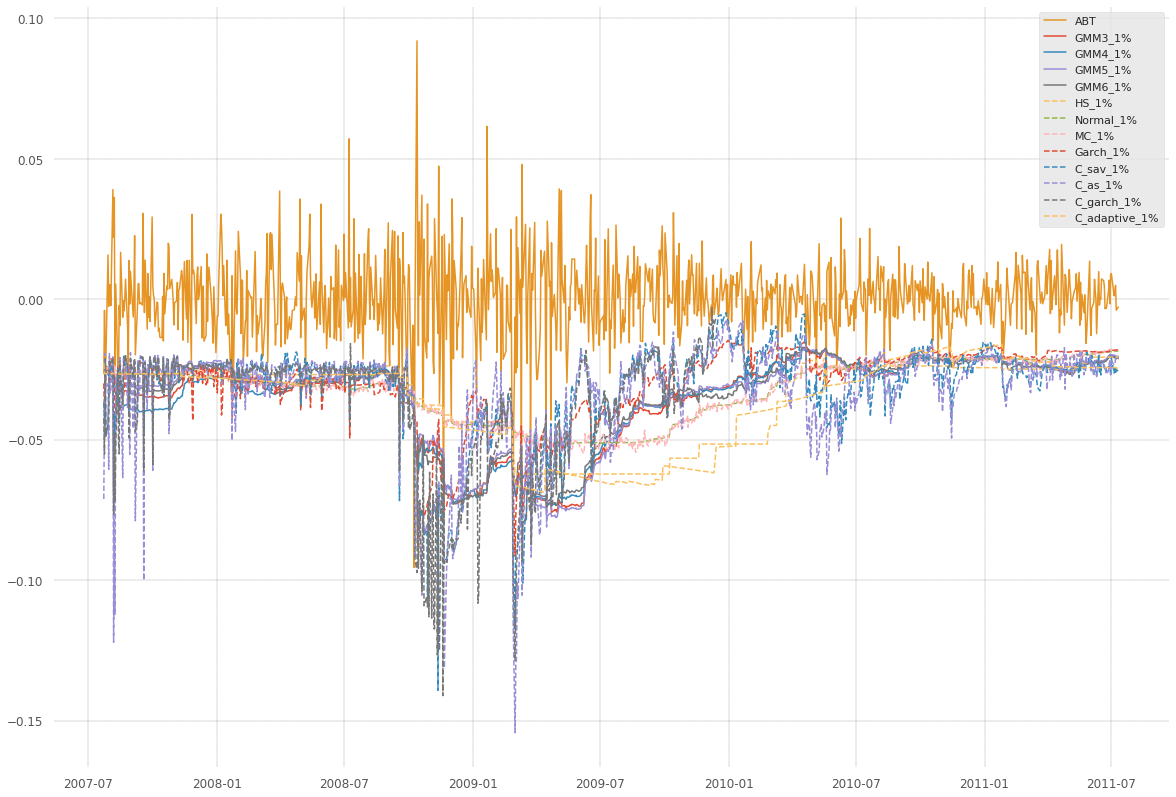}}}
\subfigure{{\includegraphics[width=7.31cm]{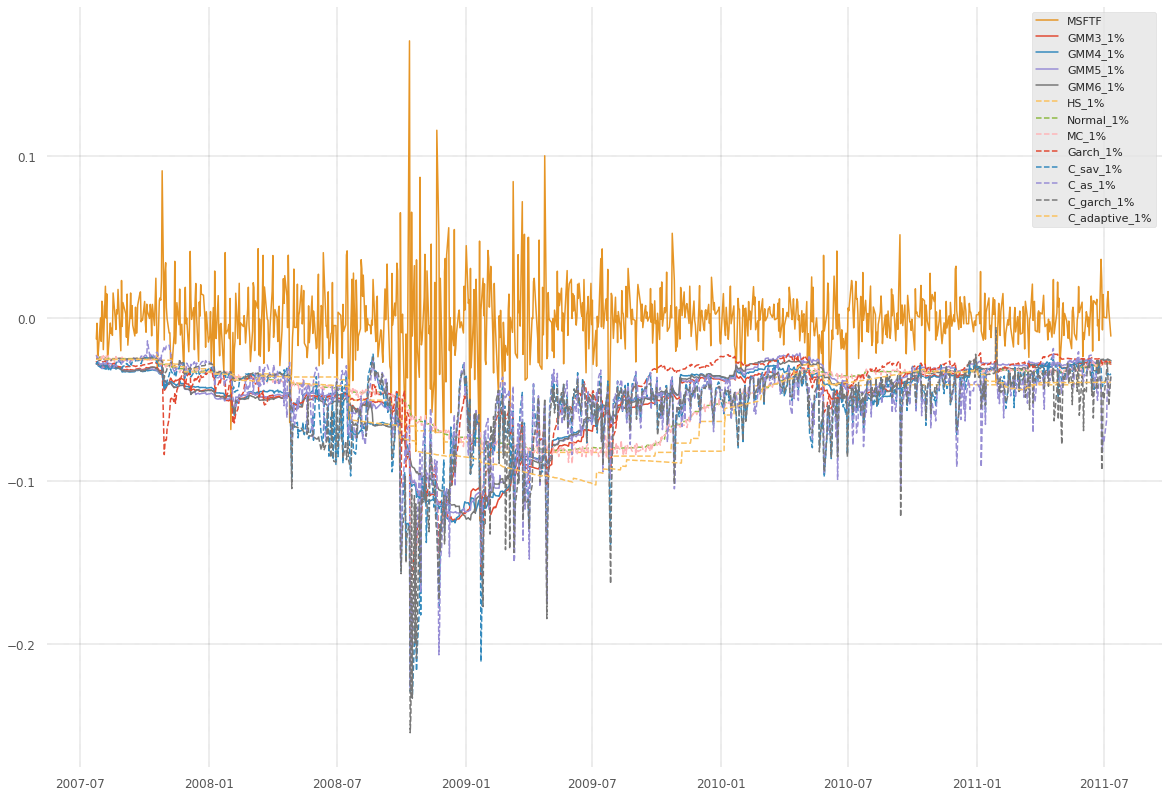}}}\\
\caption{Value at Risk 1 \% for individual stocks during a crisis period}
\label{fig:VAR1_a}
\end{figure*}

\begin{figure*}
\centering
\subfigure{{\includegraphics[width=7.31cm]{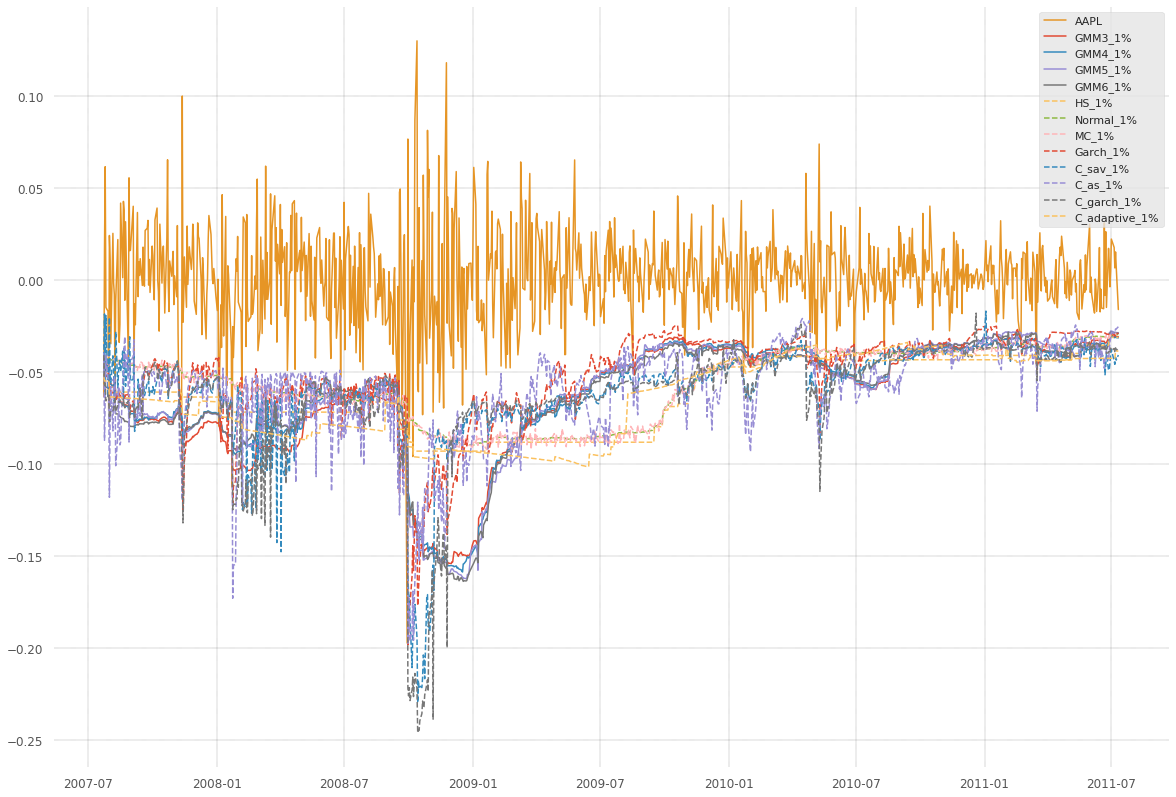}}}
\subfigure{{\includegraphics[width=7.31cm]{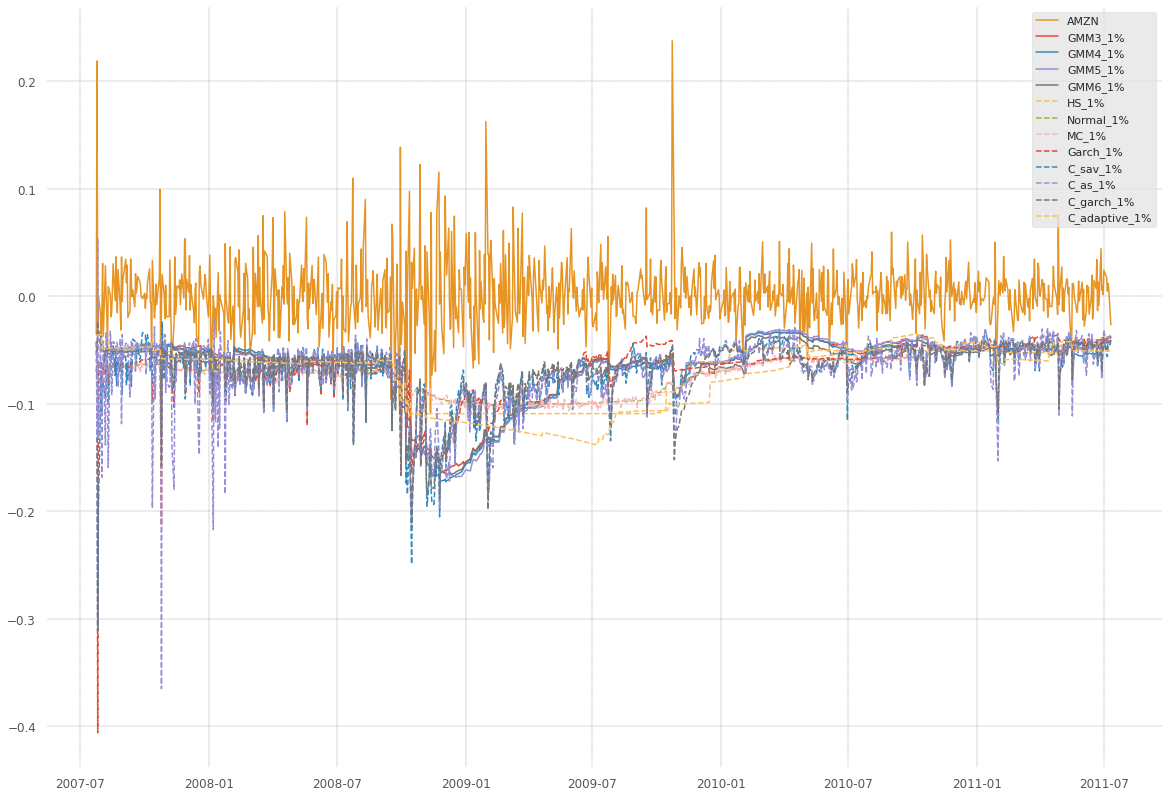}}}\\
\subfigure{{\includegraphics[width=7.31cm]{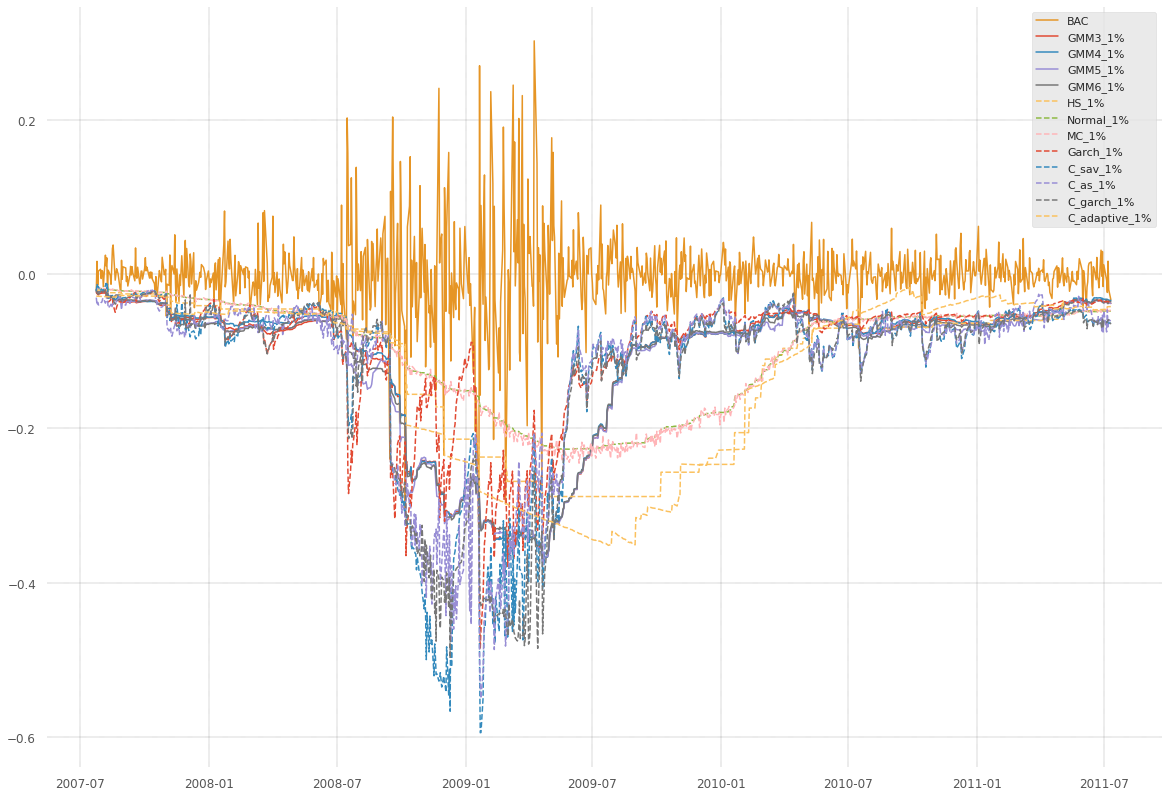}}}
\subfigure{{\includegraphics[width=7.31cm]{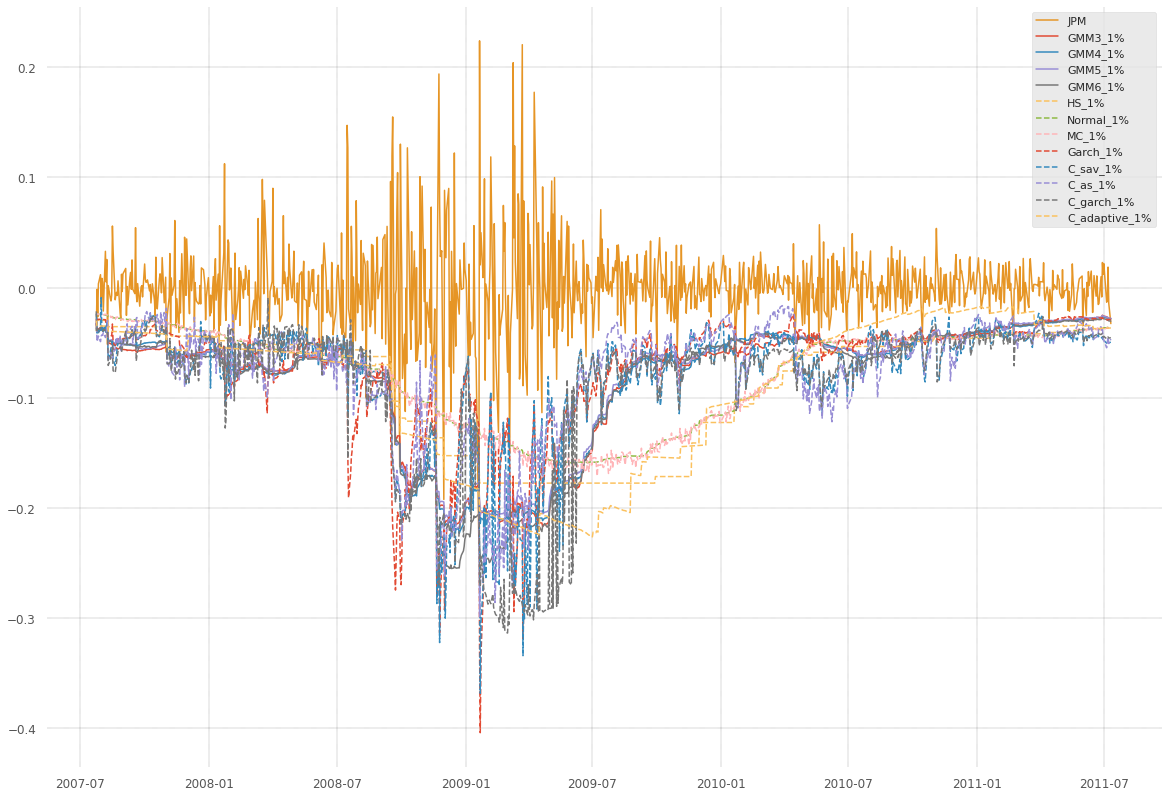}}}\\
\subfigure{{\includegraphics[width=7.31cm]{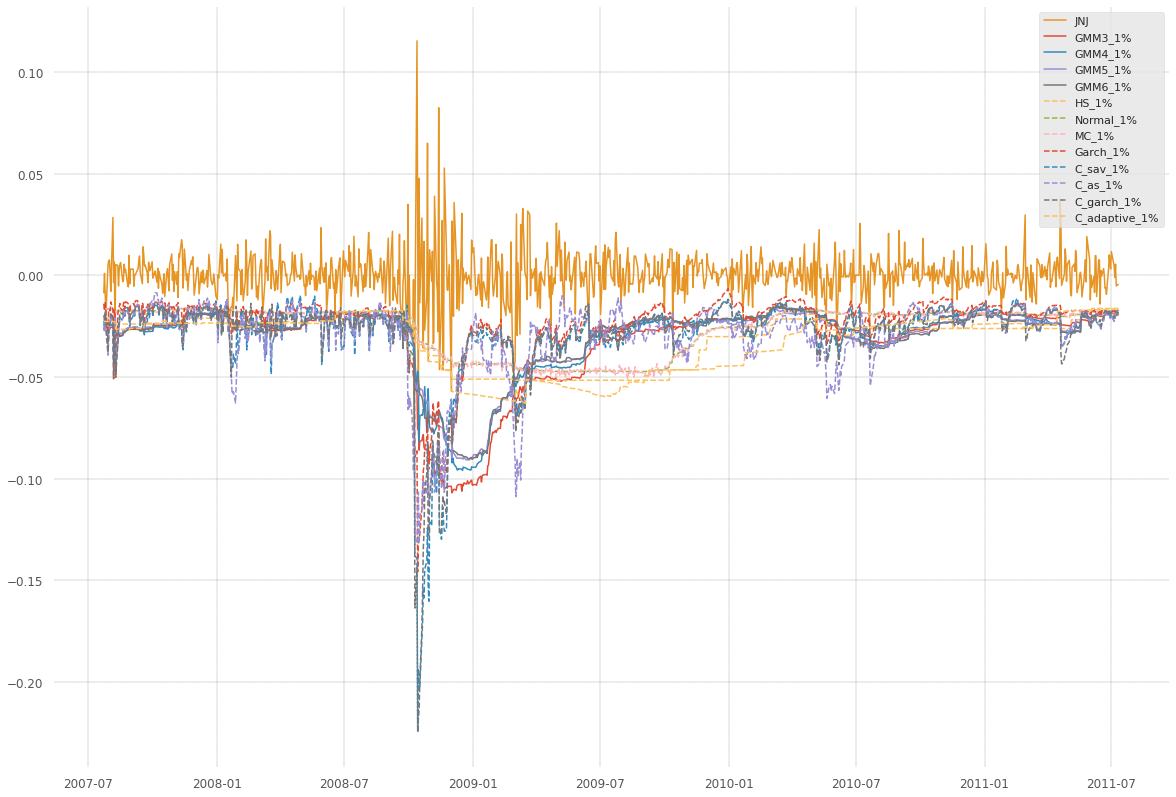}}}
\subfigure{{\includegraphics[width=7.31cm]{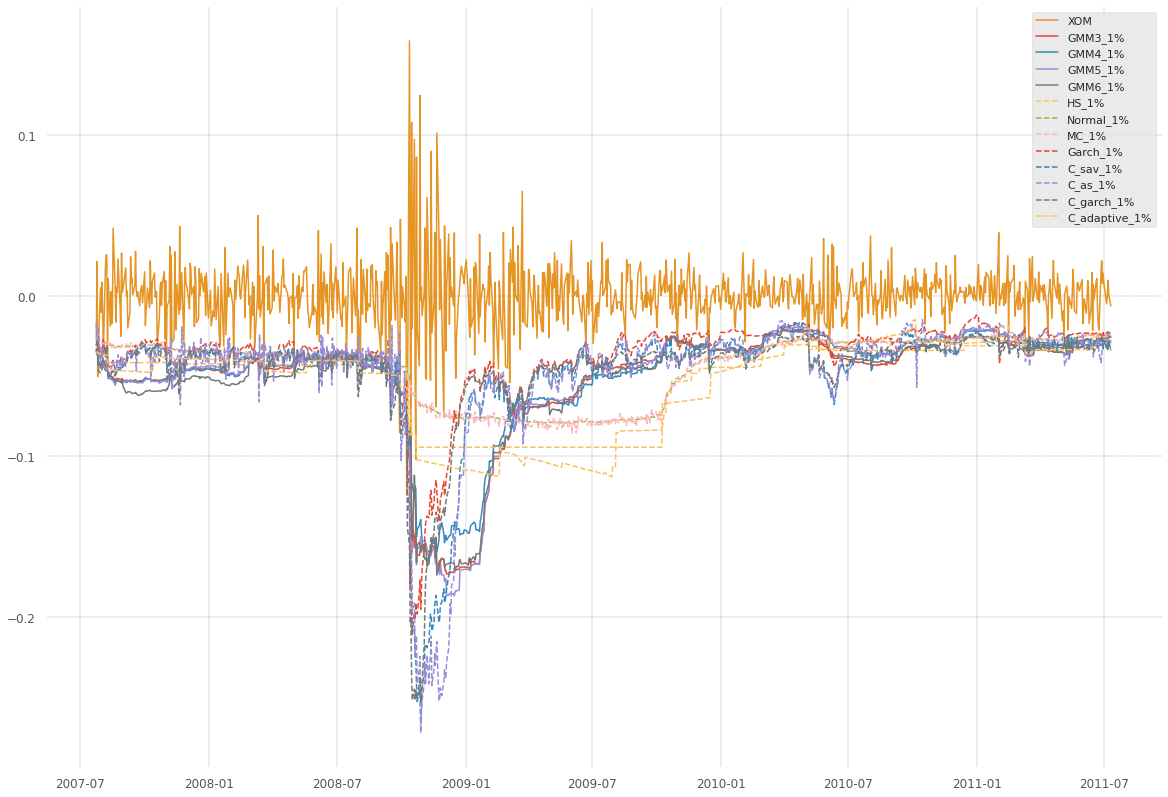}}}\\
\subfigure{{\includegraphics[width=7.31cm]{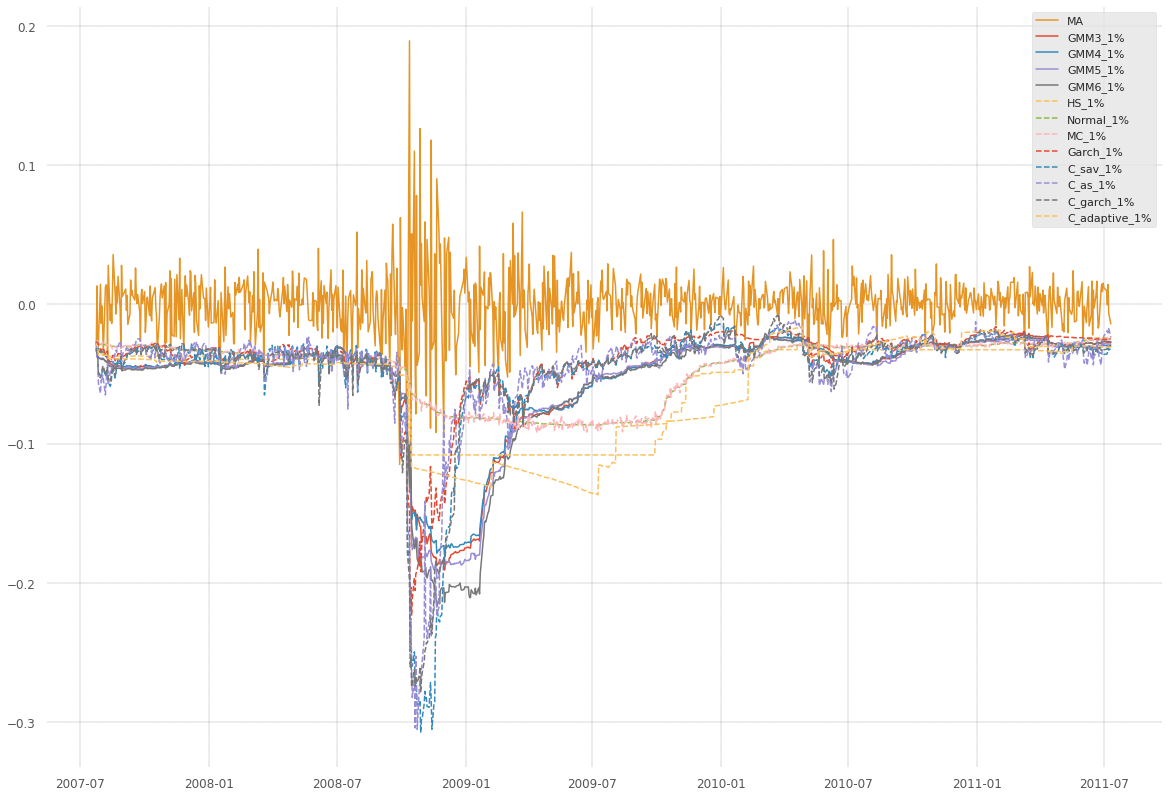}}}\\

\caption{Value at Risk 1 \% for individual stocks during a crisis period}
\label{fig:VAR1_b}
\end{figure*}

\begin{figure*}
\centering
\subfigure{{\includegraphics[width=7.31cm]{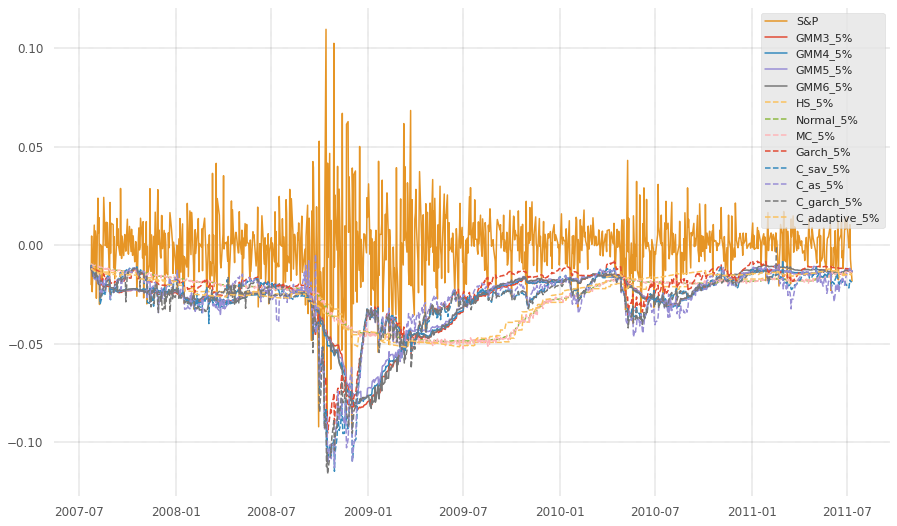}}}
\subfigure{{\includegraphics[width=7.31cm]{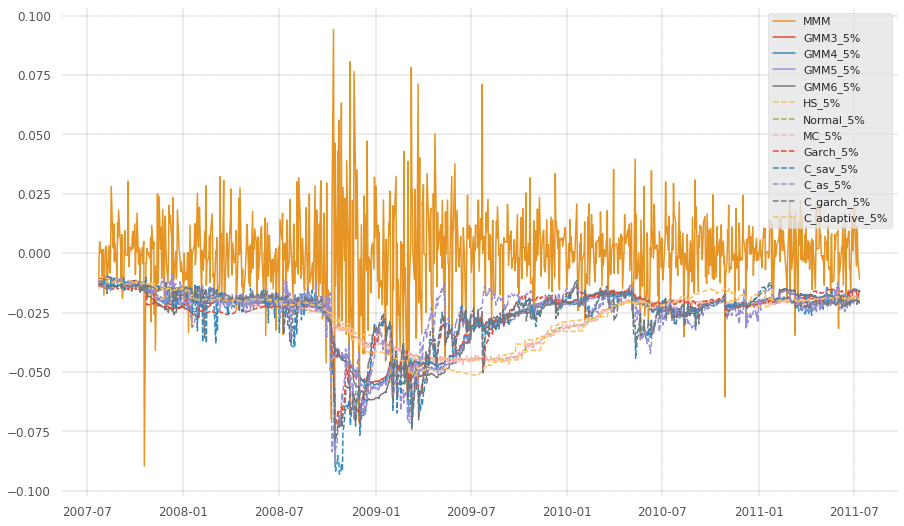}}}\\
\subfigure{{\includegraphics[width=7.31cm]{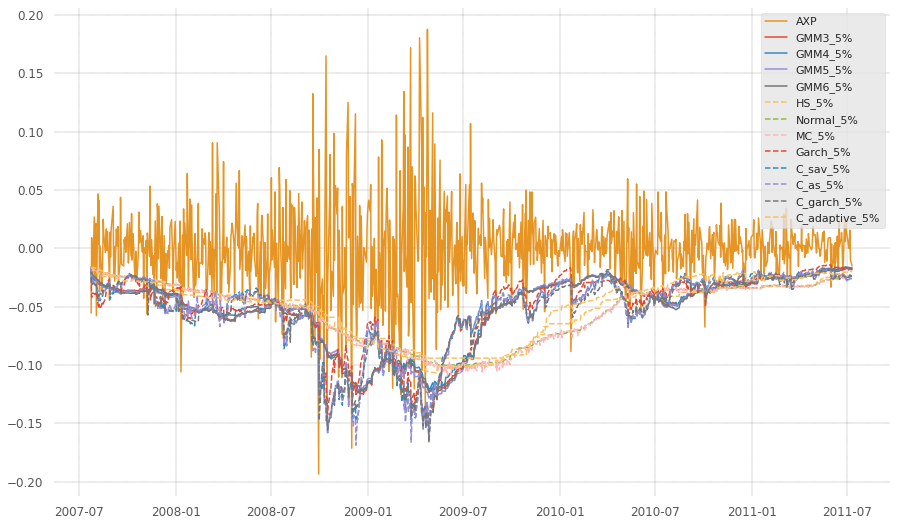}}}
\subfigure{{\includegraphics[width=7.31cm]{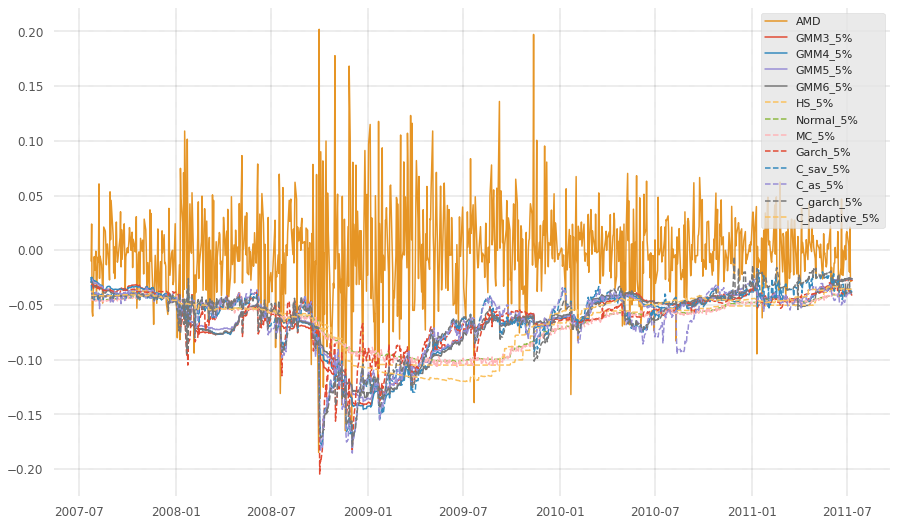}}}\\
\subfigure{{\includegraphics[width=7.31cm]{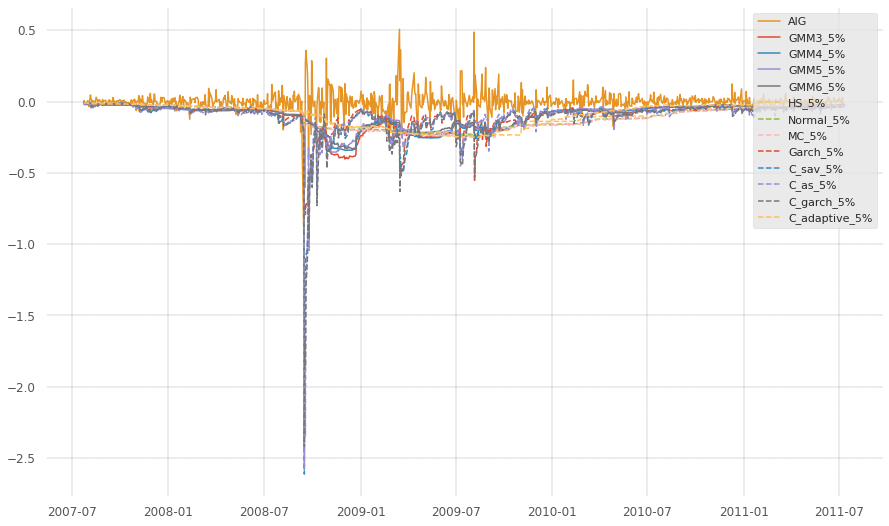}}}
\subfigure{{\includegraphics[width=7.31cm]{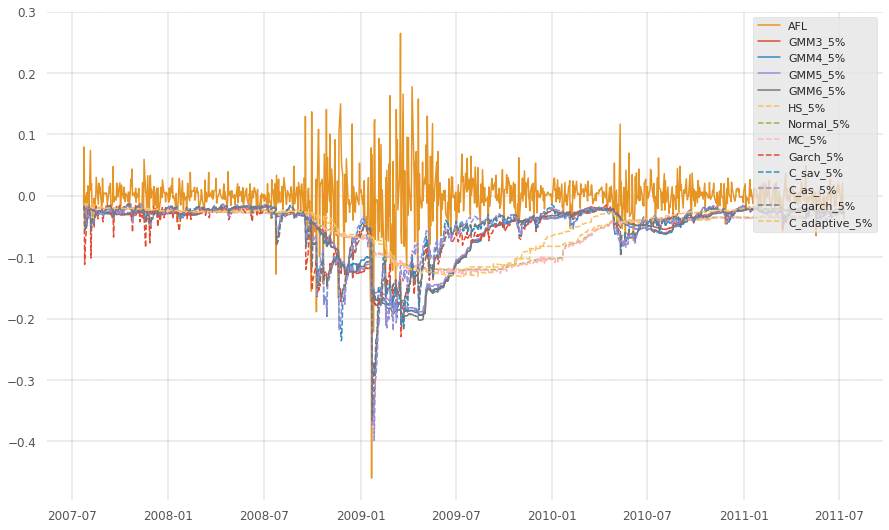}}}\\
\subfigure{{\includegraphics[width=7.31cm]{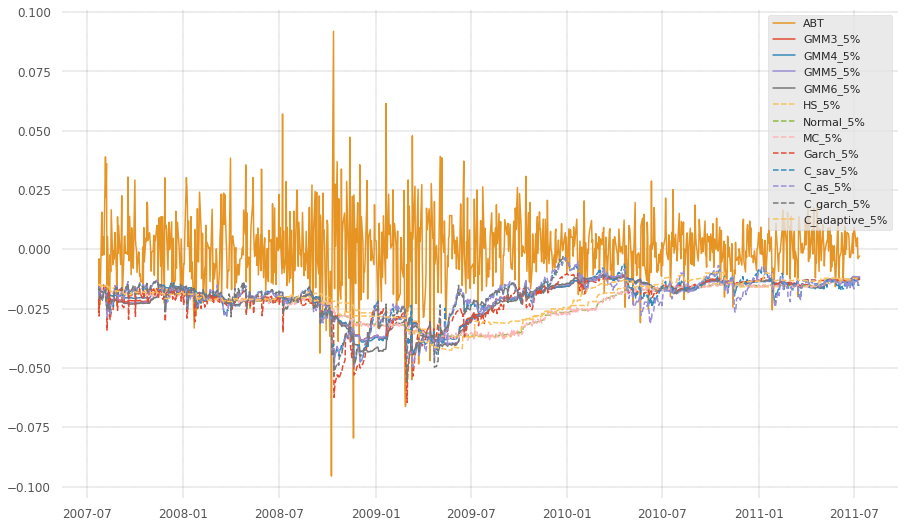}}}
\subfigure{{\includegraphics[width=7.31cm]{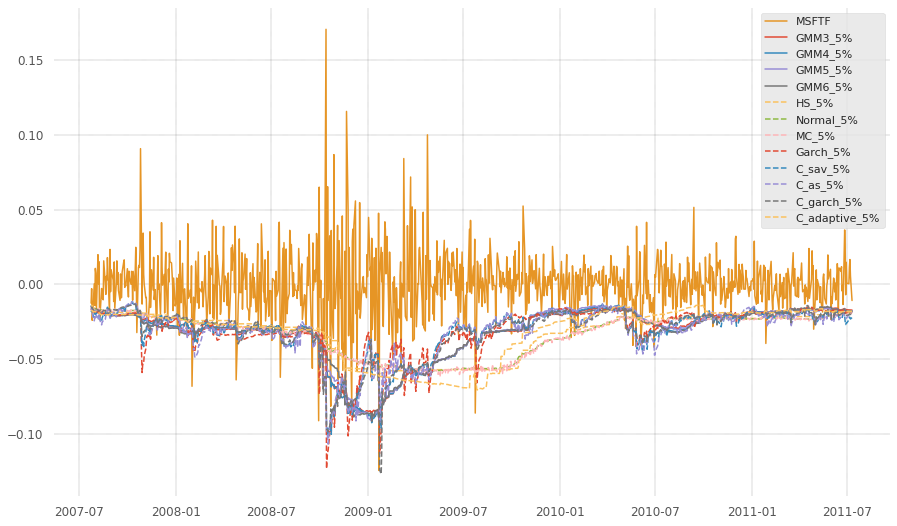}}}\\
\caption{Value at Risk 5 \% for individual stocks during a crisis period}
\label{fig:VAR5_a}
\end{figure*}

\begin{figure*}
\centering
\subfigure{{\includegraphics[width=7.31cm]{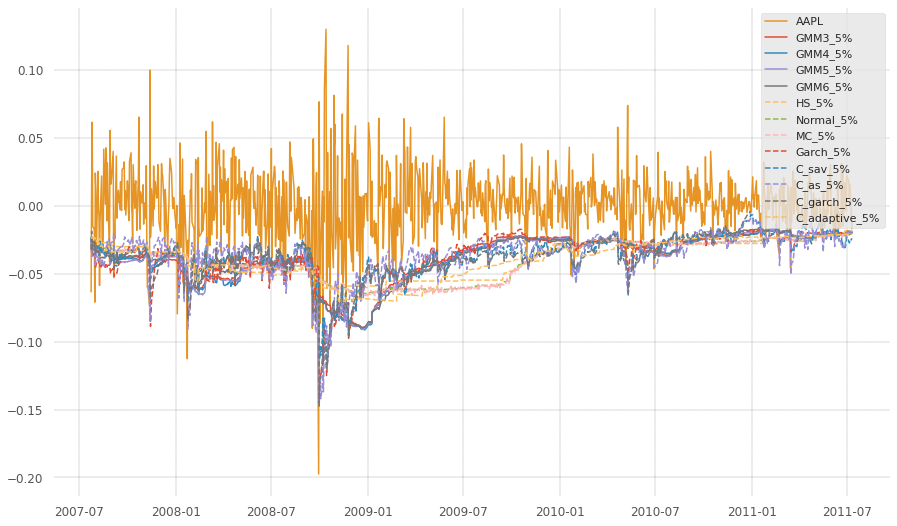}}}
\subfigure{{\includegraphics[width=7.31cm]{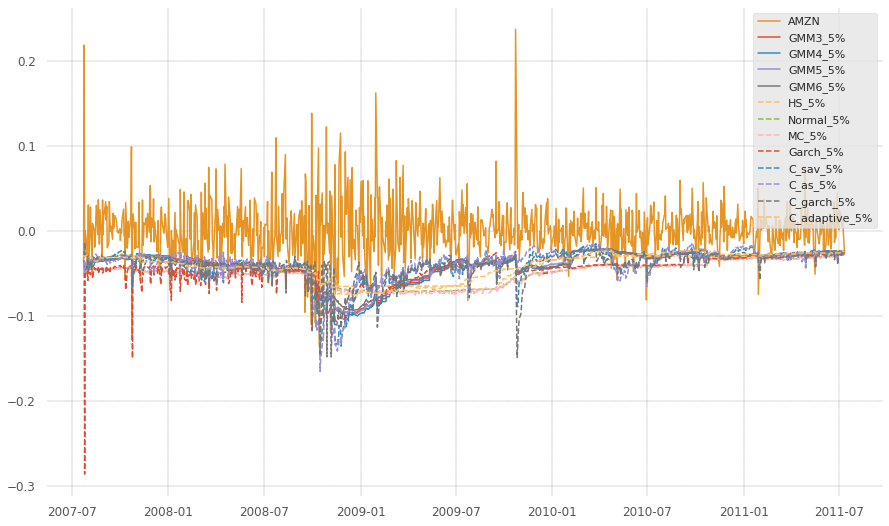}}}\\
\subfigure{{\includegraphics[width=7.31cm]{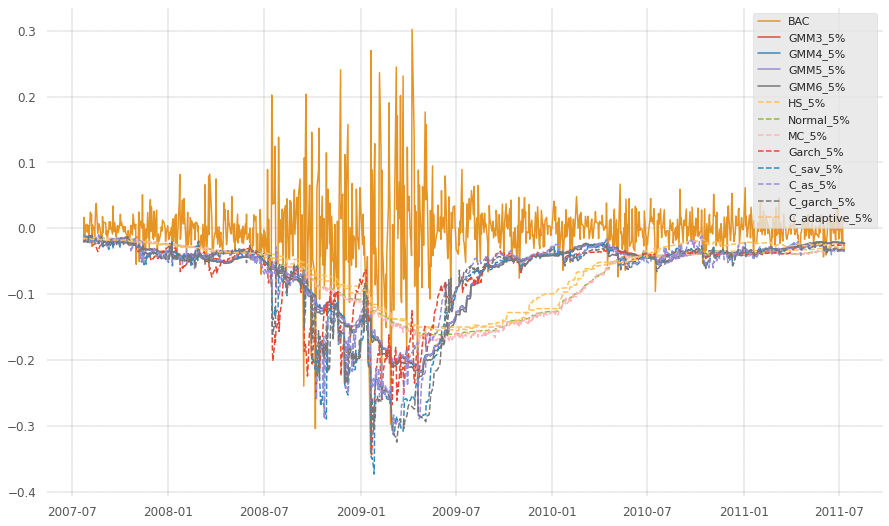}}}
\subfigure{{\includegraphics[width=7.31cm]{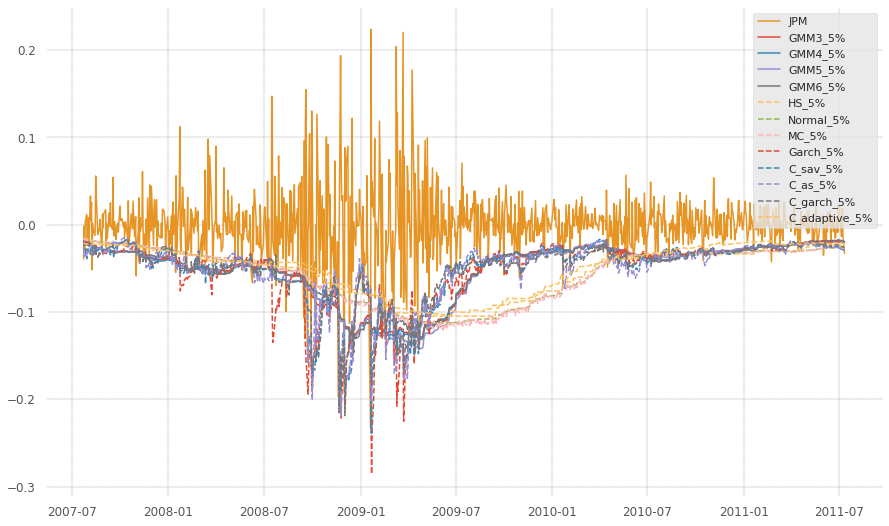}}}\\
\subfigure{{\includegraphics[width=7.31cm]{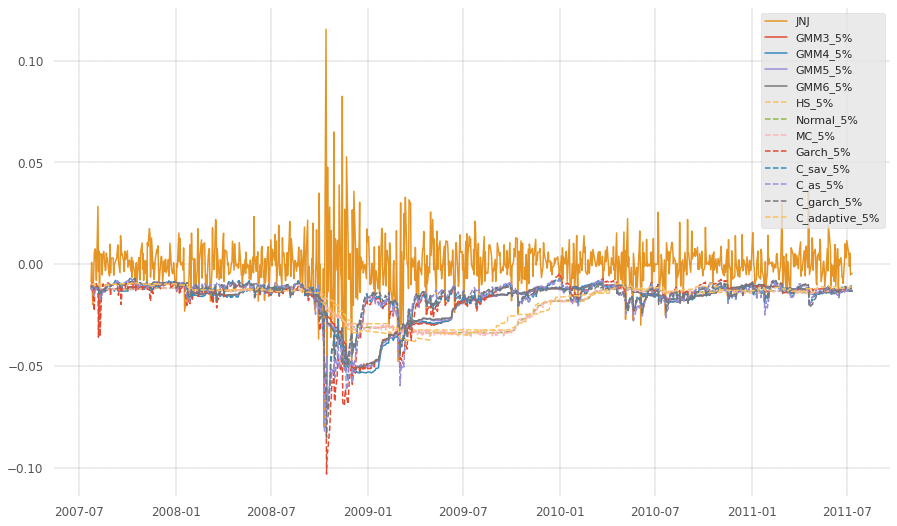}}}
\subfigure{{\includegraphics[width=7.31cm]{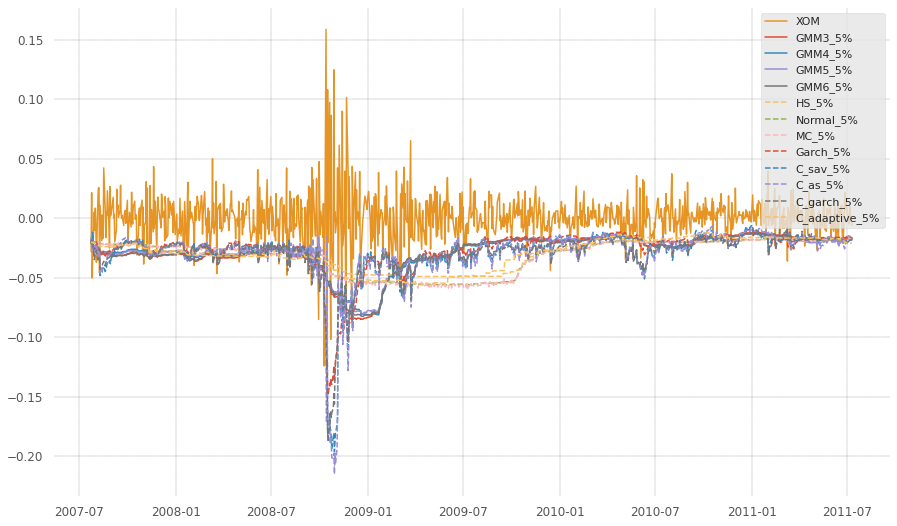}}}\\
\subfigure{{\includegraphics[width=7.31cm]{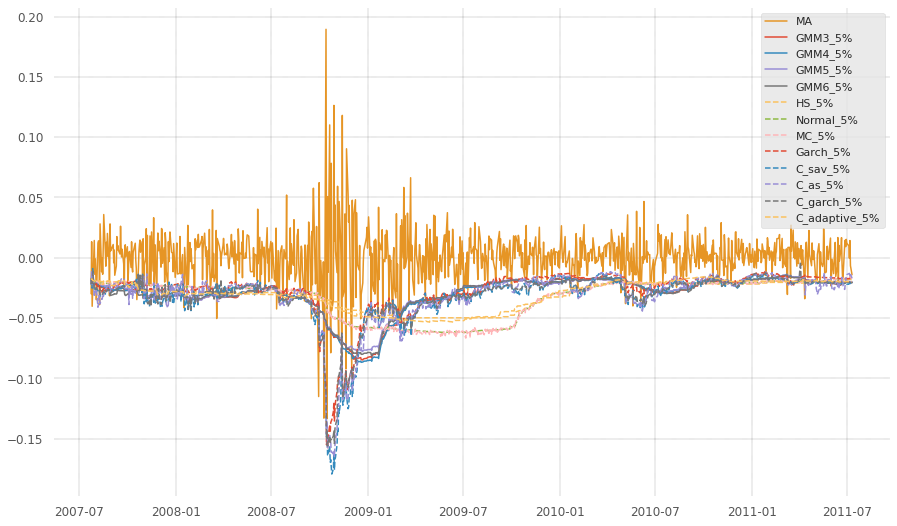}}}\\

\caption{Value at Risk 5 \% for individual stocks during a crisis period}
\label{fig:VAR5_b}
\end{figure*}

\begin{table}[h!]
% [width=1.0\linewidth,cols=16,pos=h]
  \centering
  \caption{Christofferson results for different $\sigma_{\text{short}}$ with $N_c = 3$ }
\resizebox{\textwidth}{!}{
    \begin{tabular}{lccccccccccccccc}
    % {\tblwidth}{@{}LCCCCCCCCCCCCCCC@{}}
\toprule
      & \multicolumn{15}{c}{\boldmath{}\textbf{$N_c= 3$}\unboldmath{}} \\
      & \multicolumn{15}{c}{\textbf{VaR 1 \%}} \\
      & \textcolor[rgb]{ .129,  .129,  .129}{\textbf{S\&P500}} & \textcolor[rgb]{ .129,  .129,  .129}{\textbf{MMM}} & \textcolor[rgb]{ .129,  .129,  .129}{\textbf{ARP}} & \textcolor[rgb]{ .129,  .129,  .129}{\textbf{AMD}} & \textcolor[rgb]{ .129,  .129,  .129}{\textbf{AIG}} & \textcolor[rgb]{ .129,  .129,  .129}{\textbf{AFL}} & \textcolor[rgb]{ .129,  .129,  .129}{\textbf{ABT}} & \textcolor[rgb]{ .129,  .129,  .129}{\textbf{MSFTF}} & \textcolor[rgb]{ .129,  .129,  .129}{\textbf{AAPL}} & \textcolor[rgb]{ .129,  .129,  .129}{\textbf{AMZN}} & \textcolor[rgb]{ .129,  .129,  .129}{\textbf{BAC}} & \textcolor[rgb]{ .129,  .129,  .129}{\textbf{JPM}} & \textcolor[rgb]{ .129,  .129,  .129}{\textbf{JNJ}} & \textcolor[rgb]{ .129,  .129,  .129}{\textbf{ROM}} & \textcolor[rgb]{ .129,  .129,  .129}{\textbf{MA}} \\
\textcolor[rgb]{ .129,  .129,  .129}{\textbf{10}} & \textcolor[rgb]{ 1,  0,  0}{R} & \textcolor[rgb]{ 1,  0,  0}{R} & \textcolor[rgb]{ 1,  0,  0}{R} & \textcolor[rgb]{ 1,  0,  0}{R} & \textcolor[rgb]{ 1,  0,  0}{R} & \textcolor[rgb]{ 1,  0,  0}{R} & \textcolor[rgb]{ 1,  0,  0}{R} & \textcolor[rgb]{ 1,  0,  0}{R} & \textcolor[rgb]{ 1,  0,  0}{R} & \textcolor[rgb]{ 1,  0,  0}{R} & \textcolor[rgb]{ 1,  0,  0}{R} & \textcolor[rgb]{ 1,  0,  0}{R} & \textcolor[rgb]{ 0,  .502,  0}{NR} & \textcolor[rgb]{ 0,  .502,  0}{NR} & \textcolor[rgb]{ 1,  0,  0}{R} \\
\textcolor[rgb]{ .129,  .129,  .129}{\textbf{20}} & \textcolor[rgb]{ 1,  0,  0}{R} & \textcolor[rgb]{ 1,  0,  0}{R} & \textcolor[rgb]{ 0,  .502,  0}{NR} & \textcolor[rgb]{ 0,  .502,  0}{NR} & \textcolor[rgb]{ 1,  0,  0}{R} & \textcolor[rgb]{ 0,  .502,  0}{NR} & \textcolor[rgb]{ 1,  0,  0}{R} & \textcolor[rgb]{ 1,  0,  0}{R} & \textcolor[rgb]{ 0,  .502,  0}{NR} & \textcolor[rgb]{ 0,  .502,  0}{NR} & \textcolor[rgb]{ 1,  0,  0}{R} & \textcolor[rgb]{ 0,  .502,  0}{NR} & \textcolor[rgb]{ 0,  .502,  0}{NR} & \textcolor[rgb]{ 0,  .502,  0}{NR} & \textcolor[rgb]{ 0,  .502,  0}{NR} \\
\textcolor[rgb]{ .129,  .129,  .129}{\textbf{30}} & \textcolor[rgb]{ 1,  0,  0}{R} & \textcolor[rgb]{ 0,  .502,  0}{NR} & \textcolor[rgb]{ 1,  0,  0}{R} & \textcolor[rgb]{ 0,  .502,  0}{NR} & \textcolor[rgb]{ 1,  0,  0}{R} & \textcolor[rgb]{ 0,  .502,  0}{NR} & \textcolor[rgb]{ 0,  .502,  0}{NR} & \textcolor[rgb]{ 0,  .502,  0}{NR} & \textcolor[rgb]{ 0,  .502,  0}{NR} & \textcolor[rgb]{ 0,  .502,  0}{NR} & \textcolor[rgb]{ 1,  0,  0}{R} & \textcolor[rgb]{ 1,  0,  0}{R} & \textcolor[rgb]{ 0,  .502,  0}{NR} & \textcolor[rgb]{ 0,  .502,  0}{NR} & \textcolor[rgb]{ 0,  .502,  0}{NR} \\
\textcolor[rgb]{ .129,  .129,  .129}{\textbf{40}} & \textcolor[rgb]{ 0,  .502,  0}{NR} & \textcolor[rgb]{ 0,  .502,  0}{NR} & \textcolor[rgb]{ 0,  .502,  0}{NR} & \textcolor[rgb]{ 0,  .502,  0}{NR} & \textcolor[rgb]{ 0,  .502,  0}{NR} & \textcolor[rgb]{ 0,  .502,  0}{NR} & \textcolor[rgb]{ 0,  .502,  0}{NR} & \textcolor[rgb]{ 0,  .502,  0}{NR} & \textcolor[rgb]{ 1,  0,  0}{R} & \textcolor[rgb]{ 1,  0,  0}{R} & \textcolor[rgb]{ 1,  0,  0}{R} & \textcolor[rgb]{ 0,  .502,  0}{NR} & \textcolor[rgb]{ 0,  .502,  0}{NR} & \textcolor[rgb]{ 0,  .502,  0}{NR} & \textcolor[rgb]{ 0,  .502,  0}{NR} \\
\textcolor[rgb]{ .129,  .129,  .129}{\textbf{50}} & \textcolor[rgb]{ 0,  .502,  0}{NR} & \textcolor[rgb]{ 0,  .502,  0}{NR} & \textcolor[rgb]{ 0,  .502,  0}{NR} & \textcolor[rgb]{ 0,  .502,  0}{NR} & \textcolor[rgb]{ 1,  0,  0}{R} & \textcolor[rgb]{ 0,  .502,  0}{NR} & \textcolor[rgb]{ 0,  .502,  0}{NR} & \textcolor[rgb]{ 0,  .502,  0}{NR} & \textcolor[rgb]{ 0,  .502,  0}{NR} & \textcolor[rgb]{ 1,  0,  0}{R} & \textcolor[rgb]{ 1,  0,  0}{R} & \textcolor[rgb]{ 0,  .502,  0}{NR} & \textcolor[rgb]{ 0,  .502,  0}{NR} & \textcolor[rgb]{ 0,  .502,  0}{NR} & \textcolor[rgb]{ 0,  .502,  0}{NR} \\
\textcolor[rgb]{ .129,  .129,  .129}{\textbf{60}} & \textcolor[rgb]{ 0,  .502,  0}{NR} & \textcolor[rgb]{ 0,  .502,  0}{NR} & \textcolor[rgb]{ 0,  .502,  0}{NR} & \textcolor[rgb]{ 0,  .502,  0}{NR} & \textcolor[rgb]{ 1,  0,  0}{R} & \textcolor[rgb]{ 0,  .502,  0}{NR} & \textcolor[rgb]{ 0,  .502,  0}{NR} & \textcolor[rgb]{ 0,  .502,  0}{NR} & \textcolor[rgb]{ 0,  .502,  0}{NR} & \textcolor[rgb]{ 0,  .502,  0}{NR} & \textcolor[rgb]{ 1,  0,  0}{R} & \textcolor[rgb]{ 0,  .502,  0}{NR} & \textcolor[rgb]{ 0,  .502,  0}{NR} & \textcolor[rgb]{ 0,  .502,  0}{NR} & \textcolor[rgb]{ 0,  .502,  0}{NR} \\
\textcolor[rgb]{ .129,  .129,  .129}{\textbf{70}} & \textcolor[rgb]{ 0,  .502,  0}{NR} & \textcolor[rgb]{ 0,  .502,  0}{NR} & \textcolor[rgb]{ 0,  .502,  0}{NR} & \textcolor[rgb]{ 0,  .502,  0}{NR} & \textcolor[rgb]{ 0,  .502,  0}{NR} & \textcolor[rgb]{ 0,  .502,  0}{NR} & \textcolor[rgb]{ 0,  .502,  0}{NR} & \textcolor[rgb]{ 0,  .502,  0}{NR} & \textcolor[rgb]{ 0,  .502,  0}{NR} & \textcolor[rgb]{ 0,  .502,  0}{NR} & \textcolor[rgb]{ 0,  .502,  0}{NR} & \textcolor[rgb]{ 0,  .502,  0}{NR} & \textcolor[rgb]{ 0,  .502,  0}{NR} & \textcolor[rgb]{ 0,  .502,  0}{NR} & \textcolor[rgb]{ 0,  .502,  0}{NR} \\
\textcolor[rgb]{ .129,  .129,  .129}{\textbf{80}} & \textcolor[rgb]{ 0,  .502,  0}{NR} & \textcolor[rgb]{ 0,  .502,  0}{NR} & \textcolor[rgb]{ 0,  .502,  0}{NR} & \textcolor[rgb]{ 1,  0,  0}{R} & \textcolor[rgb]{ 1,  0,  0}{R} & \textcolor[rgb]{ 0,  .502,  0}{NR} & \textcolor[rgb]{ 0,  .502,  0}{NR} & \textcolor[rgb]{ 0,  .502,  0}{NR} & \textcolor[rgb]{ 0,  .502,  0}{NR} & \textcolor[rgb]{ 0,  .502,  0}{NR} & \textcolor[rgb]{ 0,  .502,  0}{NR} & \textcolor[rgb]{ 0,  .502,  0}{NR} & \textcolor[rgb]{ 1,  0,  0}{R} & \textcolor[rgb]{ 0,  .502,  0}{NR} & \textcolor[rgb]{ 0,  .502,  0}{NR} \\
\textcolor[rgb]{ .129,  .129,  .129}{\textbf{90}} & \textcolor[rgb]{ 0,  .502,  0}{NR} & \textcolor[rgb]{ 0,  .502,  0}{NR} & \textcolor[rgb]{ 0,  .502,  0}{NR} & \textcolor[rgb]{ 0,  .502,  0}{NR} & \textcolor[rgb]{ 1,  0,  0}{R} & \textcolor[rgb]{ 0,  .502,  0}{NR} & \textcolor[rgb]{ 0,  .502,  0}{NR} & \textcolor[rgb]{ 0,  .502,  0}{NR} & \textcolor[rgb]{ 0,  .502,  0}{NR} & \textcolor[rgb]{ 0,  .502,  0}{NR} & \textcolor[rgb]{ 0,  .502,  0}{NR} & \textcolor[rgb]{ 0,  .502,  0}{NR} & \textcolor[rgb]{ 0,  .502,  0}{NR} & \textcolor[rgb]{ 0,  .502,  0}{NR} & \textcolor[rgb]{ 0,  .502,  0}{NR} \\
      &  \multicolumn{15}{c}{\textbf{VaR 5 \%}} \\
\textcolor[rgb]{ .129,  .129,  .129}{} & \textcolor[rgb]{ .129,  .129,  .129}{\textbf{S\&P500}} & \textcolor[rgb]{ .129,  .129,  .129}{\textbf{MMM}} & \textcolor[rgb]{ .129,  .129,  .129}{\textbf{ARP}} & \textcolor[rgb]{ .129,  .129,  .129}{\textbf{AMD}} & \textcolor[rgb]{ .129,  .129,  .129}{\textbf{AIG}} & \textcolor[rgb]{ .129,  .129,  .129}{\textbf{AFL}} & \textcolor[rgb]{ .129,  .129,  .129}{\textbf{ABT}} & \textcolor[rgb]{ .129,  .129,  .129}{\textbf{MSFTF}} & \textcolor[rgb]{ .129,  .129,  .129}{\textbf{AAPL}} & \textcolor[rgb]{ .129,  .129,  .129}{\textbf{AMZN}} & \textcolor[rgb]{ .129,  .129,  .129}{\textbf{BAC}} & \textcolor[rgb]{ .129,  .129,  .129}{\textbf{JPM}} & \textcolor[rgb]{ .129,  .129,  .129}{\textbf{JNJ}} & \textcolor[rgb]{ .129,  .129,  .129}{\textbf{ROM}} & \textcolor[rgb]{ .129,  .129,  .129}{\textbf{MA}} \\
\textcolor[rgb]{ .129,  .129,  .129}{\textbf{10}} & \textcolor[rgb]{ 0,  .502,  0}{NR} & \textcolor[rgb]{ 1,  0,  0}{R} & \textcolor[rgb]{ 1,  0,  0}{R} & \textcolor[rgb]{ 1,  0,  0}{R} & \textcolor[rgb]{ 1,  0,  0}{R} & \textcolor[rgb]{ 1,  0,  0}{R} & \textcolor[rgb]{ 1,  0,  0}{R} & \textcolor[rgb]{ 1,  0,  0}{R} & \textcolor[rgb]{ 1,  0,  0}{R} & \textcolor[rgb]{ 1,  0,  0}{R} & \textcolor[rgb]{ 1,  0,  0}{R} & \textcolor[rgb]{ 1,  0,  0}{R} & \textcolor[rgb]{ 1,  0,  0}{R} & \textcolor[rgb]{ 1,  0,  0}{R} & \textcolor[rgb]{ 1,  0,  0}{R} \\
\textcolor[rgb]{ .129,  .129,  .129}{\textbf{20}} & \textcolor[rgb]{ 0,  .502,  0}{NR} & \textcolor[rgb]{ 0,  .502,  0}{NR} & \textcolor[rgb]{ 1,  0,  0}{R} & \textcolor[rgb]{ 1,  0,  0}{R} & \textcolor[rgb]{ 1,  0,  0}{R} & \textcolor[rgb]{ 0,  .502,  0}{NR} & \textcolor[rgb]{ 1,  0,  0}{R} & \textcolor[rgb]{ 0,  .502,  0}{NR} & \textcolor[rgb]{ 1,  0,  0}{R} & \textcolor[rgb]{ 0,  .502,  0}{NR} & \textcolor[rgb]{ 1,  0,  0}{R} & \textcolor[rgb]{ 0,  .502,  0}{NR} & \textcolor[rgb]{ 1,  0,  0}{R} & \textcolor[rgb]{ 1,  0,  0}{R} & \textcolor[rgb]{ 1,  0,  0}{R} \\
\textcolor[rgb]{ .129,  .129,  .129}{\textbf{30}} & \textcolor[rgb]{ 0,  .502,  0}{NR} & \textcolor[rgb]{ 0,  .502,  0}{NR} & \textcolor[rgb]{ 0,  .502,  0}{NR} & \textcolor[rgb]{ 1,  0,  0}{R} & \textcolor[rgb]{ 1,  0,  0}{R} & \textcolor[rgb]{ 0,  .502,  0}{NR} & \textcolor[rgb]{ 0,  .502,  0}{NR} & \textcolor[rgb]{ 0,  .502,  0}{NR} & \textcolor[rgb]{ 1,  0,  0}{R} & \textcolor[rgb]{ 0,  .502,  0}{NR} & \textcolor[rgb]{ 1,  0,  0}{R} & \textcolor[rgb]{ 0,  .502,  0}{NR} & \textcolor[rgb]{ 1,  0,  0}{R} & \textcolor[rgb]{ 1,  0,  0}{R} & \textcolor[rgb]{ 0,  .502,  0}{NR} \\
\textcolor[rgb]{ .129,  .129,  .129}{\textbf{40}} & \textcolor[rgb]{ 0,  .502,  0}{NR} & \textcolor[rgb]{ 0,  .502,  0}{NR} & \textcolor[rgb]{ 0,  .502,  0}{NR} & \textcolor[rgb]{ 0,  .502,  0}{NR} & \textcolor[rgb]{ 1,  0,  0}{R} & \textcolor[rgb]{ 0,  .502,  0}{NR} & \textcolor[rgb]{ 0,  .502,  0}{NR} & \textcolor[rgb]{ 0,  .502,  0}{NR} & \textcolor[rgb]{ 1,  0,  0}{R} & \textcolor[rgb]{ 0,  .502,  0}{NR} & \textcolor[rgb]{ 0,  .502,  0}{NR} & \textcolor[rgb]{ 0,  .502,  0}{NR} & \textcolor[rgb]{ 0,  .502,  0}{NR} & \textcolor[rgb]{ 0,  .502,  0}{NR} & \textcolor[rgb]{ 0,  .502,  0}{NR} \\
\textcolor[rgb]{ .129,  .129,  .129}{\textbf{50}} & \textcolor[rgb]{ 0,  .502,  0}{NR} & \textcolor[rgb]{ 0,  .502,  0}{NR} & \textcolor[rgb]{ 1,  0,  0}{R} & \textcolor[rgb]{ 1,  0,  0}{R} & \textcolor[rgb]{ 1,  0,  0}{R} & \textcolor[rgb]{ 0,  .502,  0}{NR} & \textcolor[rgb]{ 0,  .502,  0}{NR} & \textcolor[rgb]{ 0,  .502,  0}{NR} & \textcolor[rgb]{ 1,  0,  0}{R} & \textcolor[rgb]{ 0,  .502,  0}{NR} & \textcolor[rgb]{ 0,  .502,  0}{NR} & \textcolor[rgb]{ 0,  .502,  0}{NR} & \textcolor[rgb]{ 0,  .502,  0}{NR} & \textcolor[rgb]{ 0,  .502,  0}{NR} & \textcolor[rgb]{ 0,  .502,  0}{NR} \\
\textcolor[rgb]{ .129,  .129,  .129}{\textbf{60}} & \textcolor[rgb]{ 0,  .502,  0}{NR} & \textcolor[rgb]{ 0,  .502,  0}{NR} & \textcolor[rgb]{ 0,  .502,  0}{NR} & \textcolor[rgb]{ 1,  0,  0}{R} & \textcolor[rgb]{ 0,  .502,  0}{NR} & \textcolor[rgb]{ 0,  .502,  0}{NR} & \textcolor[rgb]{ 0,  .502,  0}{NR} & \textcolor[rgb]{ 0,  .502,  0}{NR} & \textcolor[rgb]{ 1,  0,  0}{R} & \textcolor[rgb]{ 0,  .502,  0}{NR} & \textcolor[rgb]{ 0,  .502,  0}{NR} & \textcolor[rgb]{ 0,  .502,  0}{NR} & \textcolor[rgb]{ 1,  0,  0}{R} & \textcolor[rgb]{ 0,  .502,  0}{NR} & \textcolor[rgb]{ 0,  .502,  0}{NR} \\
\textcolor[rgb]{ .129,  .129,  .129}{\textbf{70}} & \textcolor[rgb]{ 0,  .502,  0}{NR} & \textcolor[rgb]{ 0,  .502,  0}{NR} & \textcolor[rgb]{ 0,  .502,  0}{NR} & \textcolor[rgb]{ 1,  0,  0}{R} & \textcolor[rgb]{ 0,  .502,  0}{NR} & \textcolor[rgb]{ 0,  .502,  0}{NR} & \textcolor[rgb]{ 0,  .502,  0}{NR} & \textcolor[rgb]{ 0,  .502,  0}{NR} & \textcolor[rgb]{ 1,  0,  0}{R} & \textcolor[rgb]{ 0,  .502,  0}{NR} & \textcolor[rgb]{ 0,  .502,  0}{NR} & \textcolor[rgb]{ 0,  .502,  0}{NR} & \textcolor[rgb]{ 1,  0,  0}{R} & \textcolor[rgb]{ 0,  .502,  0}{NR} & \textcolor[rgb]{ 0,  .502,  0}{NR} \\
\textcolor[rgb]{ .129,  .129,  .129}{\textbf{80}} & \textcolor[rgb]{ 0,  .502,  0}{NR} & \textcolor[rgb]{ 0,  .502,  0}{NR} & \textcolor[rgb]{ 0,  .502,  0}{NR} & \textcolor[rgb]{ 1,  0,  0}{R} & \textcolor[rgb]{ 0,  .502,  0}{NR} & \textcolor[rgb]{ 0,  .502,  0}{NR} & \textcolor[rgb]{ 0,  .502,  0}{NR} & \textcolor[rgb]{ 0,  .502,  0}{NR} & \textcolor[rgb]{ 1,  0,  0}{R} & \textcolor[rgb]{ 0,  .502,  0}{NR} & \textcolor[rgb]{ 0,  .502,  0}{NR} & \textcolor[rgb]{ 0,  .502,  0}{NR} & \textcolor[rgb]{ 1,  0,  0}{R} & \textcolor[rgb]{ 0,  .502,  0}{NR} & \textcolor[rgb]{ 0,  .502,  0}{NR} \\
\textcolor[rgb]{ .129,  .129,  .129}{\textbf{90}} & \textcolor[rgb]{ 0,  .502,  0}{NR} & \textcolor[rgb]{ 0,  .502,  0}{NR} & \textcolor[rgb]{ 0,  .502,  0}{NR} & \textcolor[rgb]{ 1,  0,  0}{R} & \textcolor[rgb]{ 0,  .502,  0}{NR} & \textcolor[rgb]{ 0,  .502,  0}{NR} & \textcolor[rgb]{ 0,  .502,  0}{NR} & \textcolor[rgb]{ 1,  0,  0}{R} & \textcolor[rgb]{ 1,  0,  0}{R} & \textcolor[rgb]{ 0,  .502,  0}{NR} & \textcolor[rgb]{ 0,  .502,  0}{NR} & \textcolor[rgb]{ 0,  .502,  0}{NR} & \textcolor[rgb]{ 1,  0,  0}{R} & \textcolor[rgb]{ 0,  .502,  0}{NR} & \textcolor[rgb]{ 0,  .502,  0}{NR} \\
\bottomrule
    \end{tabular}%
}
  \label{tab_sigma_s3}%
\end{table}%

\begin{table}
% [width=1.0\linewidth,cols=16,pos=h]
  \centering
  \caption{Christofferson results for different $\sigma_{\text{short}}$ with $N_c = 4$ }
\resizebox{\textwidth}{!}{
    \begin{tabular}{lccccccccccccccc}
    % {\tblwidth}{@{}LCCCCCCCCCCCCCCC@{}}
\toprule
      & \multicolumn{15}{c}{\boldmath{}\textbf{$N_c= 4$}\unboldmath{}} \\
      & \multicolumn{15}{c}{\textbf{VaR 1 \%}} \\
      & \textcolor[rgb]{ .129,  .129,  .129}{\textbf{S\&P500}} & \textcolor[rgb]{ .129,  .129,  .129}{\textbf{MMM}} & \textcolor[rgb]{ .129,  .129,  .129}{\textbf{ARP}} & \textcolor[rgb]{ .129,  .129,  .129}{\textbf{AMD}} & \textcolor[rgb]{ .129,  .129,  .129}{\textbf{AIG}} & \textcolor[rgb]{ .129,  .129,  .129}{\textbf{AFL}} & \textcolor[rgb]{ .129,  .129,  .129}{\textbf{ABT}} & \textcolor[rgb]{ .129,  .129,  .129}{\textbf{MSFTF}} & \textcolor[rgb]{ .129,  .129,  .129}{\textbf{AAPL}} & \textcolor[rgb]{ .129,  .129,  .129}{\textbf{AMZN}} & \textcolor[rgb]{ .129,  .129,  .129}{\textbf{BAC}} & \textcolor[rgb]{ .129,  .129,  .129}{\textbf{JPM}} & \textcolor[rgb]{ .129,  .129,  .129}{\textbf{JNJ}} & \textcolor[rgb]{ .129,  .129,  .129}{\textbf{ROM}} & \textcolor[rgb]{ .129,  .129,  .129}{\textbf{MA}} \\
\textcolor[rgb]{ .129,  .129,  .129}{\textbf{10}} & \textcolor[rgb]{ 1,  0,  0}{R} & \textcolor[rgb]{ 1,  0,  0}{R} & \textcolor[rgb]{ 0,  .502,  0}{NR} & \textcolor[rgb]{ 1,  0,  0}{R} & \textcolor[rgb]{ 1,  0,  0}{R} & \textcolor[rgb]{ 1,  0,  0}{R} & \textcolor[rgb]{ 1,  0,  0}{R} & \textcolor[rgb]{ 1,  0,  0}{R} & \textcolor[rgb]{ 1,  0,  0}{R} & \textcolor[rgb]{ 1,  0,  0}{R} & \textcolor[rgb]{ 1,  0,  0}{R} & \textcolor[rgb]{ 1,  0,  0}{R} & \textcolor[rgb]{ 0,  .502,  0}{NR} & \textcolor[rgb]{ 1,  0,  0}{R} & \textcolor[rgb]{ 0,  .502,  0}{NR} \\
\textcolor[rgb]{ .129,  .129,  .129}{\textbf{20}} & \textcolor[rgb]{ 1,  0,  0}{R} & \textcolor[rgb]{ 1,  0,  0}{R} & \textcolor[rgb]{ 0,  .502,  0}{NR} & \textcolor[rgb]{ 0,  .502,  0}{NR} & \textcolor[rgb]{ 0,  .502,  0}{NR} & \textcolor[rgb]{ 0,  .502,  0}{NR} & \textcolor[rgb]{ 1,  0,  0}{R} & \textcolor[rgb]{ 1,  0,  0}{R} & \textcolor[rgb]{ 1,  0,  0}{R} & \textcolor[rgb]{ 1,  0,  0}{R} & \textcolor[rgb]{ 1,  0,  0}{R} & \textcolor[rgb]{ 0,  .502,  0}{NR} & \textcolor[rgb]{ 0,  .502,  0}{NR} & \textcolor[rgb]{ 0,  .502,  0}{NR} & \textcolor[rgb]{ 0,  .502,  0}{NR} \\
\textcolor[rgb]{ .129,  .129,  .129}{\textbf{30}} & \textcolor[rgb]{ 1,  0,  0}{R} & \textcolor[rgb]{ 1,  0,  0}{R} & \textcolor[rgb]{ 0,  .502,  0}{NR} & \textcolor[rgb]{ 0,  .502,  0}{NR} & \textcolor[rgb]{ 0,  .502,  0}{NR} & \textcolor[rgb]{ 0,  .502,  0}{NR} & \textcolor[rgb]{ 0,  .502,  0}{NR} & \textcolor[rgb]{ 0,  .502,  0}{NR} & \textcolor[rgb]{ 0,  .502,  0}{NR} & \textcolor[rgb]{ 0,  .502,  0}{NR} & \textcolor[rgb]{ 1,  0,  0}{R} & \textcolor[rgb]{ 1,  0,  0}{R} & \textcolor[rgb]{ 0,  .502,  0}{NR} & \textcolor[rgb]{ 0,  .502,  0}{NR} & \textcolor[rgb]{ 0,  .502,  0}{NR} \\
\textcolor[rgb]{ .129,  .129,  .129}{\textbf{40}} & \textcolor[rgb]{ 1,  0,  0}{R} & \textcolor[rgb]{ 1,  0,  0}{R} & \textcolor[rgb]{ 0,  .502,  0}{NR} & \textcolor[rgb]{ 0,  .502,  0}{NR} & \textcolor[rgb]{ 0,  .502,  0}{NR} & \textcolor[rgb]{ 0,  .502,  0}{NR} & \textcolor[rgb]{ 0,  .502,  0}{NR} & \textcolor[rgb]{ 0,  .502,  0}{NR} & \textcolor[rgb]{ 0,  .502,  0}{NR} & \textcolor[rgb]{ 0,  .502,  0}{NR} & \textcolor[rgb]{ 1,  0,  0}{R} & \textcolor[rgb]{ 1,  0,  0}{R} & \textcolor[rgb]{ 0,  .502,  0}{NR} & \textcolor[rgb]{ 0,  .502,  0}{NR} & \textcolor[rgb]{ 0,  .502,  0}{NR} \\
\textcolor[rgb]{ .129,  .129,  .129}{\textbf{50}} & \textcolor[rgb]{ 0,  .502,  0}{NR} & \textcolor[rgb]{ 1,  0,  0}{R} & \textcolor[rgb]{ 0,  .502,  0}{NR} & \textcolor[rgb]{ 1,  0,  0}{R} & \textcolor[rgb]{ 0,  .502,  0}{NR} & \textcolor[rgb]{ 0,  .502,  0}{NR} & \textcolor[rgb]{ 0,  .502,  0}{NR} & \textcolor[rgb]{ 0,  .502,  0}{NR} & \textcolor[rgb]{ 0,  .502,  0}{NR} & \textcolor[rgb]{ 1,  0,  0}{R} & \textcolor[rgb]{ 1,  0,  0}{R} & \textcolor[rgb]{ 1,  0,  0}{R} & \textcolor[rgb]{ 0,  .502,  0}{NR} & \textcolor[rgb]{ 0,  .502,  0}{NR} & \textcolor[rgb]{ 0,  .502,  0}{NR} \\
\textcolor[rgb]{ .129,  .129,  .129}{\textbf{60}} & \textcolor[rgb]{ 0,  .502,  0}{NR} & \textcolor[rgb]{ 1,  0,  0}{R} & \textcolor[rgb]{ 0,  .502,  0}{NR} & \textcolor[rgb]{ 0,  .502,  0}{NR} & \textcolor[rgb]{ 0,  .502,  0}{NR} & \textcolor[rgb]{ 0,  .502,  0}{NR} & \textcolor[rgb]{ 0,  .502,  0}{NR} & \textcolor[rgb]{ 0,  .502,  0}{NR} & \textcolor[rgb]{ 1,  0,  0}{R} & \textcolor[rgb]{ 0,  .502,  0}{NR} & \textcolor[rgb]{ 1,  0,  0}{R} & \textcolor[rgb]{ 0,  .502,  0}{NR} & \textcolor[rgb]{ 1,  0,  0}{R} & \textcolor[rgb]{ 0,  .502,  0}{NR} & \textcolor[rgb]{ 0,  .502,  0}{NR} \\
\textcolor[rgb]{ .129,  .129,  .129}{\textbf{70}} & \textcolor[rgb]{ 0,  .502,  0}{NR} & \textcolor[rgb]{ 0,  .502,  0}{NR} & \textcolor[rgb]{ 0,  .502,  0}{NR} & \textcolor[rgb]{ 0,  .502,  0}{NR} & \textcolor[rgb]{ 0,  .502,  0}{NR} & \textcolor[rgb]{ 0,  .502,  0}{NR} & \textcolor[rgb]{ 0,  .502,  0}{NR} & \textcolor[rgb]{ 0,  .502,  0}{NR} & \textcolor[rgb]{ 0,  .502,  0}{NR} & \textcolor[rgb]{ 0,  .502,  0}{NR} & \textcolor[rgb]{ 0,  .502,  0}{NR} & \textcolor[rgb]{ 0,  .502,  0}{NR} & \textcolor[rgb]{ 0,  .502,  0}{NR} & \textcolor[rgb]{ 0,  .502,  0}{NR} & \textcolor[rgb]{ 0,  .502,  0}{NR} \\
\textcolor[rgb]{ .129,  .129,  .129}{\textbf{80}} & \textcolor[rgb]{ 1,  0,  0}{R} & \textcolor[rgb]{ 0,  .502,  0}{NR} & \textcolor[rgb]{ 0,  .502,  0}{NR} & \textcolor[rgb]{ 1,  0,  0}{R} & \textcolor[rgb]{ 0,  .502,  0}{NR} & \textcolor[rgb]{ 0,  .502,  0}{NR} & \textcolor[rgb]{ 0,  .502,  0}{NR} & \textcolor[rgb]{ 0,  .502,  0}{NR} & \textcolor[rgb]{ 0,  .502,  0}{NR} & \textcolor[rgb]{ 0,  .502,  0}{NR} & \textcolor[rgb]{ 1,  0,  0}{R} & \textcolor[rgb]{ 0,  .502,  0}{NR} & \textcolor[rgb]{ 0,  .502,  0}{NR} & \textcolor[rgb]{ 0,  .502,  0}{NR} & \textcolor[rgb]{ 0,  .502,  0}{NR} \\
\textcolor[rgb]{ .129,  .129,  .129}{\textbf{90}} & \textcolor[rgb]{ 1,  0,  0}{R} & \textcolor[rgb]{ 1,  0,  0}{R} & \textcolor[rgb]{ 0,  .502,  0}{NR} & \textcolor[rgb]{ 1,  0,  0}{R} & \textcolor[rgb]{ 0,  .502,  0}{NR} & \textcolor[rgb]{ 0,  .502,  0}{NR} & \textcolor[rgb]{ 0,  .502,  0}{NR} & \textcolor[rgb]{ 0,  .502,  0}{NR} & \textcolor[rgb]{ 0,  .502,  0}{NR} & \textcolor[rgb]{ 0,  .502,  0}{NR} & \textcolor[rgb]{ 1,  0,  0}{R} & \textcolor[rgb]{ 0,  .502,  0}{NR} & \textcolor[rgb]{ 0,  .502,  0}{NR} & \textcolor[rgb]{ 0,  .502,  0}{NR} & \textcolor[rgb]{ 0,  .502,  0}{NR} \\
            & \multicolumn{15}{c}{\textbf{VaR 5 \%}} \\
\textcolor[rgb]{ .129,  .129,  .129}{} & \textcolor[rgb]{ .129,  .129,  .129}{\textbf{S\&P500}} & \textcolor[rgb]{ .129,  .129,  .129}{\textbf{MMM}} & \textcolor[rgb]{ .129,  .129,  .129}{\textbf{ARP}} & \textcolor[rgb]{ .129,  .129,  .129}{\textbf{AMD}} & \textcolor[rgb]{ .129,  .129,  .129}{\textbf{AIG}} & \textcolor[rgb]{ .129,  .129,  .129}{\textbf{AFL}} & \textcolor[rgb]{ .129,  .129,  .129}{\textbf{ABT}} & \textcolor[rgb]{ .129,  .129,  .129}{\textbf{MSFTF}} & \textcolor[rgb]{ .129,  .129,  .129}{\textbf{AAPL}} & \textcolor[rgb]{ .129,  .129,  .129}{\textbf{AMZN}} & \textcolor[rgb]{ .129,  .129,  .129}{\textbf{BAC}} & \textcolor[rgb]{ .129,  .129,  .129}{\textbf{JPM}} & \textcolor[rgb]{ .129,  .129,  .129}{\textbf{JNJ}} & \textcolor[rgb]{ .129,  .129,  .129}{\textbf{ROM}} & \textcolor[rgb]{ .129,  .129,  .129}{\textbf{MA}} \\
\textcolor[rgb]{ .129,  .129,  .129}{\textbf{10}} & \textcolor[rgb]{ 1,  0,  0}{R} & \textcolor[rgb]{ 1,  0,  0}{R} & \textcolor[rgb]{ 1,  0,  0}{R} & \textcolor[rgb]{ 1,  0,  0}{R} & \textcolor[rgb]{ 1,  0,  0}{R} & \textcolor[rgb]{ 1,  0,  0}{R} & \textcolor[rgb]{ 1,  0,  0}{R} & \textcolor[rgb]{ 1,  0,  0}{R} & \textcolor[rgb]{ 1,  0,  0}{R} & \textcolor[rgb]{ 1,  0,  0}{R} & \textcolor[rgb]{ 1,  0,  0}{R} & \textcolor[rgb]{ 1,  0,  0}{R} & \textcolor[rgb]{ 0,  .502,  0}{NR} & \textcolor[rgb]{ 1,  0,  0}{R} & \textcolor[rgb]{ 1,  0,  0}{R} \\
\textcolor[rgb]{ .129,  .129,  .129}{\textbf{20}} & \textcolor[rgb]{ 1,  0,  0}{R} & \textcolor[rgb]{ 0,  .502,  0}{NR} & \textcolor[rgb]{ 0,  .502,  0}{NR} & \textcolor[rgb]{ 1,  0,  0}{R} & \textcolor[rgb]{ 1,  0,  0}{R} & \textcolor[rgb]{ 0,  .502,  0}{NR} & \textcolor[rgb]{ 0,  .502,  0}{NR} & \textcolor[rgb]{ 0,  .502,  0}{NR} & \textcolor[rgb]{ 1,  0,  0}{R} & \textcolor[rgb]{ 1,  0,  0}{R} & \textcolor[rgb]{ 1,  0,  0}{R} & \textcolor[rgb]{ 0,  .502,  0}{NR} & \textcolor[rgb]{ 0,  .502,  0}{NR} & \textcolor[rgb]{ 1,  0,  0}{R} & \textcolor[rgb]{ 1,  0,  0}{R} \\
\textcolor[rgb]{ .129,  .129,  .129}{\textbf{30}} & \textcolor[rgb]{ 1,  0,  0}{R} & \textcolor[rgb]{ 0,  .502,  0}{NR} & \textcolor[rgb]{ 0,  .502,  0}{NR} & \textcolor[rgb]{ 1,  0,  0}{R} & \textcolor[rgb]{ 1,  0,  0}{R} & \textcolor[rgb]{ 1,  0,  0}{R} & \textcolor[rgb]{ 0,  .502,  0}{NR} & \textcolor[rgb]{ 1,  0,  0}{R} & \textcolor[rgb]{ 1,  0,  0}{R} & \textcolor[rgb]{ 0,  .502,  0}{NR} & \textcolor[rgb]{ 1,  0,  0}{R} & \textcolor[rgb]{ 0,  .502,  0}{NR} & \textcolor[rgb]{ 1,  0,  0}{R} & \textcolor[rgb]{ 0,  .502,  0}{NR} & \textcolor[rgb]{ 0,  .502,  0}{NR} \\
\textcolor[rgb]{ .129,  .129,  .129}{\textbf{40}} & \textcolor[rgb]{ 1,  0,  0}{R} & \textcolor[rgb]{ 0,  .502,  0}{NR} & \textcolor[rgb]{ 0,  .502,  0}{NR} & \textcolor[rgb]{ 1,  0,  0}{R} & \textcolor[rgb]{ 1,  0,  0}{R} & \textcolor[rgb]{ 0,  .502,  0}{NR} & \textcolor[rgb]{ 0,  .502,  0}{NR} & \textcolor[rgb]{ 0,  .502,  0}{NR} & \textcolor[rgb]{ 1,  0,  0}{R} & \textcolor[rgb]{ 0,  .502,  0}{NR} & \textcolor[rgb]{ 0,  .502,  0}{NR} & \textcolor[rgb]{ 0,  .502,  0}{NR} & \textcolor[rgb]{ 0,  .502,  0}{NR} & \textcolor[rgb]{ 0,  .502,  0}{NR} & \textcolor[rgb]{ 0,  .502,  0}{NR} \\
\textcolor[rgb]{ .129,  .129,  .129}{\textbf{50}} & \textcolor[rgb]{ 0,  .502,  0}{NR} & \textcolor[rgb]{ 0,  .502,  0}{NR} & \textcolor[rgb]{ 0,  .502,  0}{NR} & \textcolor[rgb]{ 1,  0,  0}{R} & \textcolor[rgb]{ 0,  .502,  0}{NR} & \textcolor[rgb]{ 0,  .502,  0}{NR} & \textcolor[rgb]{ 0,  .502,  0}{NR} & \textcolor[rgb]{ 1,  0,  0}{R} & \textcolor[rgb]{ 1,  0,  0}{R} & \textcolor[rgb]{ 0,  .502,  0}{NR} & \textcolor[rgb]{ 0,  .502,  0}{NR} & \textcolor[rgb]{ 0,  .502,  0}{NR} & \textcolor[rgb]{ 1,  0,  0}{R} & \textcolor[rgb]{ 0,  .502,  0}{NR} & \textcolor[rgb]{ 0,  .502,  0}{NR} \\
\textcolor[rgb]{ .129,  .129,  .129}{\textbf{60}} & \textcolor[rgb]{ 1,  0,  0}{R} & \textcolor[rgb]{ 0,  .502,  0}{NR} & \textcolor[rgb]{ 0,  .502,  0}{NR} & \textcolor[rgb]{ 1,  0,  0}{R} & \textcolor[rgb]{ 0,  .502,  0}{NR} & \textcolor[rgb]{ 0,  .502,  0}{NR} & \textcolor[rgb]{ 0,  .502,  0}{NR} & \textcolor[rgb]{ 0,  .502,  0}{NR} & \textcolor[rgb]{ 1,  0,  0}{R} & \textcolor[rgb]{ 0,  .502,  0}{NR} & \textcolor[rgb]{ 0,  .502,  0}{NR} & \textcolor[rgb]{ 0,  .502,  0}{NR} & \textcolor[rgb]{ 1,  0,  0}{R} & \textcolor[rgb]{ 0,  .502,  0}{NR} & \textcolor[rgb]{ 0,  .502,  0}{NR} \\
\textcolor[rgb]{ .129,  .129,  .129}{\textbf{70}} & \textcolor[rgb]{ 0,  .502,  0}{NR} & \textcolor[rgb]{ 0,  .502,  0}{NR} & \textcolor[rgb]{ 0,  .502,  0}{NR} & \textcolor[rgb]{ 0,  .502,  0}{NR} & \textcolor[rgb]{ 0,  .502,  0}{NR} & \textcolor[rgb]{ 0,  .502,  0}{NR} & \textcolor[rgb]{ 0,  .502,  0}{NR} & \textcolor[rgb]{ 0,  .502,  0}{NR} & \textcolor[rgb]{ 0,  .502,  0}{NR} & \textcolor[rgb]{ 0,  .502,  0}{NR} & \multicolumn{1}{c}{} & \textcolor[rgb]{ 0,  .502,  0}{NR} & \textcolor[rgb]{ 1,  0,  0}{R} & \textcolor[rgb]{ 0,  .502,  0}{NR} & \textcolor[rgb]{ 0,  .502,  0}{NR} \\
\textcolor[rgb]{ .129,  .129,  .129}{\textbf{80}} & \textcolor[rgb]{ 0,  .502,  0}{NR} & \textcolor[rgb]{ 0,  .502,  0}{NR} & \textcolor[rgb]{ 0,  .502,  0}{NR} & \textcolor[rgb]{ 0,  .502,  0}{NR} & \textcolor[rgb]{ 0,  .502,  0}{NR} & \textcolor[rgb]{ 0,  .502,  0}{NR} & \textcolor[rgb]{ 0,  .502,  0}{NR} & \textcolor[rgb]{ 0,  .502,  0}{NR} & \textcolor[rgb]{ 1,  0,  0}{R} & \textcolor[rgb]{ 0,  .502,  0}{NR} & \textcolor[rgb]{ 0,  .502,  0}{NR} & \textcolor[rgb]{ 0,  .502,  0}{NR} & \textcolor[rgb]{ 1,  0,  0}{R} & \textcolor[rgb]{ 0,  .502,  0}{NR} & \textcolor[rgb]{ 0,  .502,  0}{NR} \\
\textcolor[rgb]{ .129,  .129,  .129}{\textbf{90}} & \textcolor[rgb]{ 0,  .502,  0}{NR} & \textcolor[rgb]{ 0,  .502,  0}{NR} & \textcolor[rgb]{ 0,  .502,  0}{NR} & \textcolor[rgb]{ 1,  0,  0}{R} & \textcolor[rgb]{ 0,  .502,  0}{NR} & \textcolor[rgb]{ 0,  .502,  0}{NR} & \textcolor[rgb]{ 0,  .502,  0}{NR} & \textcolor[rgb]{ 0,  .502,  0}{NR} & \textcolor[rgb]{ 1,  0,  0}{R} & \textcolor[rgb]{ 0,  .502,  0}{NR} & \textcolor[rgb]{ 0,  .502,  0}{NR} & \textcolor[rgb]{ 0,  .502,  0}{NR} & \textcolor[rgb]{ 1,  0,  0}{R} & \textcolor[rgb]{ 0,  .502,  0}{NR} & \textcolor[rgb]{ 0,  .502,  0}{NR} \\
\bottomrule
    \end{tabular}%
}
  \label{tab_sigma_s4}%
\end{table}%

\begin{table}
% [width=1.0\linewidth,cols=16,pos=h]
  \centering
  \caption{Christofferson results for different $\sigma_{\text{short}}$ with $N_c = 5$ }
\resizebox{\textwidth}{!}{
    \begin{tabular}{lccccccccccccccc}
    % {\tblwidth}{@{}LCCCCCCCCCCCCCCC@{}}
\toprule
 \textcolor[rgb]{ .129,  .129,  .129}{} & \multicolumn{15}{c}{\textbf{\$N\_c= 5\$}} \\
\textcolor[rgb]{ .129,  .129,  .129}{} & \multicolumn{15}{c}{\textbf{VaR 1 \textbackslash{}\%}} \\
      & \textcolor[rgb]{ .129,  .129,  .129}{\textbf{S\&P500}} & \textcolor[rgb]{ .129,  .129,  .129}{\textbf{MMM}} & \textcolor[rgb]{ .129,  .129,  .129}{\textbf{ARP}} & \textcolor[rgb]{ .129,  .129,  .129}{\textbf{AMD}} & \textcolor[rgb]{ .129,  .129,  .129}{\textbf{AIG}} & \textcolor[rgb]{ .129,  .129,  .129}{\textbf{AFL}} & \textcolor[rgb]{ .129,  .129,  .129}{\textbf{ABT}} & \textcolor[rgb]{ .129,  .129,  .129}{\textbf{MSFTF}} & \textcolor[rgb]{ .129,  .129,  .129}{\textbf{AAPL}} & \textcolor[rgb]{ .129,  .129,  .129}{\textbf{AMZN}} & \textcolor[rgb]{ .129,  .129,  .129}{\textbf{BAC}} & \textcolor[rgb]{ .129,  .129,  .129}{\textbf{JPM}} & \textcolor[rgb]{ .129,  .129,  .129}{\textbf{JNJ}} & \textcolor[rgb]{ .129,  .129,  .129}{\textbf{ROM}} & \textcolor[rgb]{ .129,  .129,  .129}{\textbf{MA}} \\
\textcolor[rgb]{ .129,  .129,  .129}{\textbf{10}} & \textcolor[rgb]{ 1,  0,  0}{R} & \textcolor[rgb]{ 1,  0,  0}{R} & \textcolor[rgb]{ 0,  .502,  0}{NR} & \textcolor[rgb]{ 0,  .502,  0}{NR} & \textcolor[rgb]{ 1,  0,  0}{R} & \textcolor[rgb]{ 1,  0,  0}{R} & \textcolor[rgb]{ 1,  0,  0}{R} & \textcolor[rgb]{ 1,  0,  0}{R} & \textcolor[rgb]{ 1,  0,  0}{R} & \textcolor[rgb]{ 1,  0,  0}{R} & \textcolor[rgb]{ 1,  0,  0}{R} & \textcolor[rgb]{ 1,  0,  0}{R} & \textcolor[rgb]{ 1,  0,  0}{R} & \textcolor[rgb]{ 1,  0,  0}{R} & \textcolor[rgb]{ 0,  .502,  0}{NR} \\
\textcolor[rgb]{ .129,  .129,  .129}{\textbf{20}} & \textcolor[rgb]{ 0,  .502,  0}{NR} & \textcolor[rgb]{ 1,  0,  0}{R} & \textcolor[rgb]{ 0,  .502,  0}{NR} & \textcolor[rgb]{ 0,  .502,  0}{NR} & \textcolor[rgb]{ 0,  .502,  0}{NR} & \textcolor[rgb]{ 0,  .502,  0}{NR} & \textcolor[rgb]{ 0,  .502,  0}{NR} & \textcolor[rgb]{ 1,  0,  0}{R} & \textcolor[rgb]{ 0,  .502,  0}{NR} & \textcolor[rgb]{ 1,  0,  0}{R} & \textcolor[rgb]{ 1,  0,  0}{R} & \textcolor[rgb]{ 1,  0,  0}{R} & \textcolor[rgb]{ 0,  .502,  0}{NR} & \textcolor[rgb]{ 0,  .502,  0}{NR} & \textcolor[rgb]{ 0,  .502,  0}{NR} \\
\textcolor[rgb]{ .129,  .129,  .129}{\textbf{30}} & \textcolor[rgb]{ 0,  .502,  0}{NR} & \textcolor[rgb]{ 0,  .502,  0}{NR} & \textcolor[rgb]{ 0,  .502,  0}{NR} & \textcolor[rgb]{ 0,  .502,  0}{NR} & \textcolor[rgb]{ 0,  .502,  0}{NR} & \textcolor[rgb]{ 0,  .502,  0}{NR} & \textcolor[rgb]{ 0,  .502,  0}{NR} & \textcolor[rgb]{ 1,  0,  0}{R} & \textcolor[rgb]{ 0,  .502,  0}{NR} & \textcolor[rgb]{ 0,  .502,  0}{NR} & \textcolor[rgb]{ 1,  0,  0}{R} & \textcolor[rgb]{ 1,  0,  0}{R} & \textcolor[rgb]{ 0,  .502,  0}{NR} & \textcolor[rgb]{ 0,  .502,  0}{NR} & \textcolor[rgb]{ 0,  .502,  0}{NR} \\
\textcolor[rgb]{ .129,  .129,  .129}{\textbf{40}} & \textcolor[rgb]{ 0,  .502,  0}{NR} & \textcolor[rgb]{ 0,  .502,  0}{NR} & \textcolor[rgb]{ 0,  .502,  0}{NR} & \textcolor[rgb]{ 0,  .502,  0}{NR} & \textcolor[rgb]{ 0,  .502,  0}{NR} & \textcolor[rgb]{ 0,  .502,  0}{NR} & \textcolor[rgb]{ 0,  .502,  0}{NR} & \textcolor[rgb]{ 1,  0,  0}{R} & \textcolor[rgb]{ 0,  .502,  0}{NR} & \textcolor[rgb]{ 1,  0,  0}{R} & \textcolor[rgb]{ 0,  .502,  0}{NR} & \textcolor[rgb]{ 1,  0,  0}{R} & \textcolor[rgb]{ 0,  .502,  0}{NR} & \textcolor[rgb]{ 0,  .502,  0}{NR} & \textcolor[rgb]{ 0,  .502,  0}{NR} \\
\textcolor[rgb]{ .129,  .129,  .129}{\textbf{50}} & \textcolor[rgb]{ 0,  .502,  0}{NR} & \textcolor[rgb]{ 0,  .502,  0}{NR} & \textcolor[rgb]{ 0,  .502,  0}{NR} & \textcolor[rgb]{ 0,  .502,  0}{NR} & \textcolor[rgb]{ 0,  .502,  0}{NR} & \textcolor[rgb]{ 0,  .502,  0}{NR} & \textcolor[rgb]{ 0,  .502,  0}{NR} & \textcolor[rgb]{ 0,  .502,  0}{NR} & \textcolor[rgb]{ 0,  .502,  0}{NR} & \textcolor[rgb]{ 1,  0,  0}{R} & \textcolor[rgb]{ 0,  .502,  0}{NR} & \textcolor[rgb]{ 1,  0,  0}{R} & \textcolor[rgb]{ 0,  .502,  0}{NR} & \textcolor[rgb]{ 0,  .502,  0}{NR} & \textcolor[rgb]{ 0,  .502,  0}{NR} \\
\textcolor[rgb]{ .129,  .129,  .129}{\textbf{60}} & \textcolor[rgb]{ 1,  0,  0}{R} & \textcolor[rgb]{ 0,  .502,  0}{NR} & \textcolor[rgb]{ 0,  .502,  0}{NR} & \textcolor[rgb]{ 0,  .502,  0}{NR} & \textcolor[rgb]{ 0,  .502,  0}{NR} & \textcolor[rgb]{ 0,  .502,  0}{NR} & \textcolor[rgb]{ 0,  .502,  0}{NR} & \textcolor[rgb]{ 0,  .502,  0}{NR} & \textcolor[rgb]{ 0,  .502,  0}{NR} & \textcolor[rgb]{ 1,  0,  0}{R} & \textcolor[rgb]{ 0,  .502,  0}{NR} & \textcolor[rgb]{ 0,  .502,  0}{NR} & \textcolor[rgb]{ 0,  .502,  0}{NR} & \textcolor[rgb]{ 0,  .502,  0}{NR} & \textcolor[rgb]{ 0,  .502,  0}{NR} \\
\textcolor[rgb]{ .129,  .129,  .129}{\textbf{70}} & \textcolor[rgb]{ 0,  .502,  0}{NR} & \textcolor[rgb]{ 0,  .502,  0}{NR} & \textcolor[rgb]{ 0,  .502,  0}{NR} & \textcolor[rgb]{ 0,  .502,  0}{NR} & \textcolor[rgb]{ 0,  .502,  0}{NR} & \textcolor[rgb]{ 0,  .502,  0}{NR} & \textcolor[rgb]{ 0,  .502,  0}{NR} & \textcolor[rgb]{ 0,  .502,  0}{NR} & \textcolor[rgb]{ 0,  .502,  0}{NR} & \textcolor[rgb]{ 0,  .502,  0}{NR} & \textcolor[rgb]{ 0,  .502,  0}{NR} & \textcolor[rgb]{ 0,  .502,  0}{NR} & \textcolor[rgb]{ 0,  .502,  0}{NR} & \textcolor[rgb]{ 0,  .502,  0}{NR} & \textcolor[rgb]{ 0,  .502,  0}{NR} \\
\textcolor[rgb]{ .129,  .129,  .129}{\textbf{80}} & \textcolor[rgb]{ 1,  0,  0}{R} & \textcolor[rgb]{ 0,  .502,  0}{NR} & \textcolor[rgb]{ 0,  .502,  0}{NR} & \textcolor[rgb]{ 0,  .502,  0}{NR} & \textcolor[rgb]{ 0,  .502,  0}{NR} & \textcolor[rgb]{ 0,  .502,  0}{NR} & \textcolor[rgb]{ 0,  .502,  0}{NR} & \textcolor[rgb]{ 1,  0,  0}{R} & \textcolor[rgb]{ 0,  .502,  0}{NR} & \textcolor[rgb]{ 0,  .502,  0}{NR} & \textcolor[rgb]{ 0,  .502,  0}{NR} & \textcolor[rgb]{ 1,  0,  0}{R} & \textcolor[rgb]{ 0,  .502,  0}{NR} & \textcolor[rgb]{ 0,  .502,  0}{NR} & \textcolor[rgb]{ 0,  .502,  0}{NR} \\
\textcolor[rgb]{ .129,  .129,  .129}{\textbf{90}} & \textcolor[rgb]{ 1,  0,  0}{R} & \textcolor[rgb]{ 0,  .502,  0}{NR} & \textcolor[rgb]{ 0,  .502,  0}{NR} & \textcolor[rgb]{ 0,  .502,  0}{NR} & \textcolor[rgb]{ 0,  .502,  0}{NR} & \textcolor[rgb]{ 0,  .502,  0}{NR} & \textcolor[rgb]{ 0,  .502,  0}{NR} & \textcolor[rgb]{ 1,  0,  0}{R} & \textcolor[rgb]{ 0,  .502,  0}{NR} & \textcolor[rgb]{ 0,  .502,  0}{NR} & \textcolor[rgb]{ 0,  .502,  0}{NR} & \textcolor[rgb]{ 1,  0,  0}{R} & \textcolor[rgb]{ 0,  .502,  0}{NR} & \textcolor[rgb]{ 0,  .502,  0}{NR} & \textcolor[rgb]{ 0,  .502,  0}{NR} \\
            & \multicolumn{15}{c}{\textbf{VaR 5 \%}} \\
\textcolor[rgb]{ .129,  .129,  .129}{} & \textcolor[rgb]{ .129,  .129,  .129}{\textbf{S\&P500}} & \textcolor[rgb]{ .129,  .129,  .129}{\textbf{MMM}} & \textcolor[rgb]{ .129,  .129,  .129}{\textbf{ARP}} & \textcolor[rgb]{ .129,  .129,  .129}{\textbf{AMD}} & \textcolor[rgb]{ .129,  .129,  .129}{\textbf{AIG}} & \textcolor[rgb]{ .129,  .129,  .129}{\textbf{AFL}} & \textcolor[rgb]{ .129,  .129,  .129}{\textbf{ABT}} & \textcolor[rgb]{ .129,  .129,  .129}{\textbf{MSFTF}} & \textcolor[rgb]{ .129,  .129,  .129}{\textbf{AAPL}} & \textcolor[rgb]{ .129,  .129,  .129}{\textbf{AMZN}} & \textcolor[rgb]{ .129,  .129,  .129}{\textbf{BAC}} & \textcolor[rgb]{ .129,  .129,  .129}{\textbf{JPM}} & \textcolor[rgb]{ .129,  .129,  .129}{\textbf{JNJ}} & \textcolor[rgb]{ .129,  .129,  .129}{\textbf{ROM}} & \textcolor[rgb]{ .129,  .129,  .129}{\textbf{MA}} \\
\textcolor[rgb]{ .129,  .129,  .129}{\textbf{10}} & \textcolor[rgb]{ 1,  0,  0}{R} & \textcolor[rgb]{ 1,  0,  0}{R} & \textcolor[rgb]{ 0,  .502,  0}{NR} & \textcolor[rgb]{ 1,  0,  0}{R} & \textcolor[rgb]{ 1,  0,  0}{R} & \textcolor[rgb]{ 1,  0,  0}{R} & \textcolor[rgb]{ 1,  0,  0}{R} & \textcolor[rgb]{ 1,  0,  0}{R} & \textcolor[rgb]{ 0,  .502,  0}{NR} & \textcolor[rgb]{ 1,  0,  0}{R} & \textcolor[rgb]{ 1,  0,  0}{R} & \textcolor[rgb]{ 1,  0,  0}{R} & \textcolor[rgb]{ 0,  .502,  0}{NR} & \textcolor[rgb]{ 0,  .502,  0}{NR} & \textcolor[rgb]{ 1,  0,  0}{R} \\
\textcolor[rgb]{ .129,  .129,  .129}{\textbf{20}} & \textcolor[rgb]{ 0,  .502,  0}{NR} & \textcolor[rgb]{ 0,  .502,  0}{NR} & \textcolor[rgb]{ 0,  .502,  0}{NR} & \textcolor[rgb]{ 0,  .502,  0}{NR} & \textcolor[rgb]{ 1,  0,  0}{R} & \textcolor[rgb]{ 0,  .502,  0}{NR} & \textcolor[rgb]{ 0,  .502,  0}{NR} & \textcolor[rgb]{ 0,  .502,  0}{NR} & \textcolor[rgb]{ 1,  0,  0}{R} & \textcolor[rgb]{ 0,  .502,  0}{NR} & \textcolor[rgb]{ 1,  0,  0}{R} & \textcolor[rgb]{ 0,  .502,  0}{NR} & \textcolor[rgb]{ 0,  .502,  0}{NR} & \textcolor[rgb]{ 1,  0,  0}{R} & \textcolor[rgb]{ 1,  0,  0}{R} \\
\textcolor[rgb]{ .129,  .129,  .129}{\textbf{30}} & \textcolor[rgb]{ 0,  .502,  0}{NR} & \textcolor[rgb]{ 1,  0,  0}{R} & \textcolor[rgb]{ 0,  .502,  0}{NR} & \textcolor[rgb]{ 0,  .502,  0}{NR} & \textcolor[rgb]{ 1,  0,  0}{R} & \textcolor[rgb]{ 0,  .502,  0}{NR} & \textcolor[rgb]{ 0,  .502,  0}{NR} & \textcolor[rgb]{ 1,  0,  0}{R} & \textcolor[rgb]{ 1,  0,  0}{R} & \textcolor[rgb]{ 0,  .502,  0}{NR} & \textcolor[rgb]{ 1,  0,  0}{R} & \textcolor[rgb]{ 0,  .502,  0}{NR} & \textcolor[rgb]{ 1,  0,  0}{R} & \textcolor[rgb]{ 0,  .502,  0}{NR} & \textcolor[rgb]{ 0,  .502,  0}{NR} \\
\textcolor[rgb]{ .129,  .129,  .129}{\textbf{40}} & \textcolor[rgb]{ 0,  .502,  0}{NR} & \textcolor[rgb]{ 0,  .502,  0}{NR} & \textcolor[rgb]{ 0,  .502,  0}{NR} & \textcolor[rgb]{ 0,  .502,  0}{NR} & \textcolor[rgb]{ 1,  0,  0}{R} & \textcolor[rgb]{ 0,  .502,  0}{NR} & \textcolor[rgb]{ 0,  .502,  0}{NR} & \textcolor[rgb]{ 0,  .502,  0}{NR} & \textcolor[rgb]{ 1,  0,  0}{R} & \textcolor[rgb]{ 0,  .502,  0}{NR} & \textcolor[rgb]{ 1,  0,  0}{R} & \textcolor[rgb]{ 0,  .502,  0}{NR} & \textcolor[rgb]{ 0,  .502,  0}{NR} & \textcolor[rgb]{ 0,  .502,  0}{NR} & \textcolor[rgb]{ 0,  .502,  0}{NR} \\
\textcolor[rgb]{ .129,  .129,  .129}{\textbf{50}} & \textcolor[rgb]{ 0,  .502,  0}{NR} & \textcolor[rgb]{ 0,  .502,  0}{NR} & \textcolor[rgb]{ 0,  .502,  0}{NR} & \textcolor[rgb]{ 1,  0,  0}{R} & \textcolor[rgb]{ 0,  .502,  0}{NR} & \textcolor[rgb]{ 0,  .502,  0}{NR} & \textcolor[rgb]{ 0,  .502,  0}{NR} & \textcolor[rgb]{ 1,  0,  0}{R} & \textcolor[rgb]{ 1,  0,  0}{R} & \textcolor[rgb]{ 0,  .502,  0}{NR} & \textcolor[rgb]{ 0,  .502,  0}{NR} & \textcolor[rgb]{ 0,  .502,  0}{NR} & \textcolor[rgb]{ 1,  0,  0}{R} & \textcolor[rgb]{ 0,  .502,  0}{NR} & \textcolor[rgb]{ 0,  .502,  0}{NR} \\
\textcolor[rgb]{ .129,  .129,  .129}{\textbf{60}} & \textcolor[rgb]{ 0,  .502,  0}{NR} & \textcolor[rgb]{ 0,  .502,  0}{NR} & \textcolor[rgb]{ 0,  .502,  0}{NR} & \textcolor[rgb]{ 1,  0,  0}{R} & \textcolor[rgb]{ 0,  .502,  0}{NR} & \textcolor[rgb]{ 0,  .502,  0}{NR} & \textcolor[rgb]{ 0,  .502,  0}{NR} & \textcolor[rgb]{ 1,  0,  0}{R} & \textcolor[rgb]{ 1,  0,  0}{R} & \textcolor[rgb]{ 0,  .502,  0}{NR} & \textcolor[rgb]{ 0,  .502,  0}{NR} & \textcolor[rgb]{ 0,  .502,  0}{NR} & \textcolor[rgb]{ 1,  0,  0}{R} & \textcolor[rgb]{ 0,  .502,  0}{NR} & \textcolor[rgb]{ 0,  .502,  0}{NR} \\
\textcolor[rgb]{ .129,  .129,  .129}{\textbf{70}} & \textcolor[rgb]{ 0,  .502,  0}{NR} & \textcolor[rgb]{ 0,  .502,  0}{NR} & \textcolor[rgb]{ 0,  .502,  0}{NR} & \textcolor[rgb]{ 0,  .502,  0}{NR} & \textcolor[rgb]{ 0,  .502,  0}{NR} & \textcolor[rgb]{ 0,  .502,  0}{NR} & \textcolor[rgb]{ 0,  .502,  0}{NR} & \textcolor[rgb]{ 1,  0,  0}{R} & \textcolor[rgb]{ 1,  0,  0}{R} & \textcolor[rgb]{ 0,  .502,  0}{NR} & \textcolor[rgb]{ 0,  .502,  0}{NR} & \textcolor[rgb]{ 0,  .502,  0}{NR} & \textcolor[rgb]{ 1,  0,  0}{R} & \textcolor[rgb]{ 0,  .502,  0}{NR} & \textcolor[rgb]{ 0,  .502,  0}{NR} \\
\textcolor[rgb]{ .129,  .129,  .129}{\textbf{80}} & \textcolor[rgb]{ 0,  .502,  0}{NR} & \textcolor[rgb]{ 0,  .502,  0}{NR} & \textcolor[rgb]{ 0,  .502,  0}{NR} & \textcolor[rgb]{ 1,  0,  0}{R} & \textcolor[rgb]{ 0,  .502,  0}{NR} & \textcolor[rgb]{ 0,  .502,  0}{NR} & \textcolor[rgb]{ 0,  .502,  0}{NR} & \textcolor[rgb]{ 0,  .502,  0}{NR} & \textcolor[rgb]{ 1,  0,  0}{R} & \textcolor[rgb]{ 0,  .502,  0}{NR} & \textcolor[rgb]{ 0,  .502,  0}{NR} & \textcolor[rgb]{ 0,  .502,  0}{NR} & \textcolor[rgb]{ 1,  0,  0}{R} & \textcolor[rgb]{ 0,  .502,  0}{NR} & \textcolor[rgb]{ 0,  .502,  0}{NR} \\
\textcolor[rgb]{ .129,  .129,  .129}{\textbf{90}} & \textcolor[rgb]{ 0,  .502,  0}{NR} & \textcolor[rgb]{ 0,  .502,  0}{NR} & \textcolor[rgb]{ 0,  .502,  0}{NR} & \textcolor[rgb]{ 1,  0,  0}{R} & \textcolor[rgb]{ 0,  .502,  0}{NR} & \textcolor[rgb]{ 0,  .502,  0}{NR} & \textcolor[rgb]{ 0,  .502,  0}{NR} & \textcolor[rgb]{ 1,  0,  0}{R} & \textcolor[rgb]{ 1,  0,  0}{R} & \textcolor[rgb]{ 0,  .502,  0}{NR} & \textcolor[rgb]{ 0,  .502,  0}{NR} & \textcolor[rgb]{ 0,  .502,  0}{NR} & \textcolor[rgb]{ 1,  0,  0}{R} & \textcolor[rgb]{ 0,  .502,  0}{NR} & \textcolor[rgb]{ 0,  .502,  0}{NR} \\
\bottomrule
    \end{tabular}%
}
  \label{tab_sigma_s5}%
\end{table}%

\begin{table}
% [width=1.0\linewidth,cols=16,pos=h]
  \centering
  \caption{Christofferson results for different $\sigma_{\text{short}}$ with $N_c = 6$ }
\resizebox{\textwidth}{!}{
    \begin{tabular}{lccccccccccccccc}
    % {\tblwidth}{@{}LCCCCCCCCCCCCCCC@{}}
\toprule
      & \multicolumn{15}{c}{\textbf{VaR 1 \textbackslash{}\%}} \\
\textcolor[rgb]{ .129,  .129,  .129}{} & \textcolor[rgb]{ .129,  .129,  .129}{\textbf{S\&P500}} & \textcolor[rgb]{ .129,  .129,  .129}{\textbf{MMM}} & \textcolor[rgb]{ .129,  .129,  .129}{\textbf{ARP}} & \textcolor[rgb]{ .129,  .129,  .129}{\textbf{AMD}} & \textcolor[rgb]{ .129,  .129,  .129}{\textbf{AIG}} & \textcolor[rgb]{ .129,  .129,  .129}{\textbf{AFL}} & \textcolor[rgb]{ .129,  .129,  .129}{\textbf{ABT}} & \textcolor[rgb]{ .129,  .129,  .129}{\textbf{MSFTF}} & \textcolor[rgb]{ .129,  .129,  .129}{\textbf{AAPL}} & \textcolor[rgb]{ .129,  .129,  .129}{\textbf{AMZN}} & \textcolor[rgb]{ .129,  .129,  .129}{\textbf{BAC}} & \textcolor[rgb]{ .129,  .129,  .129}{\textbf{JPM}} & \textcolor[rgb]{ .129,  .129,  .129}{\textbf{JNJ}} & \textcolor[rgb]{ .129,  .129,  .129}{\textbf{ROM}} & \textcolor[rgb]{ .129,  .129,  .129}{\textbf{MA}} \\
\textcolor[rgb]{ .129,  .129,  .129}{\textbf{10}} & \textcolor[rgb]{ 1,  0,  0}{R} & \textcolor[rgb]{ 1,  0,  0}{R} & \textcolor[rgb]{ 0,  .502,  0}{NR} & \textcolor[rgb]{ 1,  0,  0}{R} & \textcolor[rgb]{ 1,  0,  0}{R} & \textcolor[rgb]{ 1,  0,  0}{R} & \textcolor[rgb]{ 1,  0,  0}{R} & \textcolor[rgb]{ 1,  0,  0}{R} & \textcolor[rgb]{ 1,  0,  0}{R} & \textcolor[rgb]{ 1,  0,  0}{R} & \textcolor[rgb]{ 1,  0,  0}{R} & \textcolor[rgb]{ 1,  0,  0}{R} & \textcolor[rgb]{ 0,  .502,  0}{NR} & \textcolor[rgb]{ 1,  0,  0}{R} & \textcolor[rgb]{ 0,  .502,  0}{NR} \\
\textcolor[rgb]{ .129,  .129,  .129}{\textbf{20}} & \textcolor[rgb]{ 0,  .502,  0}{NR} & \textcolor[rgb]{ 1,  0,  0}{R} & \textcolor[rgb]{ 0,  .502,  0}{NR} & \textcolor[rgb]{ 0,  .502,  0}{NR} & \textcolor[rgb]{ 0,  .502,  0}{NR} & \textcolor[rgb]{ 0,  .502,  0}{NR} & \textcolor[rgb]{ 1,  0,  0}{R} & \textcolor[rgb]{ 1,  0,  0}{R} & \textcolor[rgb]{ 0,  .502,  0}{NR} & \textcolor[rgb]{ 1,  0,  0}{R} & \textcolor[rgb]{ 1,  0,  0}{R} & \textcolor[rgb]{ 0,  .502,  0}{NR} & \textcolor[rgb]{ 0,  .502,  0}{NR} & \textcolor[rgb]{ 0,  .502,  0}{NR} & \textcolor[rgb]{ 0,  .502,  0}{NR} \\
\textcolor[rgb]{ .129,  .129,  .129}{\textbf{30}} & \textcolor[rgb]{ 0,  .502,  0}{NR} & \textcolor[rgb]{ 0,  .502,  0}{NR} & \textcolor[rgb]{ 0,  .502,  0}{NR} & \textcolor[rgb]{ 0,  .502,  0}{NR} & \textcolor[rgb]{ 0,  .502,  0}{NR} & \textcolor[rgb]{ 0,  .502,  0}{NR} & \textcolor[rgb]{ 0,  .502,  0}{NR} & \textcolor[rgb]{ 0,  .502,  0}{NR} & \textcolor[rgb]{ 0,  .502,  0}{NR} & \textcolor[rgb]{ 0,  .502,  0}{NR} & \textcolor[rgb]{ 1,  0,  0}{R} & \textcolor[rgb]{ 1,  0,  0}{R} & \textcolor[rgb]{ 0,  .502,  0}{NR} & \textcolor[rgb]{ 0,  .502,  0}{NR} & \textcolor[rgb]{ 0,  .502,  0}{NR} \\
\textcolor[rgb]{ .129,  .129,  .129}{\textbf{40}} & \textcolor[rgb]{ 0,  .502,  0}{NR} & \textcolor[rgb]{ 0,  .502,  0}{NR} & \textcolor[rgb]{ 0,  .502,  0}{NR} & \textcolor[rgb]{ 0,  .502,  0}{NR} & \textcolor[rgb]{ 0,  .502,  0}{NR} & \textcolor[rgb]{ 0,  .502,  0}{NR} & \textcolor[rgb]{ 0,  .502,  0}{NR} & \textcolor[rgb]{ 0,  .502,  0}{NR} & \textcolor[rgb]{ 0,  .502,  0}{NR} & \textcolor[rgb]{ 1,  0,  0}{R} & \textcolor[rgb]{ 1,  0,  0}{R} & \textcolor[rgb]{ 1,  0,  0}{R} & \textcolor[rgb]{ 0,  .502,  0}{NR} & \textcolor[rgb]{ 0,  .502,  0}{NR} & \textcolor[rgb]{ 0,  .502,  0}{NR} \\
\textcolor[rgb]{ .129,  .129,  .129}{\textbf{50}} & \textcolor[rgb]{ 0,  .502,  0}{NR} & \textcolor[rgb]{ 0,  .502,  0}{NR} & \textcolor[rgb]{ 0,  .502,  0}{NR} & \textcolor[rgb]{ 0,  .502,  0}{NR} & \textcolor[rgb]{ 0,  .502,  0}{NR} & \textcolor[rgb]{ 0,  .502,  0}{NR} & \textcolor[rgb]{ 0,  .502,  0}{NR} & \textcolor[rgb]{ 0,  .502,  0}{NR} & \textcolor[rgb]{ 0,  .502,  0}{NR} & \textcolor[rgb]{ 1,  0,  0}{R} & \textcolor[rgb]{ 1,  0,  0}{R} & \textcolor[rgb]{ 0,  .502,  0}{NR} & \textcolor[rgb]{ 0,  .502,  0}{NR} & \textcolor[rgb]{ 0,  .502,  0}{NR} & \textcolor[rgb]{ 0,  .502,  0}{NR} \\
\textcolor[rgb]{ .129,  .129,  .129}{\textbf{60}} & \textcolor[rgb]{ 0,  .502,  0}{NR} & \textcolor[rgb]{ 0,  .502,  0}{NR} & \textcolor[rgb]{ 0,  .502,  0}{NR} & \textcolor[rgb]{ 0,  .502,  0}{NR} & \textcolor[rgb]{ 0,  .502,  0}{NR} & \textcolor[rgb]{ 0,  .502,  0}{NR} & \textcolor[rgb]{ 0,  .502,  0}{NR} & \textcolor[rgb]{ 0,  .502,  0}{NR} & \textcolor[rgb]{ 0,  .502,  0}{NR} & \textcolor[rgb]{ 1,  0,  0}{R} & \textcolor[rgb]{ 1,  0,  0}{R} & \textcolor[rgb]{ 0,  .502,  0}{NR} & \textcolor[rgb]{ 0,  .502,  0}{NR} & \textcolor[rgb]{ 0,  .502,  0}{NR} & \textcolor[rgb]{ 0,  .502,  0}{NR} \\
\textcolor[rgb]{ .129,  .129,  .129}{\textbf{70}} & \textcolor[rgb]{ 0,  .502,  0}{NR} & \textcolor[rgb]{ 0,  .502,  0}{NR} & \textcolor[rgb]{ 0,  .502,  0}{NR} & \textcolor[rgb]{ 0,  .502,  0}{NR} & \textcolor[rgb]{ 0,  .502,  0}{NR} & \textcolor[rgb]{ 0,  .502,  0}{NR} & \textcolor[rgb]{ 0,  .502,  0}{NR} & \textcolor[rgb]{ 0,  .502,  0}{NR} & \textcolor[rgb]{ 0,  .502,  0}{NR} & \textcolor[rgb]{ 0,  .502,  0}{NR} & \textcolor[rgb]{ 0,  .502,  0}{NR} & \textcolor[rgb]{ 0,  .502,  0}{NR} & \textcolor[rgb]{ 0,  .502,  0}{NR} & \textcolor[rgb]{ 0,  .502,  0}{NR} & \textcolor[rgb]{ 0,  .502,  0}{NR} \\
\textcolor[rgb]{ .129,  .129,  .129}{\textbf{80}} & \textcolor[rgb]{ 0,  .502,  0}{NR} & \textcolor[rgb]{ 0,  .502,  0}{NR} & \textcolor[rgb]{ 0,  .502,  0}{NR} & \textcolor[rgb]{ 0,  .502,  0}{NR} & \textcolor[rgb]{ 0,  .502,  0}{NR} & \textcolor[rgb]{ 0,  .502,  0}{NR} & \textcolor[rgb]{ 0,  .502,  0}{NR} & \textcolor[rgb]{ 0,  .502,  0}{NR} & \textcolor[rgb]{ 0,  .502,  0}{NR} & \textcolor[rgb]{ 0,  .502,  0}{NR} & \textcolor[rgb]{ 0,  .502,  0}{NR} & \textcolor[rgb]{ 0,  .502,  0}{NR} & \textcolor[rgb]{ 0,  .502,  0}{NR} & \textcolor[rgb]{ 0,  .502,  0}{NR} & \textcolor[rgb]{ 0,  .502,  0}{NR} \\
\textcolor[rgb]{ .129,  .129,  .129}{\textbf{90}} & \textcolor[rgb]{ 0,  .502,  0}{NR} & \textcolor[rgb]{ 0,  .502,  0}{NR} & \textcolor[rgb]{ 0,  .502,  0}{NR} & \textcolor[rgb]{ 0,  .502,  0}{NR} & \textcolor[rgb]{ 1,  0,  0}{R} & \textcolor[rgb]{ 0,  .502,  0}{NR} & \textcolor[rgb]{ 0,  .502,  0}{NR} & \textcolor[rgb]{ 0,  .502,  0}{NR} & \textcolor[rgb]{ 0,  .502,  0}{NR} & \textcolor[rgb]{ 1,  0,  0}{R} & \textcolor[rgb]{ 0,  .502,  0}{NR} & \textcolor[rgb]{ 0,  .502,  0}{NR} & \textcolor[rgb]{ 0,  .502,  0}{NR} & \textcolor[rgb]{ 0,  .502,  0}{NR} & \textcolor[rgb]{ 0,  .502,  0}{NR} \\
            & \multicolumn{15}{c}{\textbf{VaR 5 \%}} \\
\textcolor[rgb]{ .129,  .129,  .129}{} & \textcolor[rgb]{ .129,  .129,  .129}{\textbf{S\&P500}} & \textcolor[rgb]{ .129,  .129,  .129}{\textbf{MMM}} & \textcolor[rgb]{ .129,  .129,  .129}{\textbf{ARP}} & \textcolor[rgb]{ .129,  .129,  .129}{\textbf{AMD}} & \textcolor[rgb]{ .129,  .129,  .129}{\textbf{AIG}} & \textcolor[rgb]{ .129,  .129,  .129}{\textbf{AFL}} & \textcolor[rgb]{ .129,  .129,  .129}{\textbf{ABT}} & \textcolor[rgb]{ .129,  .129,  .129}{\textbf{MSFTF}} & \textcolor[rgb]{ .129,  .129,  .129}{\textbf{AAPL}} & \textcolor[rgb]{ .129,  .129,  .129}{\textbf{AMZN}} & \textcolor[rgb]{ .129,  .129,  .129}{\textbf{BAC}} & \textcolor[rgb]{ .129,  .129,  .129}{\textbf{JPM}} & \textcolor[rgb]{ .129,  .129,  .129}{\textbf{JNJ}} & \textcolor[rgb]{ .129,  .129,  .129}{\textbf{ROM}} & \textcolor[rgb]{ .129,  .129,  .129}{\textbf{MA}} \\
\textcolor[rgb]{ .129,  .129,  .129}{\textbf{10}} & \textcolor[rgb]{ 1,  0,  0}{R} & \textcolor[rgb]{ 1,  0,  0}{R} & \textcolor[rgb]{ 0,  .502,  0}{NR} & \textcolor[rgb]{ 0,  .502,  0}{NR} & \textcolor[rgb]{ 1,  0,  0}{R} & \textcolor[rgb]{ 1,  0,  0}{R} & \textcolor[rgb]{ 1,  0,  0}{R} & \textcolor[rgb]{ 1,  0,  0}{R} & \textcolor[rgb]{ 1,  0,  0}{R} & \textcolor[rgb]{ 1,  0,  0}{R} & \textcolor[rgb]{ 1,  0,  0}{R} & \textcolor[rgb]{ 1,  0,  0}{R} & \textcolor[rgb]{ 0,  .502,  0}{NR} & \textcolor[rgb]{ 1,  0,  0}{R} & \textcolor[rgb]{ 1,  0,  0}{R} \\
\textcolor[rgb]{ .129,  .129,  .129}{\textbf{20}} & \textcolor[rgb]{ 0,  .502,  0}{NR} & \textcolor[rgb]{ 0,  .502,  0}{NR} & \textcolor[rgb]{ 0,  .502,  0}{NR} & \textcolor[rgb]{ 0,  .502,  0}{NR} & \textcolor[rgb]{ 1,  0,  0}{R} & \textcolor[rgb]{ 0,  .502,  0}{NR} & \textcolor[rgb]{ 0,  .502,  0}{NR} & \textcolor[rgb]{ 0,  .502,  0}{NR} & \textcolor[rgb]{ 1,  0,  0}{R} & \textcolor[rgb]{ 0,  .502,  0}{NR} & \textcolor[rgb]{ 1,  0,  0}{R} & \textcolor[rgb]{ 0,  .502,  0}{NR} & \textcolor[rgb]{ 0,  .502,  0}{NR} & \textcolor[rgb]{ 1,  0,  0}{R} & \textcolor[rgb]{ 1,  0,  0}{R} \\
\textcolor[rgb]{ .129,  .129,  .129}{\textbf{30}} & \textcolor[rgb]{ 0,  .502,  0}{NR} & \textcolor[rgb]{ 0,  .502,  0}{NR} & \textcolor[rgb]{ 0,  .502,  0}{NR} & \textcolor[rgb]{ 0,  .502,  0}{NR} & \textcolor[rgb]{ 1,  0,  0}{R} & \textcolor[rgb]{ 0,  .502,  0}{NR} & \textcolor[rgb]{ 0,  .502,  0}{NR} & \textcolor[rgb]{ 0,  .502,  0}{NR} & \textcolor[rgb]{ 1,  0,  0}{R} & \textcolor[rgb]{ 0,  .502,  0}{NR} & \textcolor[rgb]{ 1,  0,  0}{R} & \textcolor[rgb]{ 0,  .502,  0}{NR} & \textcolor[rgb]{ 1,  0,  0}{R} & \textcolor[rgb]{ 0,  .502,  0}{NR} & \textcolor[rgb]{ 0,  .502,  0}{NR} \\
\textcolor[rgb]{ .129,  .129,  .129}{\textbf{40}} & \textcolor[rgb]{ 0,  .502,  0}{NR} & \textcolor[rgb]{ 0,  .502,  0}{NR} & \textcolor[rgb]{ 0,  .502,  0}{NR} & \textcolor[rgb]{ 0,  .502,  0}{NR} & \textcolor[rgb]{ 0,  .502,  0}{NR} & \textcolor[rgb]{ 0,  .502,  0}{NR} & \textcolor[rgb]{ 0,  .502,  0}{NR} & \textcolor[rgb]{ 0,  .502,  0}{NR} & \textcolor[rgb]{ 1,  0,  0}{R} & \textcolor[rgb]{ 0,  .502,  0}{NR} & \textcolor[rgb]{ 0,  .502,  0}{NR} & \textcolor[rgb]{ 0,  .502,  0}{NR} & \textcolor[rgb]{ 0,  .502,  0}{NR} & \textcolor[rgb]{ 0,  .502,  0}{NR} & \textcolor[rgb]{ 0,  .502,  0}{NR} \\
\textcolor[rgb]{ .129,  .129,  .129}{\textbf{50}} & \textcolor[rgb]{ 0,  .502,  0}{NR} & \textcolor[rgb]{ 0,  .502,  0}{NR} & \textcolor[rgb]{ 0,  .502,  0}{NR} & \textcolor[rgb]{ 1,  0,  0}{R} & \textcolor[rgb]{ 0,  .502,  0}{NR} & \textcolor[rgb]{ 0,  .502,  0}{NR} & \textcolor[rgb]{ 0,  .502,  0}{NR} & \textcolor[rgb]{ 1,  0,  0}{R} & \textcolor[rgb]{ 1,  0,  0}{R} & \textcolor[rgb]{ 0,  .502,  0}{NR} & \textcolor[rgb]{ 0,  .502,  0}{NR} & \textcolor[rgb]{ 0,  .502,  0}{NR} & \textcolor[rgb]{ 1,  0,  0}{R} & \textcolor[rgb]{ 0,  .502,  0}{NR} & \textcolor[rgb]{ 0,  .502,  0}{NR} \\
\textcolor[rgb]{ .129,  .129,  .129}{\textbf{60}} & \textcolor[rgb]{ 0,  .502,  0}{NR} & \textcolor[rgb]{ 0,  .502,  0}{NR} & \textcolor[rgb]{ 0,  .502,  0}{NR} & \textcolor[rgb]{ 0,  .502,  0}{NR} & \textcolor[rgb]{ 0,  .502,  0}{NR} & \textcolor[rgb]{ 0,  .502,  0}{NR} & \textcolor[rgb]{ 0,  .502,  0}{NR} & \textcolor[rgb]{ 0,  .502,  0}{NR} & \textcolor[rgb]{ 1,  0,  0}{R} & \textcolor[rgb]{ 0,  .502,  0}{NR} & \textcolor[rgb]{ 0,  .502,  0}{NR} & \textcolor[rgb]{ 0,  .502,  0}{NR} & \textcolor[rgb]{ 1,  0,  0}{R} & \textcolor[rgb]{ 0,  .502,  0}{NR} & \textcolor[rgb]{ 0,  .502,  0}{NR} \\
\textcolor[rgb]{ .129,  .129,  .129}{\textbf{70}} & \textcolor[rgb]{ 0,  .502,  0}{NR} & \textcolor[rgb]{ 0,  .502,  0}{NR} & \textcolor[rgb]{ 0,  .502,  0}{NR} & \textcolor[rgb]{ 0,  .502,  0}{NR} & \textcolor[rgb]{ 0,  .502,  0}{NR} & \textcolor[rgb]{ 0,  .502,  0}{NR} & \textcolor[rgb]{ 0,  .502,  0}{NR} & \textcolor[rgb]{ 0,  .502,  0}{NR} & \textcolor[rgb]{ 1,  0,  0}{R} & \textcolor[rgb]{ 0,  .502,  0}{NR} & \textcolor[rgb]{ 0,  .502,  0}{NR} & \textcolor[rgb]{ 0,  .502,  0}{NR} & \textcolor[rgb]{ 1,  0,  0}{R} & \textcolor[rgb]{ 0,  .502,  0}{NR} & \textcolor[rgb]{ 0,  .502,  0}{NR} \\
\textcolor[rgb]{ .129,  .129,  .129}{\textbf{80}} & \textcolor[rgb]{ 0,  .502,  0}{NR} & \textcolor[rgb]{ 0,  .502,  0}{NR} & \textcolor[rgb]{ 0,  .502,  0}{NR} & \textcolor[rgb]{ 1,  0,  0}{R} & \textcolor[rgb]{ 0,  .502,  0}{NR} & \textcolor[rgb]{ 0,  .502,  0}{NR} & \textcolor[rgb]{ 0,  .502,  0}{NR} & \textcolor[rgb]{ 0,  .502,  0}{NR} & \textcolor[rgb]{ 1,  0,  0}{R} & \textcolor[rgb]{ 0,  .502,  0}{NR} & \textcolor[rgb]{ 0,  .502,  0}{NR} & \textcolor[rgb]{ 0,  .502,  0}{NR} & \textcolor[rgb]{ 1,  0,  0}{R} & \textcolor[rgb]{ 0,  .502,  0}{NR} & \textcolor[rgb]{ 0,  .502,  0}{NR} \\
\textcolor[rgb]{ .129,  .129,  .129}{\textbf{90}} & \textcolor[rgb]{ 0,  .502,  0}{NR} & \textcolor[rgb]{ 0,  .502,  0}{NR} & \textcolor[rgb]{ 0,  .502,  0}{NR} & \textcolor[rgb]{ 1,  0,  0}{R} & \textcolor[rgb]{ 0,  .502,  0}{NR} & \textcolor[rgb]{ 0,  .502,  0}{NR} & \textcolor[rgb]{ 0,  .502,  0}{NR} & \textcolor[rgb]{ 0,  .502,  0}{NR} & \textcolor[rgb]{ 1,  0,  0}{R} & \textcolor[rgb]{ 0,  .502,  0}{NR} & \textcolor[rgb]{ 0,  .502,  0}{NR} & \textcolor[rgb]{ 0,  .502,  0}{NR} & \textcolor[rgb]{ 1,  0,  0}{R} & \textcolor[rgb]{ 0,  .502,  0}{NR} & \textcolor[rgb]{ 0,  .502,  0}{NR} \\
\bottomrule
    \end{tabular}%
}    
  \label{tab_sigma_s6}%
\end{table}%

\end{document}